\documentclass[fleqn,usenatbib]{mnras}

\usepackage[T1]{fontenc}
\usepackage{ae,aecompl}

\usepackage{makecell}
\usepackage{subfig}

\usepackage{graphicx}	
\usepackage{amsmath}	
\usepackage{amssymb}	

\usepackage{newtxtext,newtxmath}

\title[Investigating GSJ]{Investigating the spectra and physical nature of galaxy scale jets}

\author[B. Webster et al.]{
B. Webster$^{1}$\thanks{E-mail: brendan.webster@open.ac.uk}, J. H. Croston$^{1}$, J. J. Harwood$^{2}$, R. D. Baldi$^{3,4}$, M. J. Hardcastle$^{2}$, \newauthor{ B. Mingo$^{1}$ and H. J. A. R{\"o}ttgering$^{5}$}
\\
$^{1}$School of Physical Sciences, The Open University, Walton Hall, Milton Keynes, MK7 6AA, UK\\
$^{2}$Centre for Astrophysics Research, School of Physics, Astronomy and Mathematics, University of Hertfordshire, Hatfield,\\AL10 9AB, UK\\
$^{3}$INAF - Istituto di Radioastronomia, via P. Gobetti 101, I-40129, Bologna, Italy\\
$^{4}$School of Physics and Astronomy, University of Southampton, Southampton, SO17 1BJ, UK\\
$^{5}$Leiden Observatory, Leiden University, PO Box 9513, 2300 RA Leiden, The Netherlands\\
}

\date{Accepted XXX. Received YYY; in original form ZZZ}

\pubyear{2021}

\begin{document}
\label{firstpage}
\pagerange{\pageref{firstpage}--\pageref{lastpage}}
\maketitle

\begin{abstract}
Previous studies have shown that physically small, low-luminosity radio galaxies, which we refer to as galaxy scale jets (GSJ), could potentially have a significant effect upon the host galaxy's evolution. Using $6\arcsec$ resolution images taken from the first release of the LOFAR Two Metre Sky Survey (LoTSS DR1), we identified a representative sample of nine potential GSJ for which we obtained high-resolution, 2-4 GHz data using the Karl G. Jansky Very Large Array (VLA). Using these data we aim to verify the GSJ nature of these sources as well as investigating the potential role of feedback. Our VLA images reveal a diversity of structures, confirm the hosts for four of the sources and find that a fifth is the first known example of a galaxy-scale remnant showing that some radio galaxies never grow beyond the GSJ stage. We also derive spectral ages and the first estimates of the lobe expansion speeds of GSJ. We find our GSJ have maximum spectral ages of $60$ Myr with most between about $5$ and $20$ Myr, consistent with being located along an evolutionary path joining compact sources and larger radio galaxies. We find lobe advance speeds a few times the local sound speed, with most GSJ predicted to be driving strong shocks into their environment and having a significant impact upon the host's evolution. Our discovery of a remnant GSJ, which will eventually transfer all of its energy directly into the local environment, represents an important and previously hidden aspect of AGN life cycles.
\end{abstract}

\begin{keywords}
galaxies: active--galaxies: evolution--galaxies: radio continuum--galaxies: jets--galaxies
\end{keywords}



\section{Introduction}
\label{sec:Introduction}

Feedback from radio-loud galaxies is believed to play an important role in galactic evolution \citep[e.g.][]{Bower2006BreakingFormation,Croton2006TheGalaxies,Fabian2012ObservationalFeedback,Hardcastle2020RadioJets}. According to this theory, energy is transferred from the radio jets, heating the local environment and reducing the rate at which material is able to cool and be accreted by the host galaxy. This reduced flow of material limits both the rate at which the host is able to form stars and the accretion rate of the central Active Galactic Nucleus (AGN).

At present, whilst there is robust evidence of radio galaxies heating their surrounding environments \citep[e.g.][]{Hardcastle2020RadioJets}, there is no direct evidence of the effect this has on star formation rates. Recently, some authors have started looking into physically small radio galaxies \citep[e.g.][]{Jarvis2019PrevalenceQuasars,Jimenez-Gallardo2019COMP2CAT:Universe,Webster2021ALOFAR} where observations of shock fronts \citep[][]{Croston2009High-energyA,Mingo2011MarkarianSeyfert,Mingo2012ShocksGalaxy} provide evidence of direct feedback between jets and the host ISM. These sources may provide an opportunity for better understanding how feedback from radio galaxies affects galaxy evolution.

There are also a growing number of studies looking into the effects of feedback from compact radio sources such as Gigahertz-Peaked Spectrum (GPS) and Compact Steep Spectrum (CSS) sources \citep[e.g.][]{Tadhunter2016TheObservations,Bicknell2018RelativisticGalaxies,ODea2021CompactSources}. In addition, there is a growing interest in the population of FR0 galaxies \citep{Baldi2015ARadio-galaxies,Baldi2018FR0Galaxies} and their potential role in feedback \cite[][]{Ubertosi2021ACenter}.

Using data from the first release of the LOFAR Two-metre Sky Survey \citep[LoTSS DR1, ][]{Shimwell2019TheRelease,Williams2019TheSurvey}, \citealt{Webster2021ALOFAR} (hereafter W21) recently discovered a population of 195 physically small, low-luminosity ($L_{150\text{ MHz}}\lesssim10^{25}\text{ W Hz}^{-1}$) radio galaxies with total linear sizes of 80 kpc or less. Known as Galaxy Scale Jets (GSJ), these sources have an average redshift of $\sim0.2$ with a maximum of 0.5 and are hosted by both spiral and elliptical galaxies whose properties were shown by W21 to be typical of the hosts of larger radio galaxies. 

Numerical simulations have shown that low power radio galaxies, like those identified in the W21 sample, have jets that decelerate quickly, entraining material and transferring most of their energy into the surrounding ISM \citep[][]{Mukherjee2020SimulatingDynamics,Rossi2020TheDeceleration,Massaglia2016MakingJets}. Simulations show that low-power jets are likely to deposit their energy throughout large regions of the host galaxy \citep[][]{Mukherjee2018The5063}.

Morphologically the W21 sample is a mix of FRI and FRII-type sources. Whilst many have integrated spectral indices typical of larger sources, some have relatively flat spectral indices. Whilst W21 could not derive ages for individual sources, the range of spectral indices were interpreted as showing that a wide range of ages were present in the sample. Despite being smaller with lower luminosity than the majority of previously studied radio galaxies, GSJ were found to contain enough energy to potentially have a significant effect on the evolution of the host. 

The low frequencies used by LOFAR mean that any core radio emission from the AGN is less prominent. This, combined with the 6 arcsec resolution of LOFAR \citep[][]{Shimwell2017TheRelease} mean that some of the optical host galaxies may have been misidentified by W21. The resolution of LOFAR also means that, due to the small size of these sources, the source morphology of many GSJ is often ill-defined.

Unlike the population of FR0, the W21 sample are all resolved, extended sources with radio emission a few tens of kpc in size. The sample is also distinct from the population of compact CSS/GPS sources, though an overlap is expected once smaller GSJ are discovered. At present, the relationship between these populations is uncertain. Though not universally accepted, the spectral turnover seen in CSS and GPS sources is often interpreted as these sources being the young progenitors of larger radio galaxies \citep[][]{ODea2021CompactSources}. The small physical size of GSJ therefore means that these sources could be located along any evolutionary path between compact sources and larger radio galaxies.

In W21, since it was not possible to account for the contribution of shocks, only lower limits on the potential energetic impact could be obtained. Powerful sources are frequently seen to be driving shocks that transfer energy to their surroundings \citep[e.g.][]{Fabian2003ARipples,Croston2007Shock3801,Forman2007FilamentsM87,Croston2009High-energyA,Randall2015AHistory}. However, it is currently unknown whether less-powerful GSJ expand fast enough to do the same. Related to this, the amount of time a source will spend in the GSJ stage depends upon its growth rate. Slowly expanding sources will spend more time in the GSJ stage, increasing the opportunity for the energy contained in the lobes to be transferred into the host's immediate environment.

Combining archival data with new, high resolution VLA images of a representative sample of GSJ, this paper aims to address some of these outstanding issues by:
\begin{itemize}
    \item Verifying the LoTSS DR1 optical host IDs and confirming the AGN nature of the radio emission.
    \item Better constraining the source morphology, including FR class, so as to relate the radio structures of small jets to those of traditional hundred-kpc scale radio galaxies.
    \item Using spectral information for each of our sources to measure the integrated spectral index and, where possible, constrain the break frequency so as to be able to draw conclusions about age and expansion speeds.
\end{itemize}

This paper is structured as follows. Section~\ref{sec:SampleSelection} describes the method used to find our representative sample of GSJ. Section~\ref{sec:Observations} describes the image processing. Section~\ref{sec:OpticalHosts} examines the images, looking for confirmation of the previously identified optical hosts. Section~\ref{sec:SpectralIndices} looks at the integrated spectral indices for our sample. Section~\ref{sec:SpectralIndexMaps} presents the spectral maps used in the ageing analysis in Section~\ref{sec:SpectralModels}. The lobe advance speeds are derived in section~\ref{sec:AverageAdvanceSpeeds} and there is a discussion of our results in Section~\ref{sec:Discussion}. Section~\ref{sec:Conclusions} summarises our findings. Throughout this paper we assume cosmological parameters of $\Omega_m=0.3$, $\Omega_{\Lambda} = 0.7$ and $H_0 = 70 \text{ km s}^{-1} \text{ Mpc}^{-1}$. We define the spectral index, $\alpha$, using the definition of radio flux density, $\text{S}_{\nu}\propto \nu^{-\alpha}$.

\section{Sample Selection}
\label{sec:SampleSelection}

\begin{figure}
    \centering
    \includegraphics[trim={10 0 40 40},clip=true,width=0.47\textwidth]{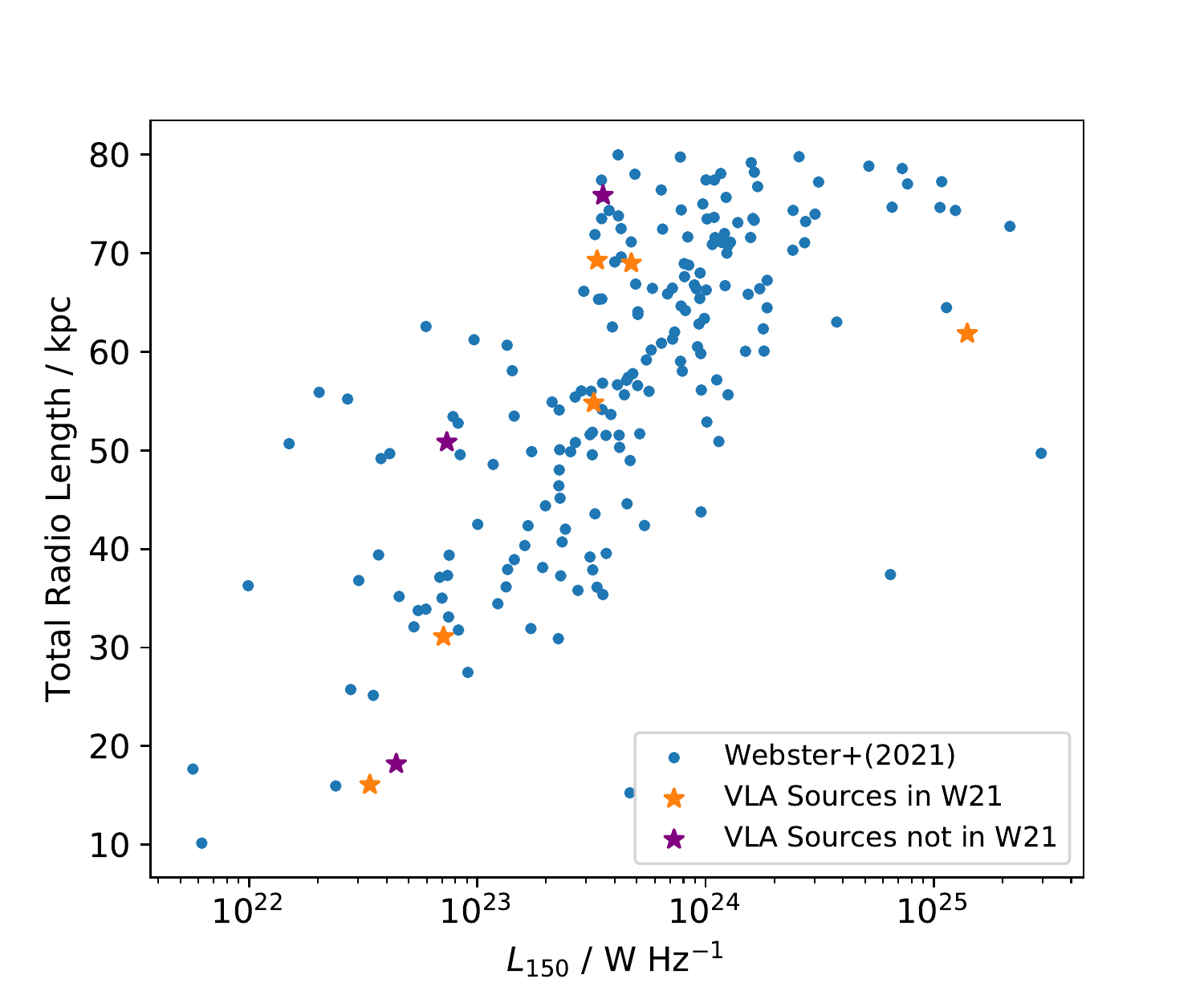}
    \caption{The distribution in the size-luminosity plane of the nine sources for which VLA measurements are presented in this paper compared to the GSJ sample of \citet{Webster2021ALOFAR}. Sources that are part of the W21 sample are highlighted in orange. Sources that are not in the W21 sample are highlighted in purple.}
    \label{fig:P-D_SourceSelection}
\end{figure}

This work aims to use a representative sub-sample of GSJ for high-resolution follow-up. To identify our sub-sample we adopted the more stringent visual selection process of W21, as it combines both automated and manual selection criteria to identify GSJ whose jets could be directly influencing the evolution of the host. In summary, W21 identify GSJ as being sources that are smaller than $80$ kpc, have a ratio of jet to host sizes (referred to as the jet:galaxy ratio) between 2 and 5 and are either classified as radio-loud according to \citet{Hardcastle2019Radio-loudSources} or are visually identified as radio
galaxies. Applying this selection process to $6\arcsec$ resolution images from a pre-release version of the LoTSS DR1 catalogue \citep[][]{Shimwell2019TheRelease},we identified a sub-sample of nine objects (Table~\ref{tab:SelectedSources}) that, in order to be representative of the wider GSJ population:
\begin{itemize}
    \item Covers the range of observed radio sizes (up to our $80$ kpc limit).
    \item Covers the range of observed radio luminosities (approximately $10^{22}$ to $10^{25} \text{ W Hz}^{-1}$ at $150$ MHz) 
    \item Includes both elliptical and spiral-hosted sources.
    \item Includes both FRI and FRII-like sources based on their appearance in the LoTSS DR1 images.
\end{itemize}
In addition, the limitations of the VLA had to be considered and so, assuming the canonical spectral index of $0.7$, only sources that would be both visible and resolved by the VLA were included in our sub-sample. The size and surface brightness of our sources meant that they should all be visible using the VLA's B configuration at S-band.

Whilst the nine sources were identified using a pre-release version of the LoTSS DR1 catalogue they are also representative of the 195 GSJ identified by W21 using the final DR1 catalogue. Figure~\ref{fig:P-D_SourceSelection} shows that the nine sources are spread across the range of sizes and luminosities observed in the W21 sample.

Six of the nine sources appear in the W21 sample. Subsequent optimisation of the W21 criteria to exclude star-forming galaxies and larger objects meant that the other three sources identified using the pre-release LoTSS DR1 catalogue, ILT J122037.67+473857.6, ILT J124627.85+520222.1 and ILT J125453.46+542923.4 are not included in the W21 sample. However, all three sources have combined jet lengths less than our 80 kpc limit with the first two sources also appearing in the radio-loud AGN catalogue of \citet{Hardcastle2019Radio-loudSources}. Both sources can therefore be considered as GSJ.

The third source, ILT J125453.46+542923.4, was initially included because its elongated morphology in the LoTSS DR1 images suggested that this was a genuine GSJ. However, subsequent improvements to the LoTSS image processing pipeline, included in the DR2 release \citep[][]{Tasse2021TheImaging}, no longer showed the elongated radio emission so that we no longer consider this source to be a genuine GSJ. The VLA data (see Figure~\ref{fig:VLAImages}) confirmed that this is not a genuine GSJ.

Of the nine sources sent for high-resolution follow-up, ILT J121847.41+520148.4 and ILT J112543.06+533112.4 are both highly unusual with spiral hosts and FRII-like morphologies. Consequently, both the $9$ per cent of spiral-hosted sources and the $\sim${}$10$ per cent of FRII-like sources in the W21 sample are marginally over-represented. However, in both cases the over-representation is small and we consider it useful to have multiple examples of these unusual sources, allowing us to better understand them.

\begin{table*}
\centering
\begin{tabular}{lccccc}
    \hline
	\multicolumn{1}{c}{LoTSS}&$z$&Radio&Host&$\log_{10}(L_{150})$&Length\\
	\multicolumn{1}{c}{Source Name}&&Morphology&Type&W MHz$^{-1}$&/ kpc\\
	\hline
	ILT J112543.06+553112.4$^a$&0.010&FR II&Spiral&22.5&16\\
	ILT J120326.64+545201.5&0.050&FR I&Elliptical&22.9&31\\
	ILT J120645.20+484451.1&0.065&FR I&Elliptical&23.5&69\\
	ILT J121847.41+520148.4$^a$&0.057&FR II&Spiral&23.7&69\\
	ILT J122037.67+473857.6&0.059&FR I&Elliptical&22.9&51\\
	ILT J124627.85+520222.1&0.067&FR I&Elliptical&23.6&76\\
	ILT J125453.46+542923.4$^b$&0.069&FR I&Elliptical&22.6&18\\
	ILT J130148.36+502753.3$^b$&0.120&FR I&Elliptical&23.5&55\\
	ILT J145604.90+472712.1&0.087&FR I&Elliptical&25.1&62\\
	\hline
\end{tabular}
\caption{Details of the sources identified for VLA follow-up. Spectroscopic redshifts are taken from the value-added LoTSS DR1 catalogue of \citet{Williams2019TheSurvey}. Luminosities and total radio lengths were derived using part of the LoMorph code of \citet{Mingo2019RevisitingLoTSS}.\newline$^a$ Host ID remains uncertain, $^b$Sources subsequently identified as not being GSJ.}
\label{tab:SelectedSources}
\end{table*}

\section{Observations}
\label{sec:Observations}

We observed all nine of our sources in the S-band (2 - 4 GHz) using the VLA B configuration in order to map the sources at a resolution of $\sim$2.1 arcseconds. Except for ILT J125453.46+542923.4 and ILT J130148.36+502753.3 which both have angular extents smaller than 30 arcseconds so that the B configuration captures all of their emission, we also obtained S-band images using the VLA C-configuration for the remaining seven sources to map the full extended structure at a resolution of 7 arcseconds. This ensured that the VLA captured the largest angular scales where emission was observed by LoTSS DR1. The C-configuration observations were taken on 2018 November 18 and the B-configuration observations were taken on 2019 April 27. Both configurations used the flux density calibrator 3C 286 and were taken under project code: 18B-083. Details of both observations are given in Table~\ref{tab:VLA_Observations}.

\begin{figure*}
    \centering
    \setlength{\tabcolsep}{1pt}
    \begin{tabular}{ccc}
        ILT J112543.06+553112.4&ILT J120326.64+545201.5&ILT J120645.20+484451.1\\
        \includegraphics[width=0.325\textwidth, trim={10 0 10 20}, clip=true]{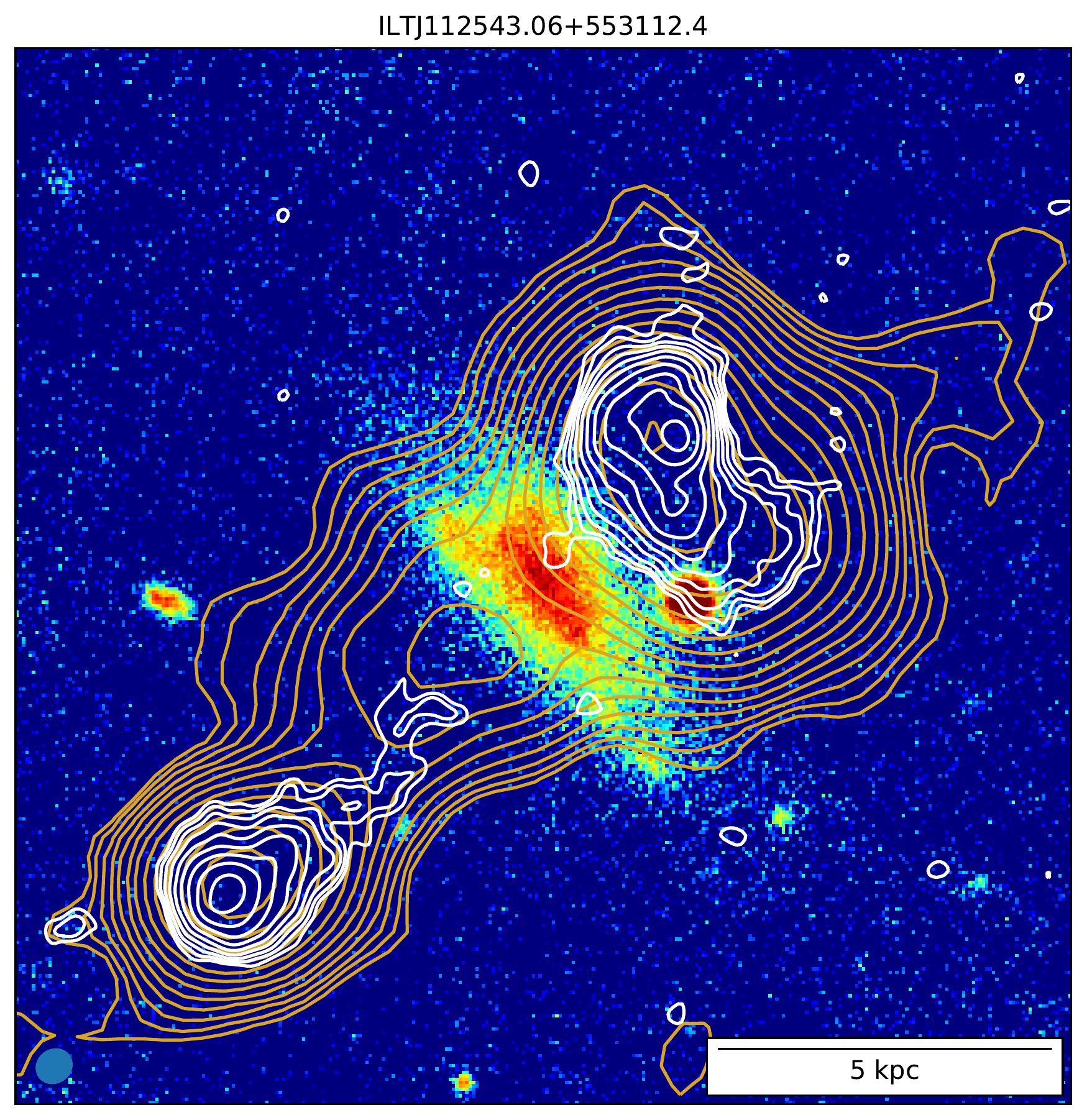} &
        \includegraphics[width=0.325\textwidth, trim={10 0 10 20}, clip=true]{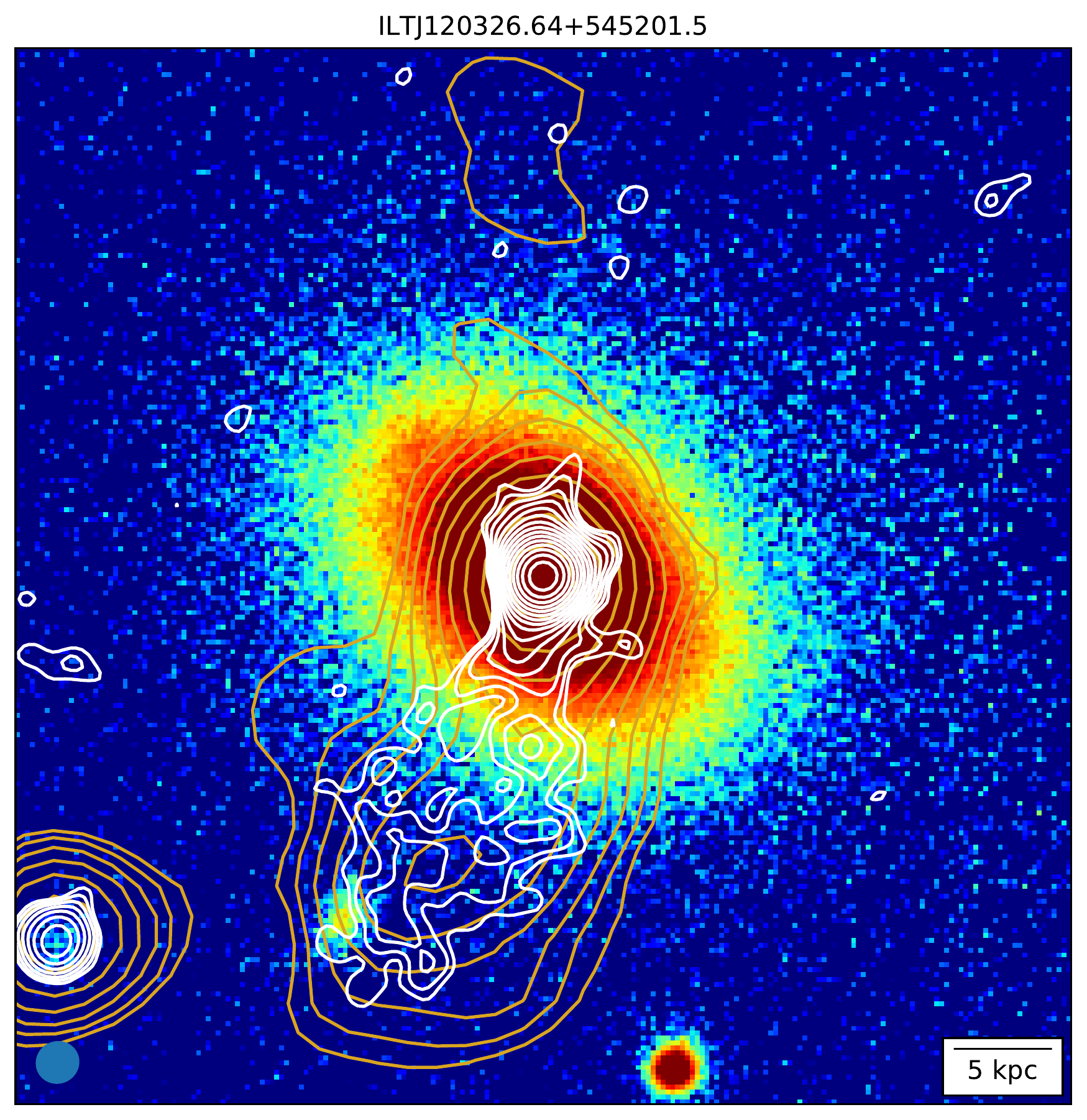} &
        \includegraphics[width=0.325\textwidth, trim={10 0 10 20}, clip=true]{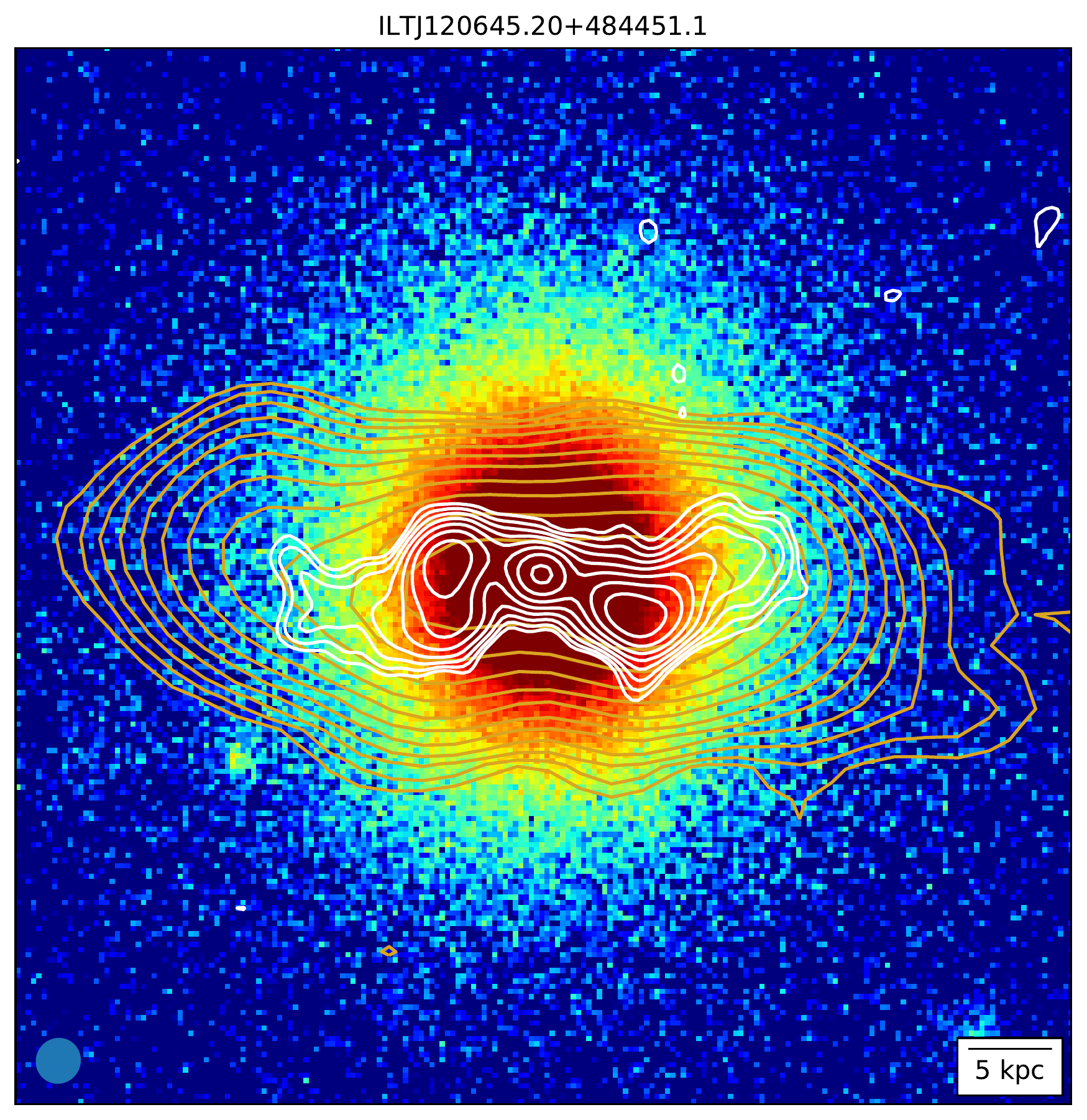}
        \\
        ILT J121847.41+520128.4&ILT J122037.67+473857.6&ILT J124627.85+520222.1\\
        \includegraphics[width=0.325\textwidth, trim={10 0 10 20}, clip=true]{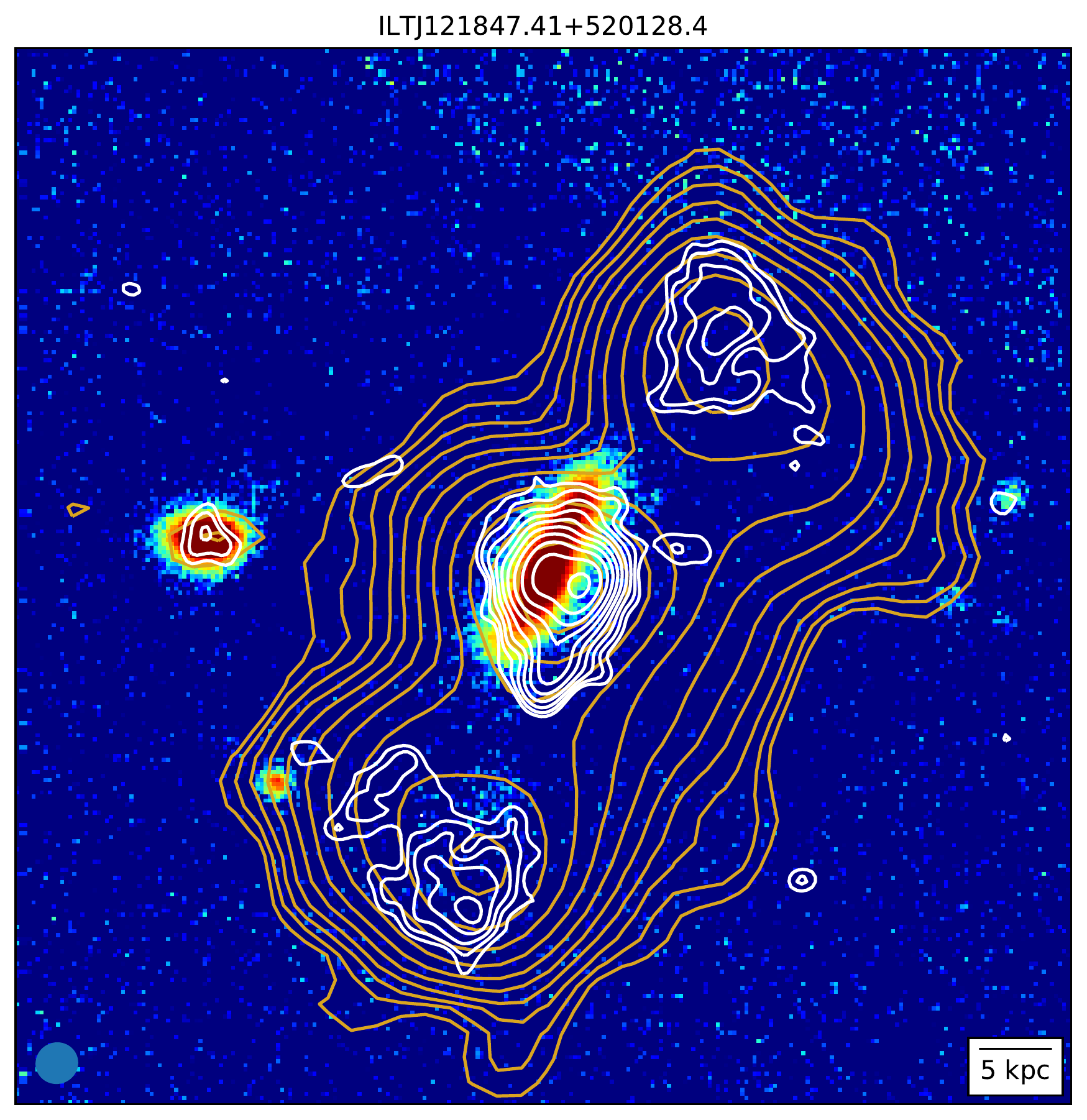} &
        \includegraphics[width=0.325\textwidth, trim={10 0 10 20}, clip=true]{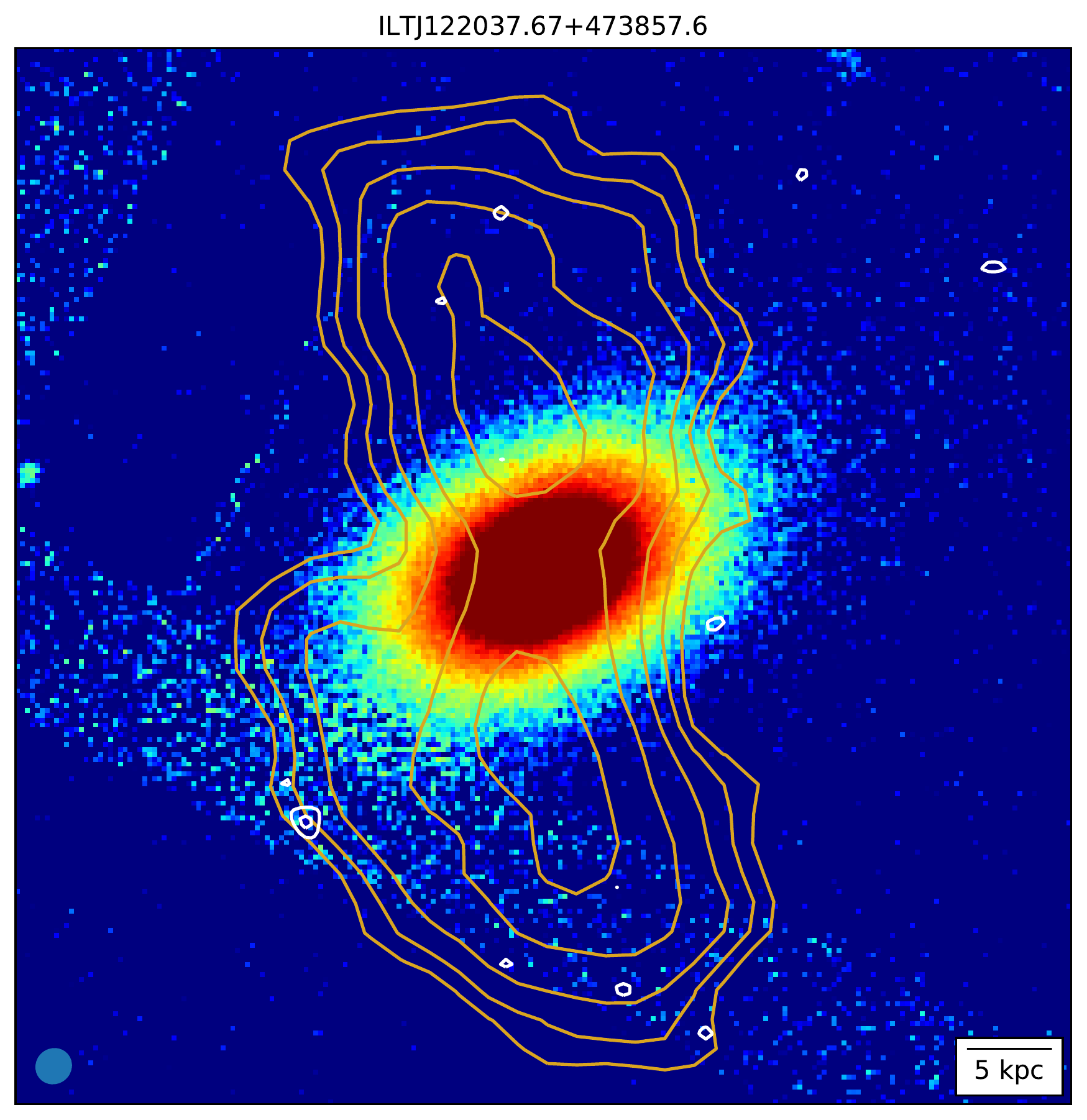} &
        \includegraphics[width=0.325\textwidth, trim={10 0 10 20}, clip=true]{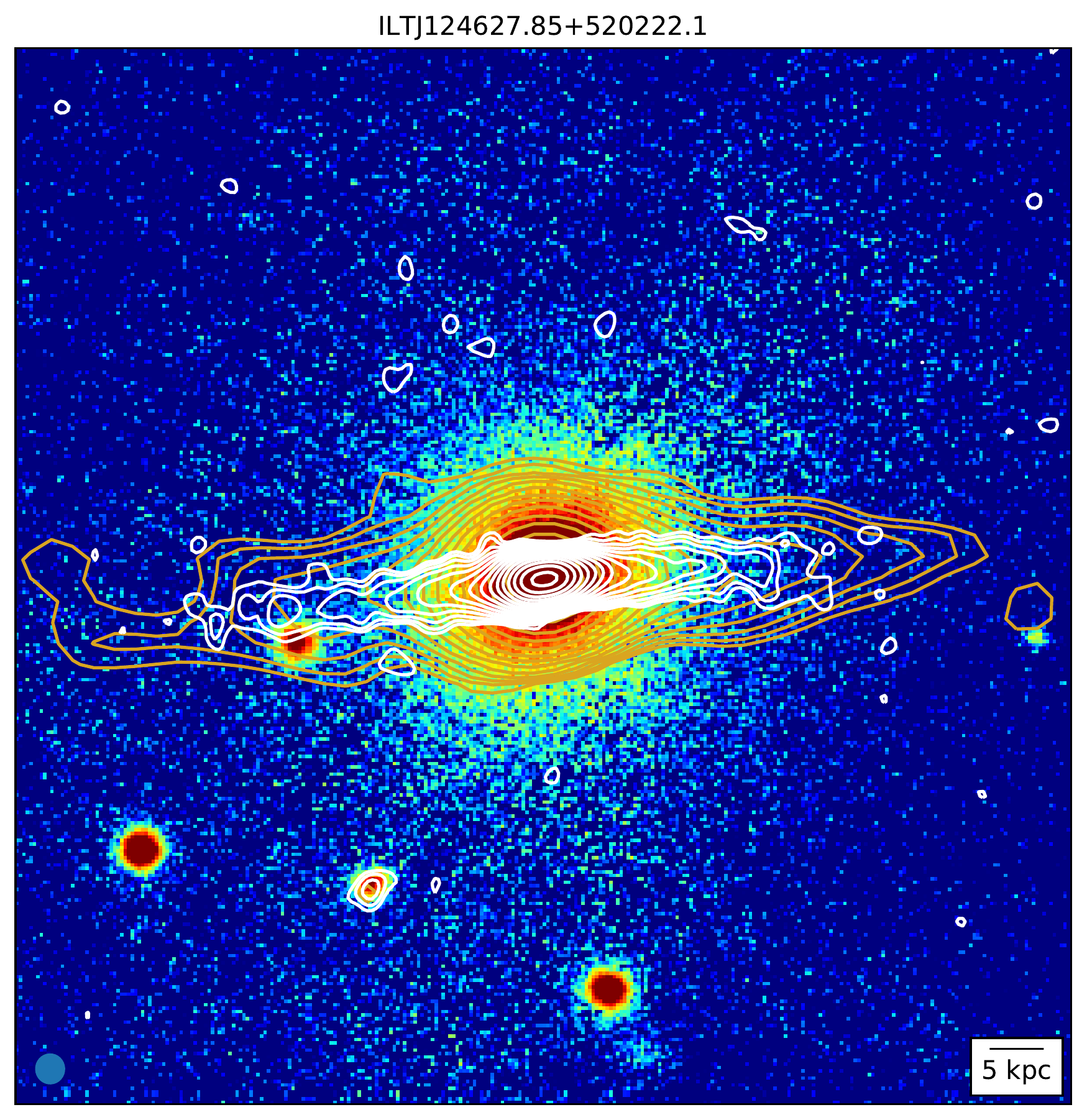}
        \\
        ILT J125453.46+542923.4&ILT J130148.36+502753.3&ILT J145604.90+472712.1\\
        \includegraphics[width=0.325\textwidth, trim={10 0 10 20},clip=true]{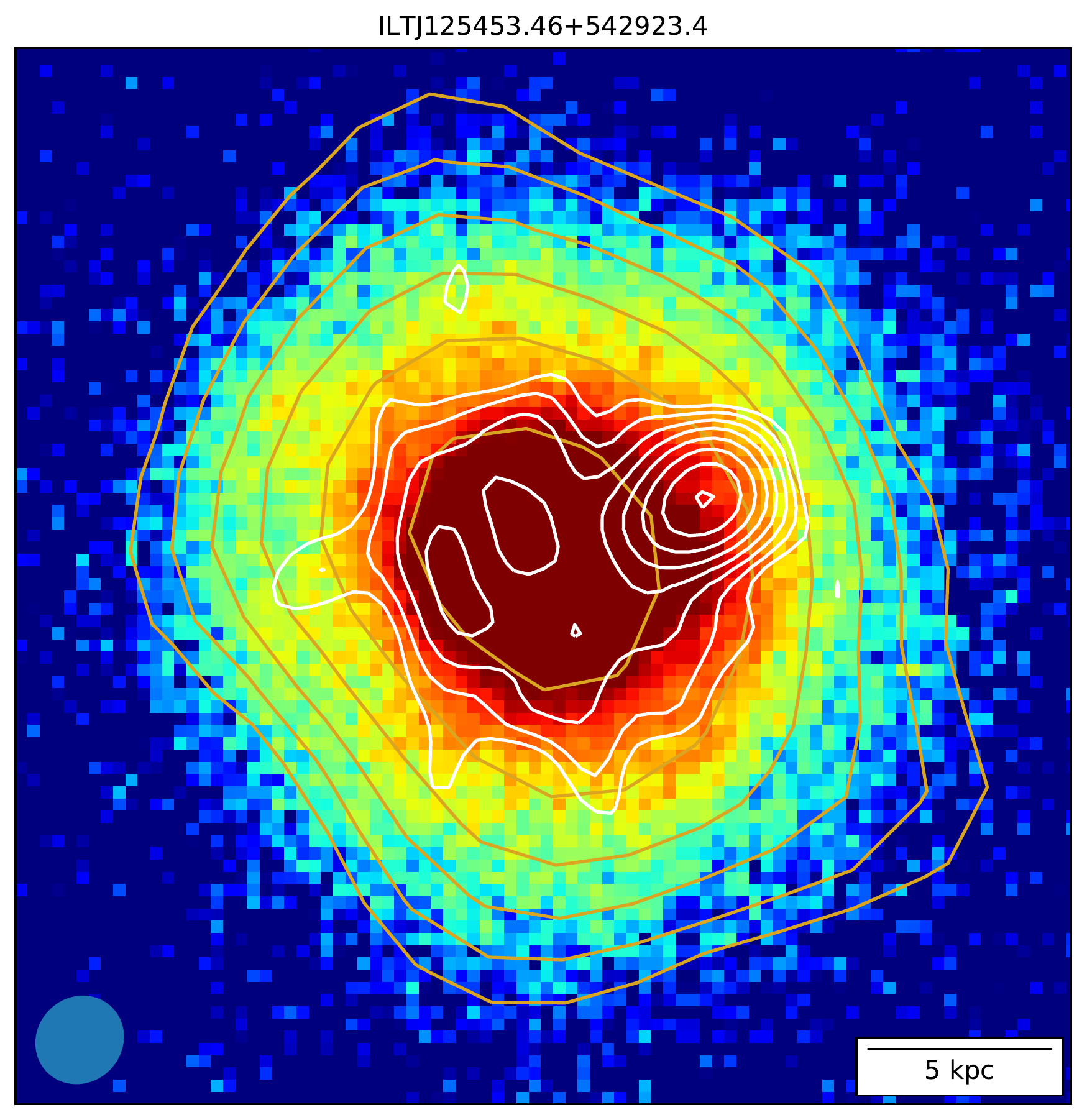} &
        \includegraphics[width=0.325\textwidth, trim={10 0 10 20}, clip=true]{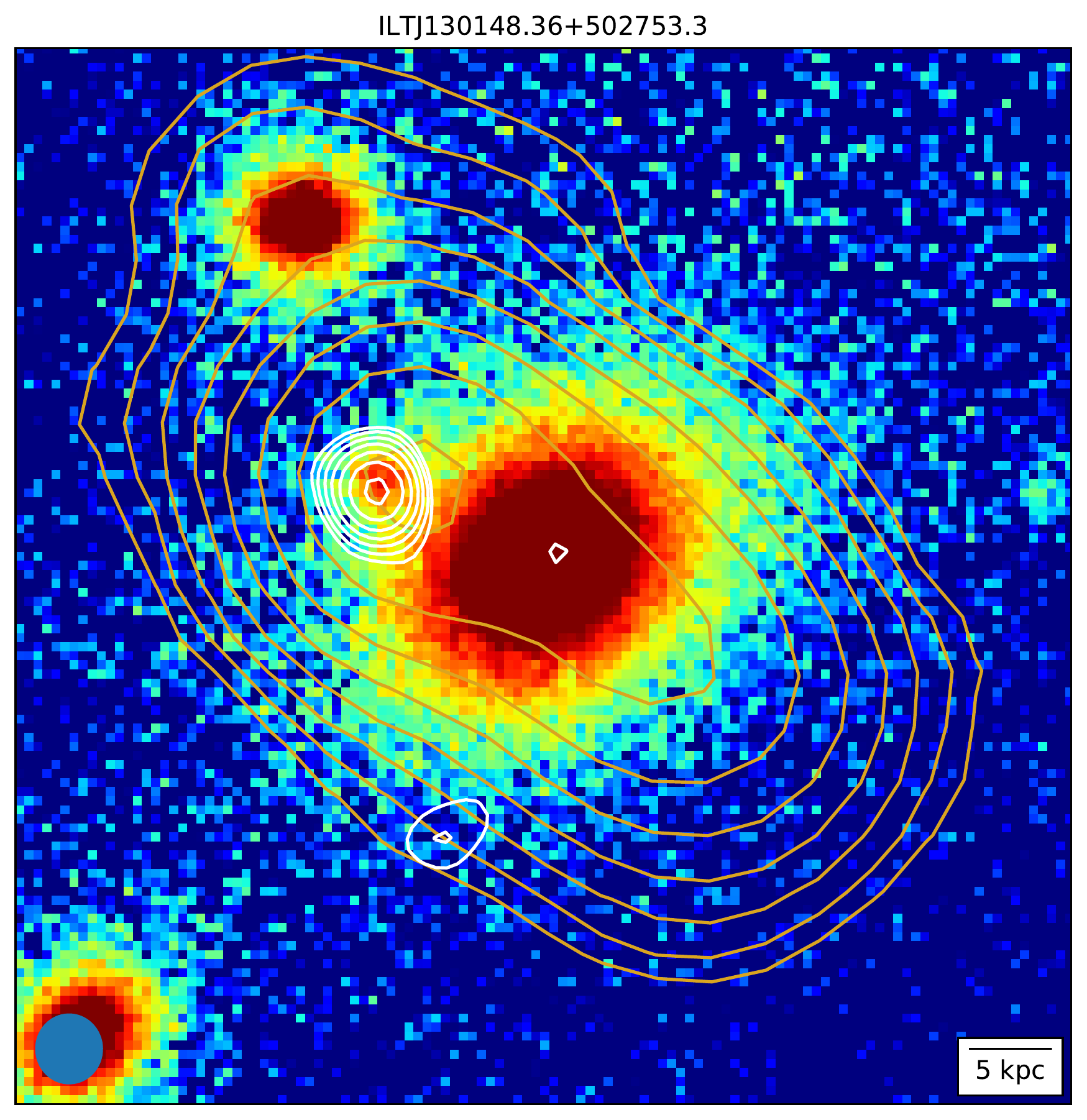} &
        \includegraphics[width=0.325\textwidth, trim={10 0 10 20}, clip=true]{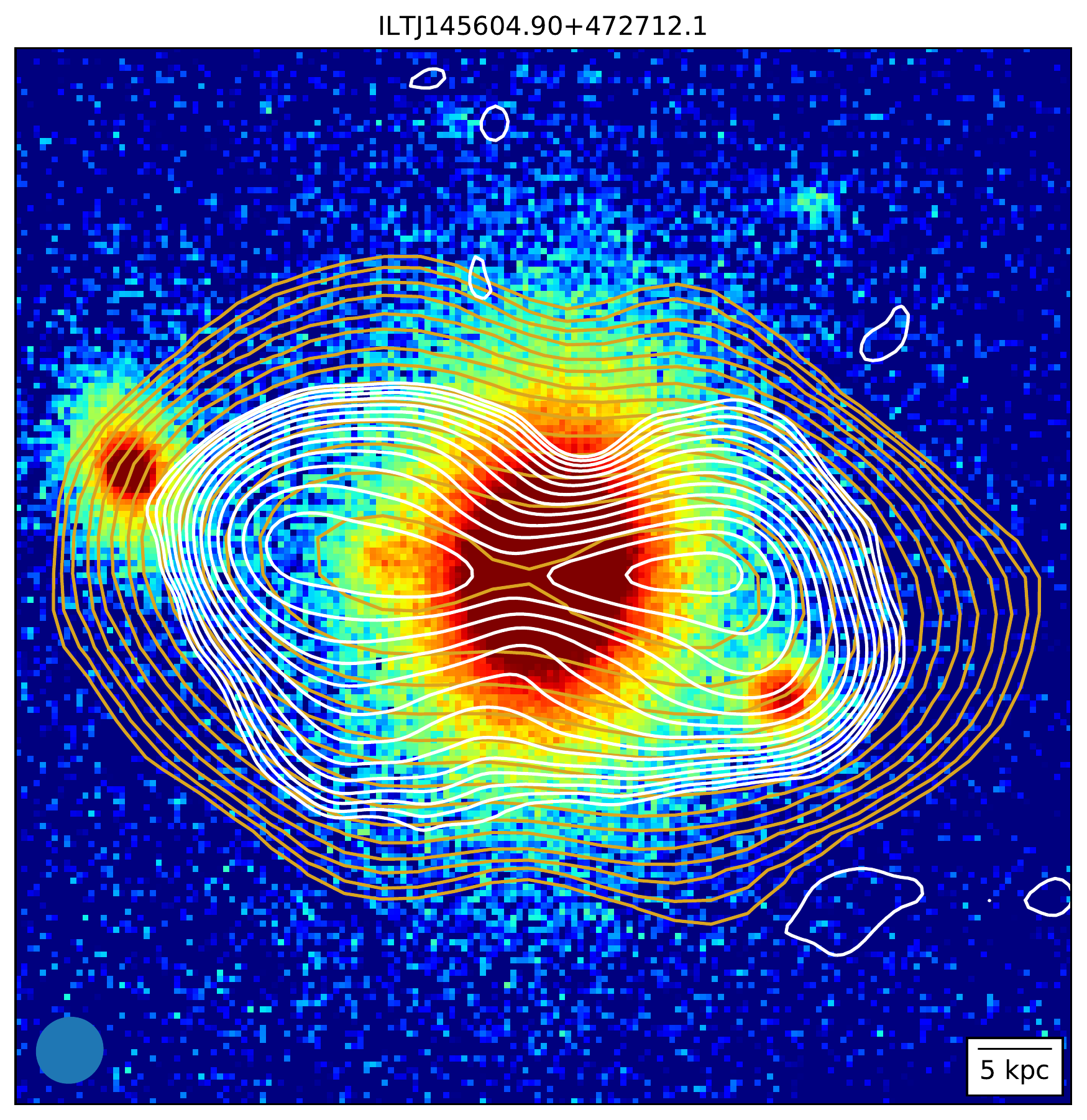}
    \end{tabular}
    \caption{Each image shows LoTSS DR2 contours in gold and VLA S-band contours from the combined B and C configuration images in white on top of the Pan-STARRS r-band optical image. The size of the VLA beam is shown in the bottom left and a scale bar is shown in the bottom right of each image.}
    \label{fig:VLAImages}
\end{figure*}

Both sets of data were reduced in the standard way using the National Radio Astronomy Observatory (NRAO) pipeline Common Observatory Software Applications (\textsc{CASA}; \citealt{McMullin2007CASAApplications}), v.5.4.1-32. Due to continuous Radio Frequency Interference (RFI) from satellite downlinks, all data in the 2.180-2.290 and 2.320-2.345 GHz ranges was flagged. Additional RFI flagging was done using the automated \emph{tfcrop} and \emph{rflag} routines as well as a manual check. Calibration was performed using a \citet{Perley2017AnGHz} model of our flux calibrator, 3C 286, to determine the flux density scale. Images were made using the \emph{tclean} algorithm with a multiscale deconvolver \citep[][]{Cornwell2008MultiscaleImages} and Briggs weighting with a robsut parameter of 0.5 \citep[][]{Briggs1995Http://www.aoc.nrao.edu/dissertations/dbriggs/}. The signal to noise ratio for our sources is low, meaning that self-calibration did not improve image quality and so we omitted this stage in our final images. We combined the B and C-configuration images using the \emph{statwt} task to compute the weightings of the images based on the RMS of the visibilities. The resulting images are shown in Figure~\ref{fig:VLAImages}.

The VLA image of ILT J125453.46+542923.4 (Figure~\ref{fig:VLAImages}) shows extended radio emission that is spatially correlated with the host suggesting a star-forming origin, though weak AGN activity is possible. The data also reveals a point source to the west of the galaxy centre. The location suggests this is not AGN-related and could be due to either a background source or a star forming region within the host. The VLA does not show the elongated emission visible in the DR1 pre-release data (see Section~\ref{sec:SampleSelection}) confirming the findings, using the LoTSS DR2 images, that this source is not a genuine GSJ. We do not consider this source any further.

Morphologically, our VLA images confirm the findings from the LoTSS DR2 images that ILT J112543.06+553112.4 and ILT J121847.41+520128.4 are FRII sources. Apart from ILT J130148.36+502743 where the VLA observes only a point source (see Section~\ref{sec:OpticalHosts-ILTJ130148}) and ILT J122037.67+473857.6 where no emission is detected, the VLA images also confirm that the remaining sources all appear FRI-like, though ILT J145604.90+472712 is somewhat borderline.

To calculate integrated flux density values we used DS9 to visually trace a polygon around the edge of the emission above the noise level. We then used the \textsc{radioflux} tool\footnote{Available at \url{https://www.github.com/mhardcastle/radioflux}} to find both the flux density and RMS noise in the image. Our images have an average RMS noise of $0.07$ mJy and an average peak signal to noise ratio of $26$. The final flux density error was found by combining the noise error in quadrature with the calibration uncertainty for the S-band, which is estimated at 5 per cent \citep[][]{Perley2017AnGHz}. We note that, within the error bounds, this gives the same flux density values as using \textsc{CASA}'s \emph{imfit} function to fit Gaussians to each separately identifiable emission region and aggregating the results. Our error values are used in Sections~\ref{sec:OpticalHosts} and \ref{sec:SpectralIndices} when comparing flux density values measured by different telescopes, but they are not used for the spectral index/ageing maps where, as described in Section~\ref{sec:SpectralIndexMaps}, a weighted least squares method is used to calculate the errors for each pixel. The results are listed in Table~\ref{tab:VLA_Observations}.

For those sources where separate core and lobe components were visible in the VLA images we used the \textsc{radioflux} tool to measure the flux densities of each component. Using these data we calculated spectral indices for each component allowing us to compare different regions from within the same image (see Section~\ref{sec:SpectralIndices-Component}).

\begin{table*}
\centering
\begin{tabular}{lccccccc}
    \hline
	\multicolumn{1}{c}{LoTSS}&WENSS&NVSS&Phase&On Source&Total Flux&$L_{\text{3GHz}}$/\\
	\multicolumn{1}{c}{Source Name}&Id&Id&Calibrator&Time (B/C)&3GHz/mJy&W Hz$^{-1}$\\
	\hline
	ILT J112543.06+553112.4$^a$&B1122.9+5547&112542+553113&J1035+5628&9m48s/9m40s&$7.74\pm0.39$&$1.6\pm0.1\times10^{21}$\\
	ILT J120326.64+545201.5&&120326+545203&J1219+4829&23m33s/9m40s&$4.47\pm0.23$&$2.7\pm0.1\times10^{22}$\\
	ILT J120645.20+484451.1&&120645+484444&J1219+4829&19m45s/9m40s&$2.80\pm0.15$&$2.8\pm0.1\times10^{22}$\\
	ILT J121847.41+520148.4$^a$&B1216.3+5218&121847+520131&J1219+4829&9m48s/8m55s&$3.71\pm0.20$&$2.9\pm0.2\times10^{22}$\\
	ILT J122037.67+473857.6&&&J1219+4829&88m36s/9m35s&$\lesssim 0.3$&\\
	ILT J124627.85+520222.1&B1244.1+5218&124627+520222&J1219+4829&19m45s/9m40s&$7.21\pm0.36$&$7.8\pm0.4\times10^{22}$\\
	ILT J125453.46+542923.4$^b$&&&J1219+4829&88m30s/-&$0.51\pm0.03$&$5.8\pm0.3\times10^{21}$\\
	ILT J130148.36+502753.3$^b$&&&J1219+4829& 11m45s/-&$0.53\pm0.04$&$2.0\pm0.2\times10^{22}$\\
	ILT J145604.90+472712.1&B1454.3+4739&145604+472712&\makecell{J1549+5038(B)\\J1438+6211(C)}&7m45s/9m40s&$138.40\pm9.92$&$2.6\pm0.1\times10^{24}$\\
	\hline
\end{tabular}
\caption{The corresponding ID's from the LoTSS, WENSS and NVSS catalogues along with details of the VLA observations taken under project code 18B-083. $^a$ Host ID remains uncertain, $^b$Sources subsequently identified as not being GSJ.}
\label{tab:VLA_Observations}
\end{table*}

The source sizes in the LoTSS catalogue were derived from the ellipses used by the Python Blob Detector and Source Finder \citep[][]{Mohan2015PyBDSF:Finder}. As discussed in \citet{Mingo2019RevisitingLoTSS}, this tends to underestimate the size and flux densities of the smallest sources, such as GSJ. Rather than using the catalogued values, we therefore applied the method above to the LoTSS DR2 images finding flux densities for both the total source as well as for any sub-components. Errors were calculated as above using the calibration error for LoTSS which is conservatively estimated at 20 per cent \citep[][]{Shimwell2017TheRelease}. Applying this technique we found flux densities for our sources that are $1-40$ per cent higher than those listed in LoTSS DR1 with the majority being $10-20$ per cent higher, confirming the findings of \citet{Mingo2019RevisitingLoTSS}. Throughout this paper we use the \textsc{radioflux} measured values.

\subsection{WENSS and NVSS Data}
\label{sec:TheData-WENSSNVSSandFIRST}

To study the spectral properties of our GSJ we combined the LoTSS and VLA data with 327 MHz and 1.4 GHz data from the Westerbork Northern Sky Survey (WENSS) \citep[][]{Rengelink1997TheWENSS} and NRAO VLA Sky Survey (NVSS) \citep[][]{Condon1998TheSurvey}, taken using the Westerbork and VLA telescopes respectively. The maximum positional uncertainty for faint sources is $7\arcsec$ in NVSS and $5-10\arcsec$ in WENSS. To account for this we used a $20\arcsec$ radius when cross-matching with both catalogues. Table~\ref{tab:VLA_Observations} lists the catalogue IDs for those sources where a match was found.

WENSS has a $54\arcsec$ resolution with a sensitivity of $18\text{ mJy}$ whilst NVSS has a $45\arcsec$ resolution with a sensitivity of $2.5\text{ mJy}$. The large beam sizes of these surveys (relative to the LoTSS and VLA images) mean it is possible that there is contamination from secondary sources. We therefore examined images from the LoTSS survey, but found that none of our sources have secondary sources close enough to affect either the NVSS or WENSS catalogued values.

\section{Optical Hosts}
\label{sec:OpticalHosts}

One of the reasons for obtaining the high resolution VLA images (Figure~\ref{fig:VLAImages}) was to verify the optical host listed in the LoTSS DR1 catalogue. The VLA data shows a radio core located within $0.15\arcsec$ of the centre of the host galaxy, confirming the host ID as well as an AGN core for three of our sources:
\begin{itemize}
    \item ILT J120326.64+545201.5
    \item ILT J120645.20+484451.1
    \item ILT J124627.85+520222.1
\end{itemize}
One of the sources, ILT J122037.67+473857.6, was undetected by the VLA, meaning that the host ID could not be confirmed, though in this particular case we are happy to retain the source as the LoTSS image leaves little doubt the host has been correctly identified. The remaining four sources are all considered individually below.

\subsection{ILT J112543.06+553112.4}
\label{sec:OpticalHosts-ILTJ112543}

\begin{figure}
    \centering
    \includegraphics[trim={10 0 0 0},clip=true,width=0.47\textwidth]{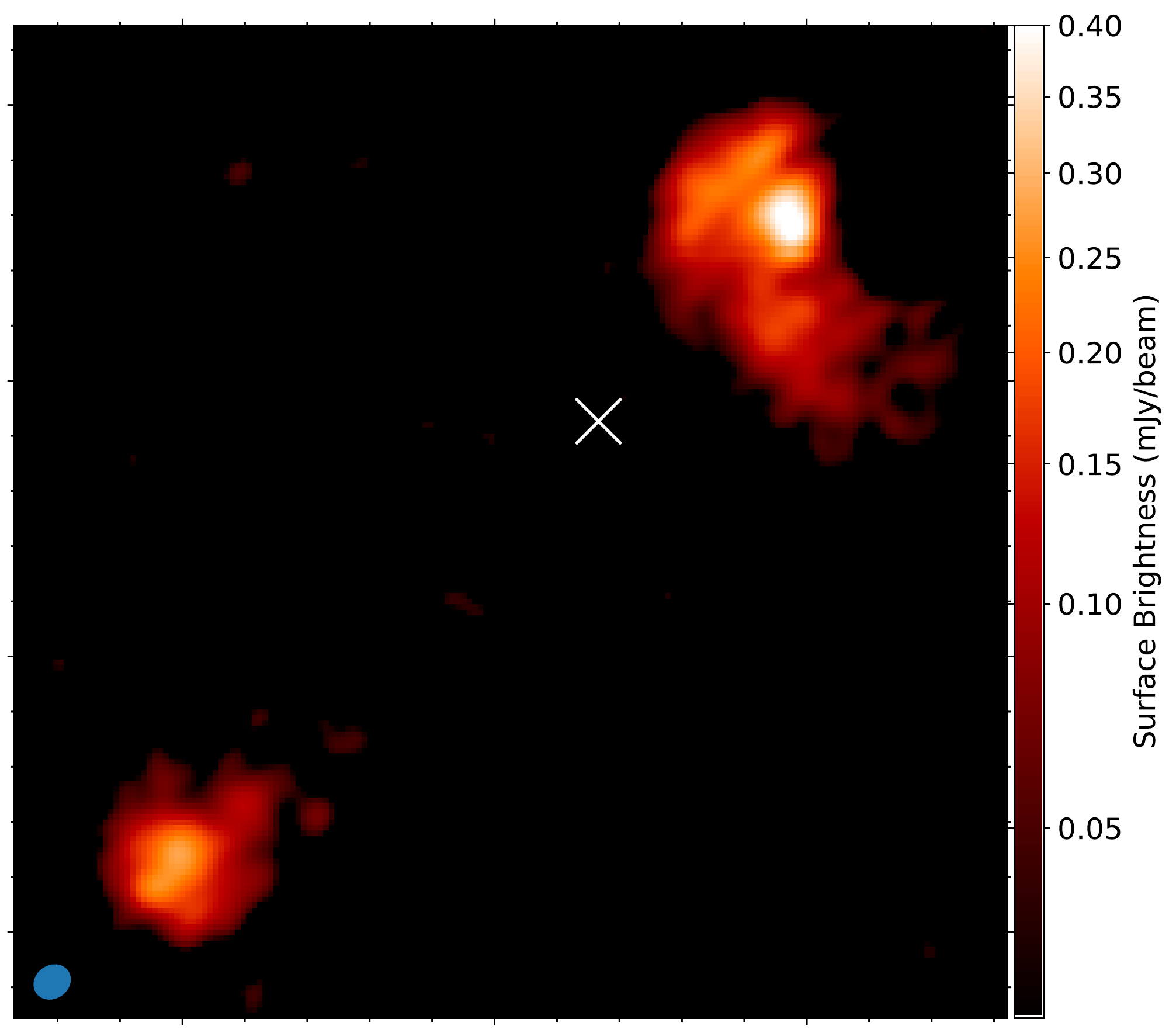}
    \caption{The source ILT J112543.06+553112.4 as imaged by the VLA in B-configuration. The RMS is $0.012$ mJy beam$^{-1}$, the VLA beam is shown in the bottom left and the position of the optical host is indicated by the white cross. A bright hotspot can be seen in the north west with a fainter hotspot visible in the south east lobe.}
    \label{fig:ILTJ112543BConfig}
\end{figure}

The VLA contours for this source appear as two disconnected islands of radio emission, one to the north west and the other to the south east. There is no observable radio core. Most easily seen using the B-configuration VLA data (Figure~\ref{fig:ILTJ112543BConfig}), the northern island contains what appears to be a bright hot spot, whilst the southern island has a less pronounced hot spot confirming the FRII nature of this source. The hotspot in the north west lobe has a surface brightness approximately 25 per cent brighter than the hotspot in the south east. Whilst the centre of the LoTSS-identified host galaxy is significantly closer to the north east lobe, it does lie directly along the line joining the two hotspots.

The FRII nature of the source leaves no doubt that we are observing radio emission caused by an AGN. The observed asymmetries in jet lengths and lobe morphology could be attributable to asymmetries in the host environment through which the jets have passed \citep[][]{Wagner2011RelativisticGalaxies,Gaibler2014AsymmetriesMedia} or to orientation \citep[see for example][]{Harwood2020UnveilingHyMoRS}. Asymmetries in jet lengths are not uncommon in FRII radio galaxies \citep[e.g.][]{Mullin2008Observed1.0}.

Unfortunately, the absence of a detected radio core means that this may be a chance line of sight effect, with the true host remaining unseen in the background. However, the absence of an alternative host means that the LoTSS DR1 host ID remains plausible, as discussed in W21.

\subsection{ILT J121847.41+520128.4}
\label{sec:OpticalHosts-ILTJ121847}

The VLA radio contours for this source show a peak `core' emission lying on a line joining the two lobes. The `core' is offset from the host galaxy centre where an AGN would most likely be located by $\sim1.4\arcsec$ suggesting that the host may have been misidentified. The `core' is also not the Gaussian shape typically associated with radio cores, being slightly elongated and extending to the south west of the host.

We therefore investigate the possibility that the offset `core' is not AGN related and is due entirely to star formation activities. Using a spectral index of $\alpha=0.7$ to convert the 1.4 GHz Radio-SFR relation of \citet{Gurkan2018LOFAR/H-ATLAS:Relation} to 3 GHz results in the relation:
\begin{equation}
    \text{log}_{10}(L_{\text{3 GHz}}) = 0.87\pm0.01 \times \log_{10}(\Psi) + 21.09\pm0.03
\end{equation}
where $\Psi$ is the star formation rate in solar masses per year. According to this relation, the largest SFR published in SDSS-DR14, calculated at $97.5$ per cent of the probability distribution and derived using the methods of \citet{Brinchmann2004TheUniverse}, would result in a 3 GHz radio luminosity of $\text{log}_{10}(L_{\text{3 GHz}})= 21.92\pm0.04 \text{W Hz}^{-1}$. Lower star formation rates give even lower predicted radio luminosities, all of which are less than the observed $22.56\pm0.02 \text{ W Hz}^{-1}$ seen in the core region.

Therefore, assuming all the observed `core' flux comes from the host galaxy, star formation alone cannot account for the observed emission, suggesting an AGN is present. This is reinforced by the elongated shape extending only to the south as, if this were due to starforming winds, a similar extension to the north would be expected. We therefore consider other explanations for the observed morphology. 

One possible explanation would be if the host galaxy were moving through a relatively dense environment causing bending of the underlying jets and elongation of the central core. Using the group catalogue of \citet{Tempel2014Flux-Estimation}, the host is the second most luminous galaxy in a small group of 4, situated at an approximate distance of $0.2\text{ Mpc}$ from the brightest group galaxy. Such a small group would imply an extremely sparse environment suggesting this is not the cause of the elongated emission.

An alternative possibility that we cannot discount is that we are observing a galaxy merger. Whilst a merging galaxy would explain the orientation of the jets, there is no suggestion in the PanSTARRS images of a secondary galaxy, though it may be obscured by the foreground host. If correct, the redshift of the two galaxies would be the same and so the GSJ nature of the source would remain.

Since spiral-hosted radio galaxies are extremely rare, we must also consider the possibility that the host has been misidentified and that the true host is an unseen background galaxy. However, since no secondary galaxy is seen, the PanSTARRS limiting apparent magnitude of 24 means the secondary galaxy would have to be extremely faint and/or obscured by the foreground galaxy. In this scenario it is also possible that the foreground galaxy also contains either a weak AGN and/or star-formation related radio emission that is resulting in the elongated radio emission of the core region. For the remainder of the paper we continue to consider this source as a GSJ, but note that future, deeper optical/IR images may identify alternative potential hosts.

\subsection{ILT J130148.36+502753.3}
\label{sec:OpticalHosts-ILTJ130148}

The image shown in Figure~\ref{fig:VLAImages} shows a large foreground galaxy identified as the host by the LoTSS DR1 survey with a smaller object located to the north-east about which the VLA has detected strong emission. The VLA data does not show any emission coming from the host identified by LoTSS. The emission observed by the VLA does not appear to be jetted in nature.

We could not find any entry for the source of the VLA emission in either NED or SIMBAD. SDSS does detect the source of the VLA emission, though the pipelines used to find spectroscopic and photometric redshifts failed for this object. PanSTARRS also detects this source, though photometric redshifts are not yet available. It is therefore unclear whether the source of the VLA emission is a background galaxy or quasar, a satellite of the larger galaxy or even a bright component of the larger galaxy. 

Irrespective of the true nature of the source of the VLA emission, it seems likely that at least some of the emission seen by LOFAR is due to this source. Equally, the elongated morphology of the emission seen by LOFAR, combined with the lack of any peak at the location of the VLA emission suggests that at least some of the emission seen by LOFAR is caused by activity in the foreground galaxy with the relative contributions of the two sources being impossible to distinguish. Whilst it remains possible there is galaxy-scale jet behaviour in the LoTSS-catalogued source, the uncertainty in the GSJ nature of this source means that we do not consider it any further.

\subsection{ILT J145604.90+472712.1}
\label{sec:OpticalHosts-ILTJ145604}

Although the VLA image does not show a radio core for this object, it does still show two separate regions of emission emanating to the west and east of the host galaxy consistent with jetted activity. The core location from which these two regions of radio emission originate is consistent with the host identification and so we are confident that the host has been correctly identified and that this is a GSJ.

\section{Spectral Indices}
\label{sec:SpectralIndices}

\subsection{Integrated Spectra}
\label{sec:SpectralIndices-Integrated}

In order to characterise the bulk properties of the radio emission, compare our sources to other samples of radio galaxies and confirm the findings of W21, we derived the integrated $150$ MHz to $3$ GHz spectral indices. For all our sources, the angular sizes of the detected LOFAR structures are smaller than the largest angular scales that our VLA observations are sensitive to and so these results are expected to be robust. The integrated spectral indices are listed in Table~\ref{tab:LoTSS_VLA_SpectralIndices}.

\begin{table}
    \centering
    \begin{tabular}{lc}
        \hline
        \multicolumn{1}{c}{Source Name}& $\alpha^{3\text{ GHz}}_{150\text{ MHz}}$\\
        \hline
        ILT J112543.06+553112.4$^a$& $1.0\pm0.1$\\
        ILT J120326.64+545201.5 & $0.4\pm0.1$\\
        ILT J120645.20+484451.1 & $0.8\pm0.1$\\
        ILT J121847.41+520128.4$^a$ & $0.9\pm0.1$\\
        ILT J122037.67+473857.6 & $>2.4$\\
        ILT J124627.85+520222.1 & $0.5\pm0.1$\\
        ILT J145604.90+472712.1 & $0.6\pm0.1$\\
        \hline
    \end{tabular}
    \caption{The integrated spectral index for our sources (excluding the two non-GSJ sources) using the 150 MHz LoTSS and 3 GHz VLA observations. $^a$ Host ID remains uncertain.}
    \label{tab:LoTSS_VLA_SpectralIndices}
\end{table}

ILT J122037.67+473857.6 is undetected in the VLA observations and so, using three times the RMS as the detection threshold, an upper limit is given. This source has an integrated spectral index far steeper than the criterion of $1.3$ traditionally used to identify remnant sources \citep[e.g.][]{Brienza2017SearchField,Cohen2007TheSurvey,Parma2007InUniverse}. The steep spectral index, combined with the lack of any visible core in the LoTSS data, shows that this is a remnant source. According to the group catalogue of \citet{Tempel2014Flux-Estimation}, this source is located within a group of 9 galaxies, suggesting a low-density environment. Since remnant sources outside of cluster environments are extremely rare \citep[][]{Brienza2016LOFARRedshift} this is a particularly interesting object.

Another object, ILT J120326.64+545201.5 has a spectral index below the value of 0.5 typically used to identify flat spectra. When observed with the VLA, this source has comparatively weak radio emission coming from the lobes (Figure~\ref{fig:VLAImages}) with over $75$ per cent of the total flux coming from the core (over $45$ per cent when viewed with LOFAR). As a result its integrated spectra, particularly at higher frequencies, is dominated by the core region. The core-dominance and flat spectra indicate that this source is strongly affected by absorption; either Free-Free Absorption (FFA) or Synchrotron Self Absorption (SSA) \citep[][]{ODea2021CompactSources}.

All the other sources in our sample have integrated spectral indices that are typical of larger radio galaxies \citep[for example the 3CRR sample of][]{Laing1983BrightGalaxies}. Our results therefore comfirm the findings of W21.

We also wish to see how the integrated spectral index varies with frequency and whether there is any low-frequency flattening as would be expected for compact sources. To do this we compared the radio flux densities from the LoTSS, WENSS and NVSS catalogues (where available) with the values derived from our VLA images. Measured using the LoTSS images, ILT J112543.06+553112.4 has the largest angular size of all our sources at $70\arcsec$, this is substantially less than the maximum angular scale of NVSS ($20\arcmin$) and WENSS (over $1\degr$) so that spatially these surveys can observe all the flux seen by LOFAR. However, the WENSS and NVSS surveys are less sensitive and may not be detecting all of the fainter flux associated with a source, potentially causing the reported total flux densities to be underestimated, though no allowance is made for this.

\begin{figure}
    \centering
    \includegraphics[trim={20 30 50 60},clip=true,width=0.47\textwidth]{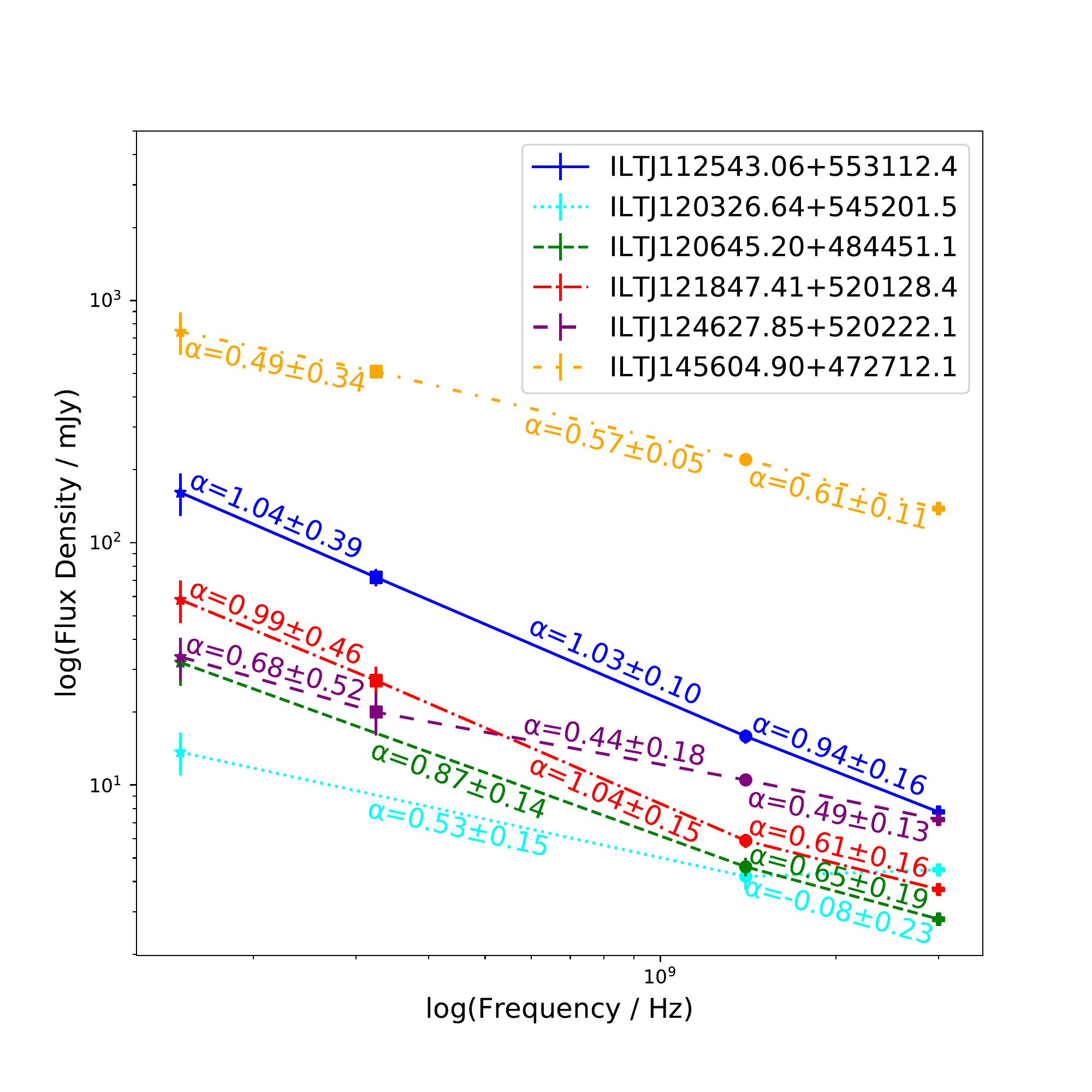}
    \caption{Plot of the total measured flux density against frequency for each source. Shown against each line are the corresponding spectral indices. Stars represent LoTSS measurements, squares WENSS, circles NVSS and crosses VLA. WENSS and NVSS are less sensitive than the other surveys, possibly causing their flux densities to be underestimated (see text for details).}
    \label{fig:TotalFluxes}
\end{figure}

The results are shown in Figure~\ref{fig:TotalFluxes}. The NVSS errors shown are those reported by the survey and the WENSS errors are calculated using the methods of \citet{Rengelink1997TheWENSS}. The errors for the LoTSS and VLA images were calculated as described in Section~\ref{sec:Observations}. ILT J122037.67+473857.6 is not detected in either the NVSS or WENSS catalogues or in our new VLA observations and is therefore not shown.

As can be seen in Figure~\ref{fig:TotalFluxes}, there is no evidence of significant low-frequency flattening in any of our sources indicating that the turnover, if present, must be at lower frequencies. For the majority of our sources the spectral index is relatively constant over the measured frequency range showing little deviation from the $3\text{ GHz}$ to $150\text{ MHz}$ measured integrated spectral index. The spectra for our sources are therefore similar to those of larger radio galaxies and not compact sources.

\subsection{Component Spectra}
\label{sec:SpectralIndices-Component}

The 3 GHz VLA image of ILT J120645.20+484451.1 (Figure~\ref{fig:VLAImages}) clearly shows a pronounced radio core. However, no radio core is seen in the 150 MHz LoTSS image. This difference in the spectral behaviour of the core is seen in another GSJ, NGC 3801 \citep[][]{Heesen2014ThePopulation}, and is also identical to larger radio galaxies which have flat core regions \citep[e.g.][]{Simpson2017ExtragalacticEra,Morganti1997ASources}.

For the three GSJ within our sample that have separately identifiable lobes and cores in both the LoTSS and VLA images, ILT J120326.64+545201.5, ILT J121847.41+520128.4 and ILT J124627.85+520222.1, we calculated the spectra of the lobe and core sub-components, using the same technique as described in Section~\ref{sec:Observations} to calculate the flux densities for both LoTSS and VLA images. In each case we found significant differences between the lobes and cores, again indicating different physical properties in these regions.

\section{Spectral Index Maps}
\label{sec:SpectralIndexMaps}

Apart from ILT J122037.67+473857.6, the other six sources identified in Table~\ref{tab:LoTSS_VLA_SpectralIndices} all have extended emission in both the LoTSS DR2 and VLA images. For these sources we produced spatially resolved spectral index maps that would then allow us to identify any structure within our sources. To do this we used the Broadband Radio Astronomy Tools\footnote{\url{http://www.askanastronomer.co.uk/brats/}} (\textsc{BRATS}) \citep[][]{Harwood2013SpectralData, Harwood2015SpectralGalaxies} which allows for detailed spectral analysis of radio images, providing a suite of model fitting, visualisation and statistical tools.

The VLA observations were carefully designed to ensure all scales of emission seen in the LoTSS images could be sampled; however, the two telescopes do have different beam sizes and resolutions. We therefore used \textsc{CASA}'s \emph{imsmooth} function to match the beam sizes and \emph{imregrid} to ensure each pixel represents the same spatial area. 

Combining our VLA images with the LoTSS DR2 images, we set the \textsc{BRATS} source detection threshold at three sigma, the flux calibration errors for the LOFAR images to $0.2$ \citep[][]{Shimwell2019TheRelease} and used the standard errors for VLA images. The spectral maps produced are shown in Figure~\ref{fig:BRATSResults-SpectralIndexMaps}.

\begin{figure*}
\centering
\begin{tabular}{cc}
\includegraphics[width=0.45\textwidth,trim = {110 25 50 10}, clip=true]{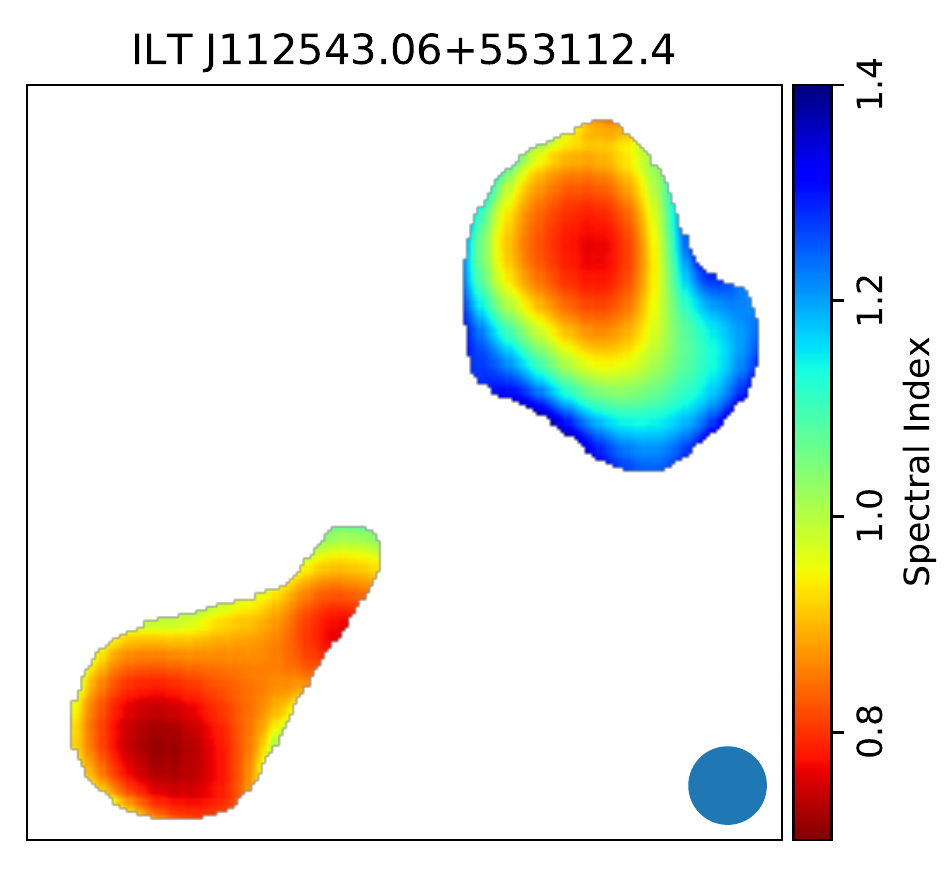} &
\includegraphics[width=0.45\textwidth,trim = {110 25 50 10}, clip=true]{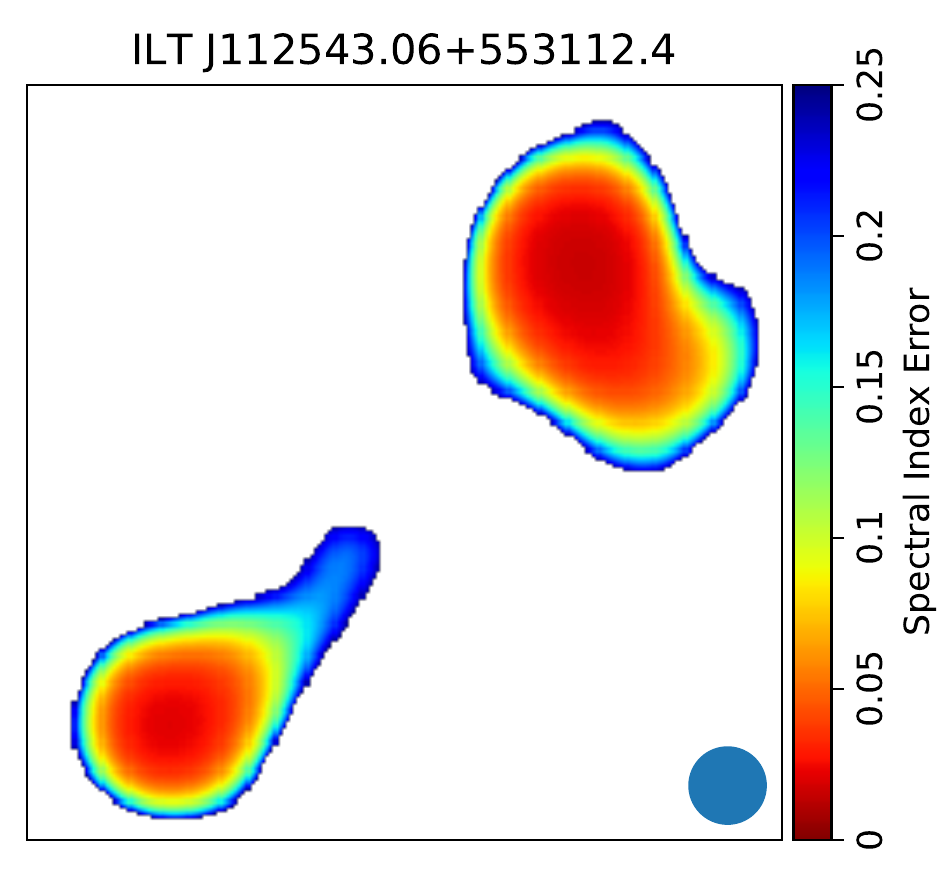} \\
\includegraphics[width=0.45\textwidth,trim = {110 25 50 10}, clip=true]{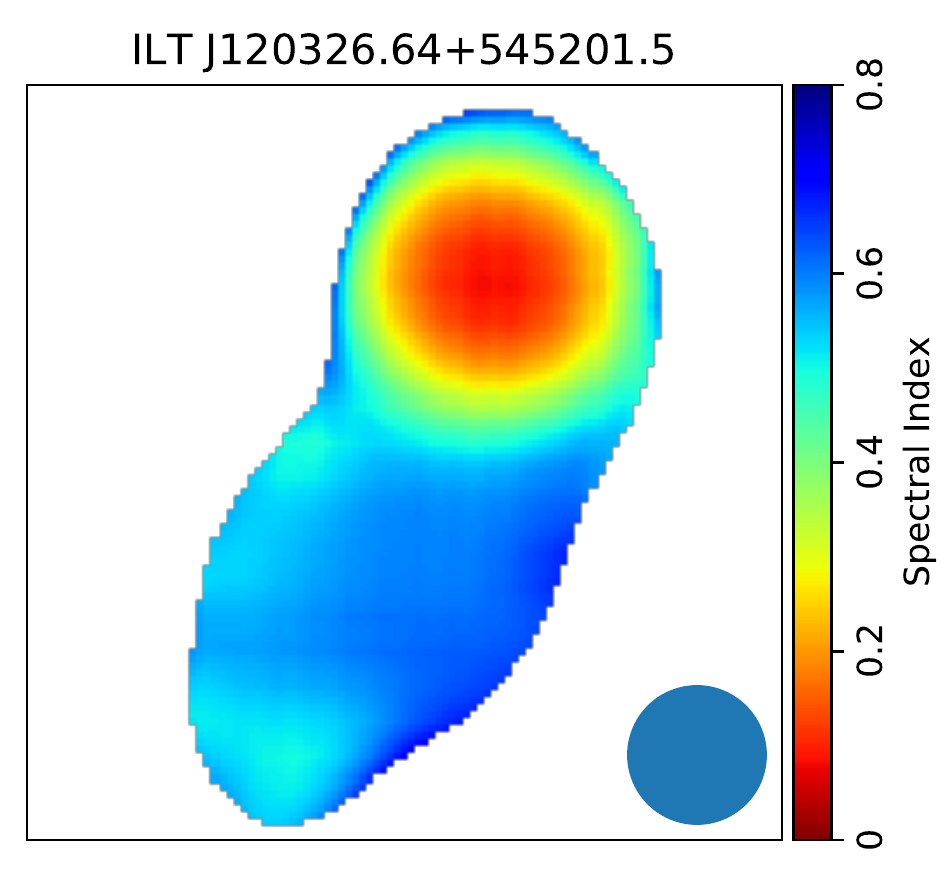} &
\includegraphics[width=0.45\textwidth,trim = {110 25 50 10}, clip=true]{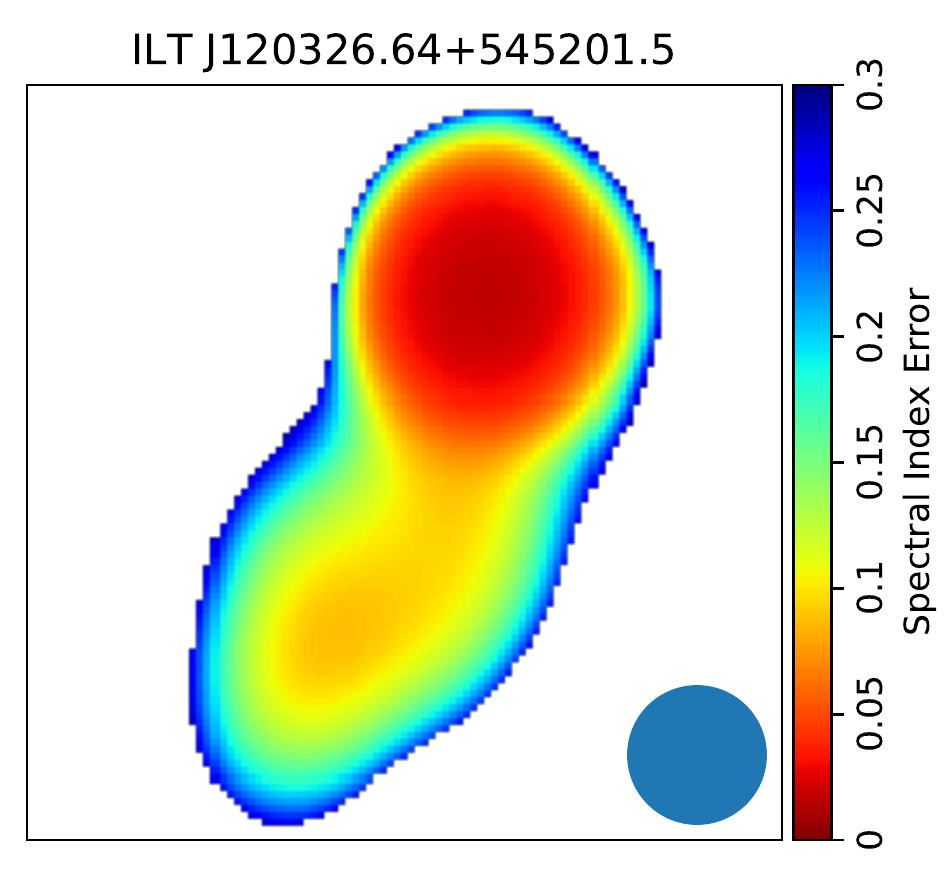} \\
\includegraphics[width=0.45\textwidth,trim = {110 25 50 10}, clip=true]{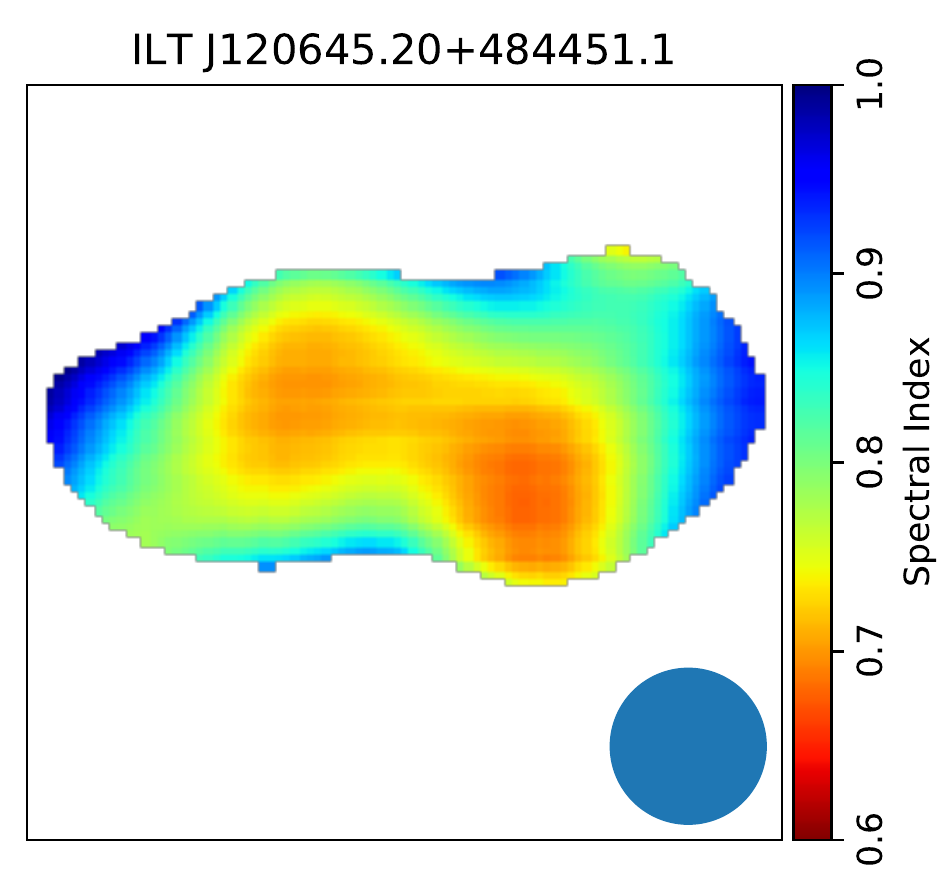} &
\includegraphics[width=0.45\textwidth,trim = {110 25 50 10}, clip=true]{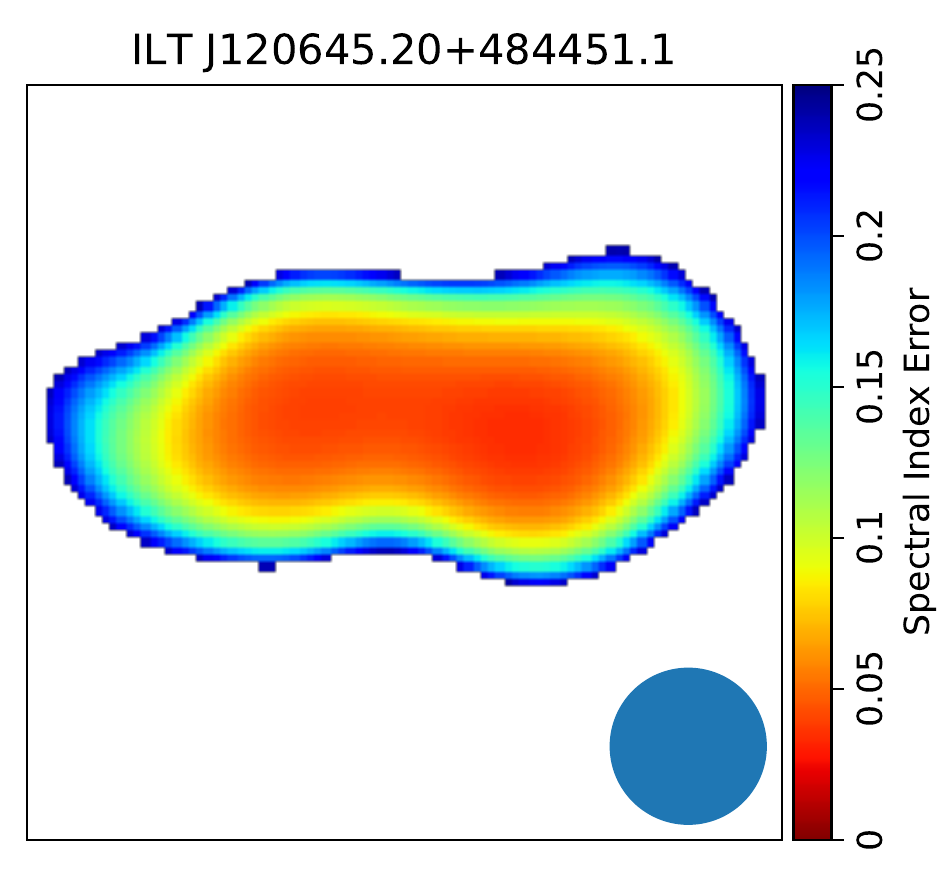} \\
\end{tabular}
\caption{Left Column shows the 150 MHz - 3 GHz spectral index maps and the right column shows the associated error maps. The errors presented are those used for the spectral ageing analysis and include flux calibration errors.}
\label{fig:BRATSResults-SpectralIndexMaps}
\end{figure*}

\begin{figure*}
\ContinuedFloat
\centering
\begin{tabular}{cc}
\includegraphics[width=0.45\textwidth,trim = {110 25 50 10}, clip=true]{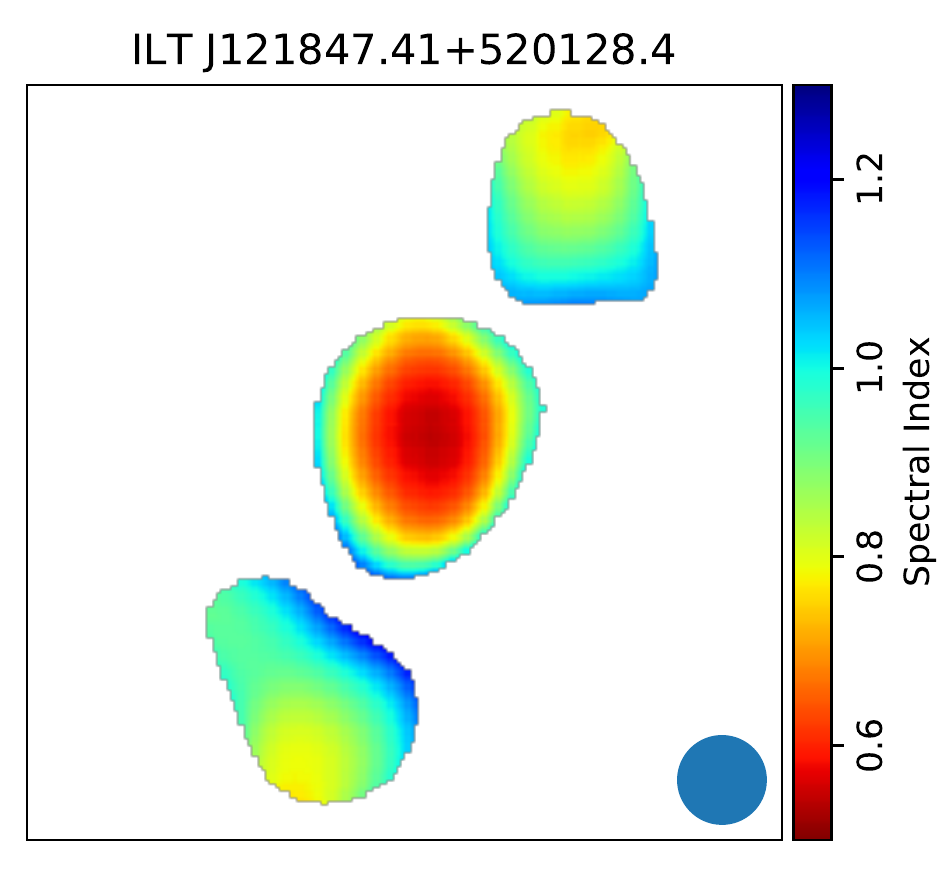} &
\includegraphics[width=0.45\textwidth,trim = {110 25 50 10}, clip=true]{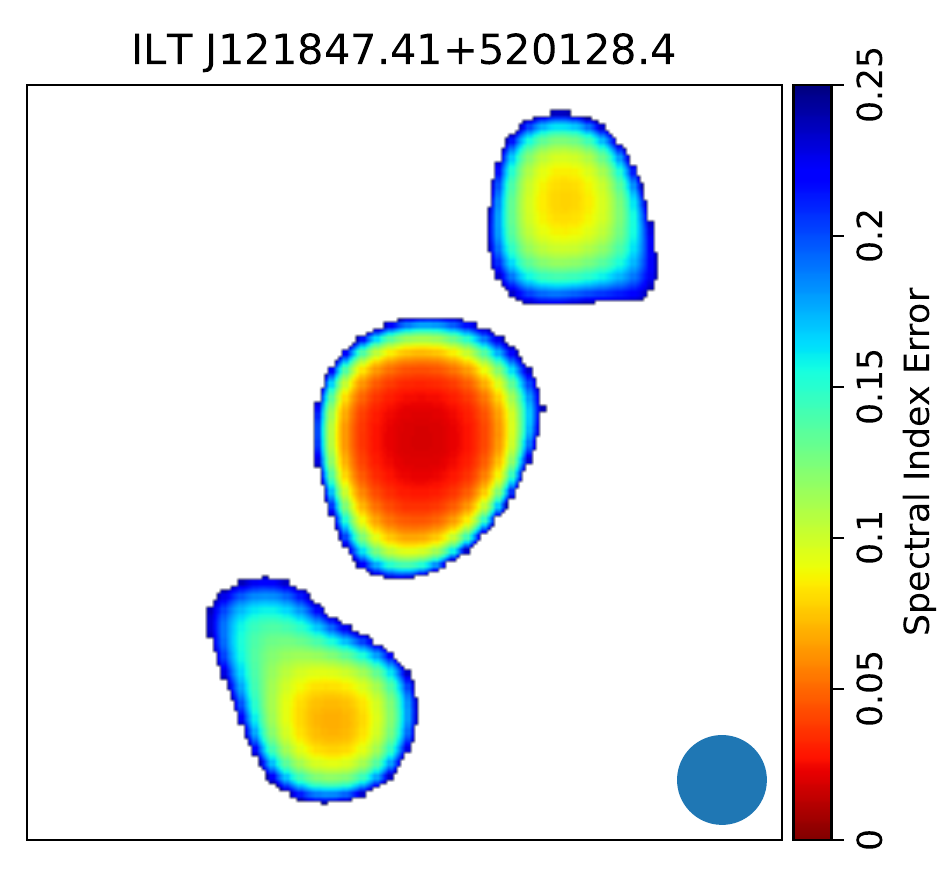} \\
\includegraphics[width=0.45\textwidth,trim = {110 25 50 10}, clip=true]{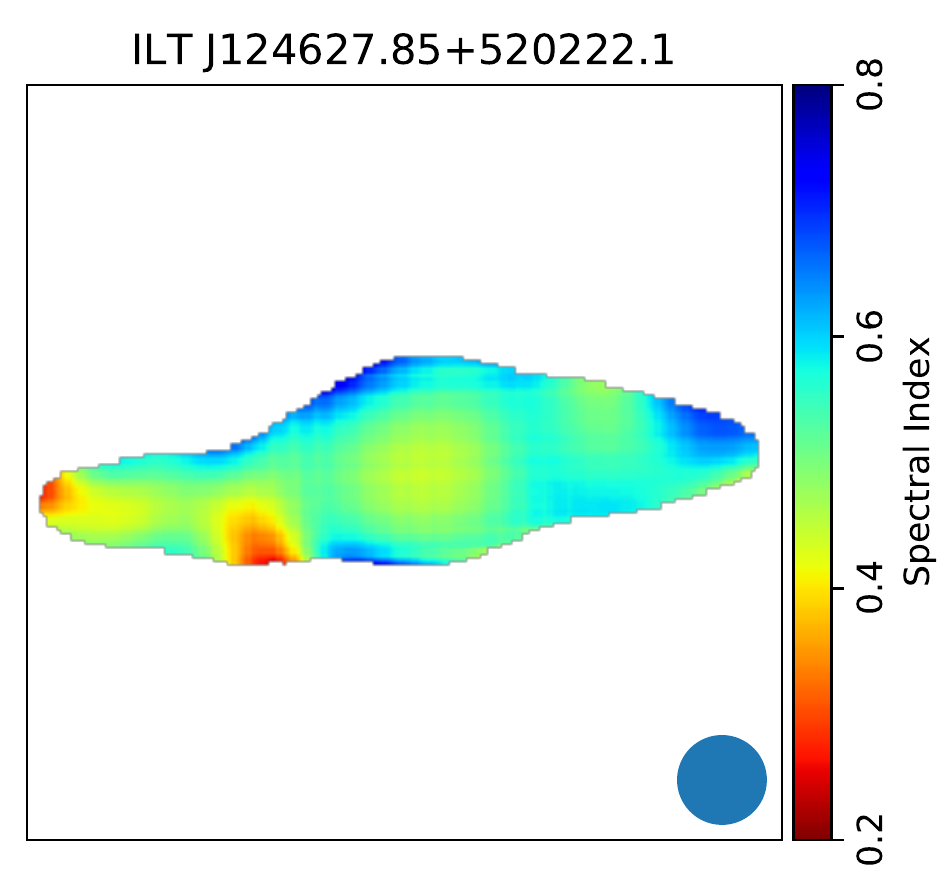} &
\includegraphics[width=0.45\textwidth,trim = {110 25 50 10}, clip=true]{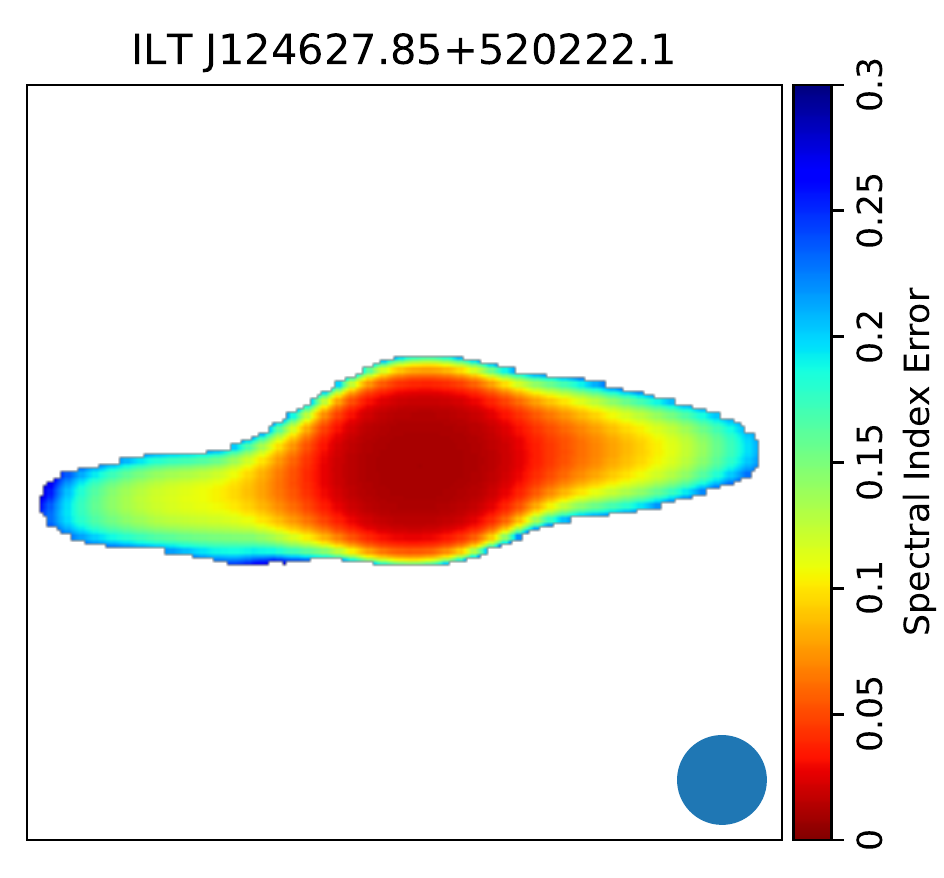} \\
\includegraphics[width=0.45\textwidth,trim = {110 25 50 10},
clip=true]{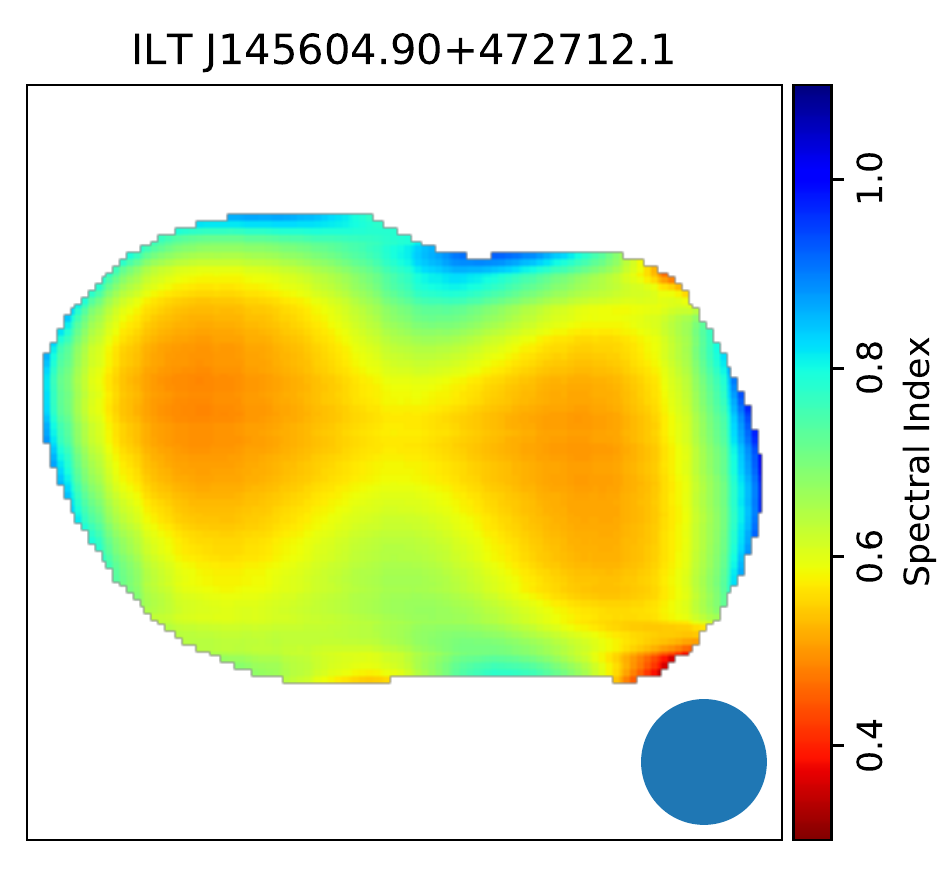} &
\includegraphics[width=0.45\textwidth,trim = {110 25 50 10},
clip=true]{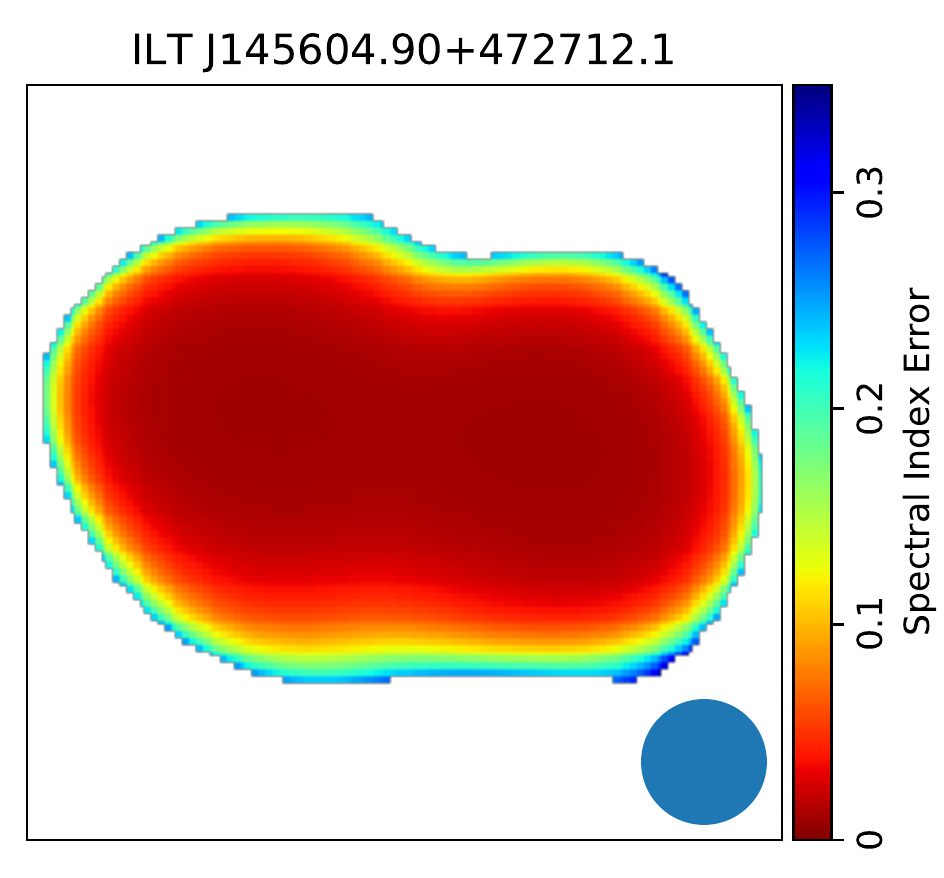} \\
\end{tabular}
\caption{(Continued) Left Column shows the 150 MHz - 3 GHz spectral index maps and the right column shows the associated error maps. The errors presented are those used for the spectral ageing analysis and include flux calibration errors.}
\end{figure*}

The spectral index error maps are calculated by BRATS using a weighted least squares method \citep[see][for details]{Harwood2015SpectralGalaxies}. They show increased uncertainties towards the edge of the emission with some sources showing high uncertainties at the very edge of the observed emission. However, overall the spectral index errors are low across the majority of the observed emission and most spectral structure shown is on resolved scales and showing gradients in the sense expected for typical radio-galaxy structures indicating that our spectral index maps are reliable except, in some cases, for the outer few pixels.

Outside of the core/hotspot emission (where visible), our sources show only a small amount of variation in the spectral index across the majority of the extended emission which is not surprising given the fairly low resolution. As expected of FRII sources, the spectral gradients that can be observed for ILT J112543.06+553112.5 and ILT J121847.41+520148.4 become overall progressively steeper with distance from the hotspot. Similarly the FRI sources in our sample show a general trend for the spectral index to increase with distance away from the core.

Although obeying this general trend, the eastern jet of ILT J124627.85+520222.1 does have two regions of flat spectral index at the tip and midway along the jet. This is discussed further in Section~\ref{sec:SpectralModels-SpectralAges}. The spectral index maps also show the different core spectral behaviour for ILT J120326.64+545201.5 and ILT J121847.41+520148.4 mentioned in Section~\ref{sec:SpectralIndices-Component}. Though less obvious, the core of ILT J124627.85+520222.1 also shows a different core spectral behaviour with a slightly flatter spectral index than the surrounding regions.

\section{Spectral Model Fitting}
\label{sec:SpectralModels}

For the six sources with extended emission in both the LoTSS DR2 and VLA images we fitted spectral models to find the age of the radio emission. When fitting spectral models, in order to constrain the spectral ages as much as possible, we divided our matched VLA images (described in Section~\ref{sec:SpectralIndexMaps}) into four frequency ranges each of  500MHz bandwidth, using the resulting images along with the 150 MHz LoTSS data to perform the model fitting.

As described in Sections~\ref{sec:SpectralIndices-Component} and \ref{sec:SpectralIndexMaps}, the cores of some of our sources have significantly different spectral behaviour from the lobes. Since we are interested in the age of the lobes we follow the methodology of \citet{Heesen2014ThePopulation} and exclude the cores from all four sources where the core region could be identified to eliminate any adverse effects they would have upon our model fitting.

We used \textsc{BRATS} to fit Jaffe-Perola (JP) models \citep[][]{Jaffe1973DynamicalGalaxies} to our data. Using the JP model allows a direct comparison with the few previous spectral ageing studies of GSJ \citep[e.g.][]{Parma1999RadiativeGalaxies,Heesen2014ThePopulation}. JP models are single-injection models where the angle of the synchrotron-emitting electrons is isotropic with respect to the magnetic field lines and is believed to provide a realistic model of the conditions in radio galaxies. In JP models the electrons are subject to both synchrotron and inverse-Compton energy losses. Along with the radio images, the other inputs required for spectral age modelling are the magnetic field strength and the injection index, described in Sections~\ref{sec:SpectralModels-MagFieldStrengths} and \ref{sec:SpectralModels-InjectionIndices} respectively. Our results are summarised in Table~\ref{tab:Spectral_Ages}. The resulting images are shown in Figures~\ref{fig:BRATSResults-ILTJ112543} to \ref{fig:BRATSResults-ILTJ145604} inclusive with the results discussed in Section~\ref{sec:SpectralModels-SpectralAges}.

\subsection{Magnetic Field Strengths}
\label{sec:SpectralModels-MagFieldStrengths}

The strength of the magnetic field in the lobes was calculated using a Python version of the \textsc{SYNCH} code first used by \citet{Hardcastle1998Magnetic111}\footnote{https://github.com/mhardcastle/pysynch} initially under equipartition energy conditions assuming no non-radiating particles (listed as $B_{eq}$ in Table~\ref{tab:Spectral_Ages}). The size and flux of the lobes was taken from the VLA images. We note that similar magnetic field strengths were obtained using the LoTSS images.

While we considered the case of equipartition field strengths assuming no non-radiating particles, we also considered more physically realistic scenarios for both FRI and FRII sources based on the conclusions of X-ray studies. For FRII-type jets independent measurements of the magnetic field strength using X-ray observations have found the average magnetic field strength is $\sim0.4B_{\text{eq}}$ \citep[][]{Ineson2017AGalaxies}, where $B_{\text{eq}}$ is the equipartition magnetic field strength assuming no proton content.

For FRIs the lack of inverse Compton X-ray detection means the magnetic field strengths are less well constrained. However, using equipartition estimates with no proton content, it has been shown that the internal lobe pressure is insufficient to balance the external medium. The extra pressure is believed to come from either entrained material, higher magnetic field strengths or a combination of both \citep[][]{Croston2003iXMM-Newton/i449,Croston2008AnGalaxies,Croston2014TheClusters}. Recently, for a population of 27 FRIs, \citet{Croston2018ParticleEqual} found that in order to achieve pressure balance, where the magnetic field and total (radiating and non-radiating) particle energy densities are in equipartition, a ratio of $\sim10$ for the non-radiating to radiating particle energy density was required.

Therefore in the discussion that follows we adopt magnetic field strength values (i) for the FRIs assuming equipartition between the magnetic and total particle energy densities with a ratio between the non-radiating and radiating particle energy density of 10 and (ii) for the FRIIs using magnetic field strengths that are 0.4 times the equipartition magnetic field strengths calculated assuming no proton content. These adjusted values are given as $B_{adj}$ in Table~\ref{tab:Spectral_Ages}. In the following discussion we also note how our conclusions would differ for the assumption of equipartition with no non-radiating particles in both cases.

\subsection{Injection Indices}
\label{sec:SpectralModels-InjectionIndices}

The injection index is the spectral index at the point of injection, corresponding to the initial electron energy distribution. The injection indices, shown in Table~\ref{tab:Spectral_Ages}, were all calculated using the \emph{findinject} function in \textsc{BRATS}. We note these values were the same no matter how the magnetic field strengths were calculated. The majority of our sources have injection indices within the range $0.5$ to $0.7$ typically assumed for radio loud AGN. The injection indices are also consistent with those found for the B2 sample of radio galaxies which are a mix of FRI and FRII radio galaxies covering a range of luminosities and physical sizes, the lower end of which are similar to our sample of GSJ \cite[][]{Parma1999RadiativeGalaxies}. The injection indices are also similar to the value of $0.5\pm0.05$ found for another GSJ, NGC3801 \citep[][]{Heesen2014ThePopulation}.

However, both FRII spirals, ILT J112543.06+553112.4 and ILT J121847.41+520148.4, have steeper injection indices closer to $0.8$. These sources support the argument that FRIIs have steeper injection indices than previously thought \citep[][]{Harwood2016FREnergetics,Harwood2017FR-IIDynamics}. Models have suggested a steeper injection index should equate to more powerful jets \citep[][]{Konar2013ParticleObservations} but, as can be seen in Table~\ref{tab:VLA_Observations}, our FRIIs are amongst the least luminous in our sample. Whilst this, again, suggests that the hosts for these sources have been misidentified, this is not conclusive as, whilst they cannot rule the model of \citet{Konar2013ParticleObservations} out, \citet{Harwood2016FREnergetics} do find two FRIIs whose hotspots, according to the model, imply a flatter injection index than is found, suggesting other factors also need to be considered.

\subsection{Spectral Ages}
\label{sec:SpectralModels-SpectralAges}

\begin{figure*}
\centering
\begin{tabular}{cc}
\includegraphics[width=0.44\textwidth, trim = {110 25 50 10}, clip=true]{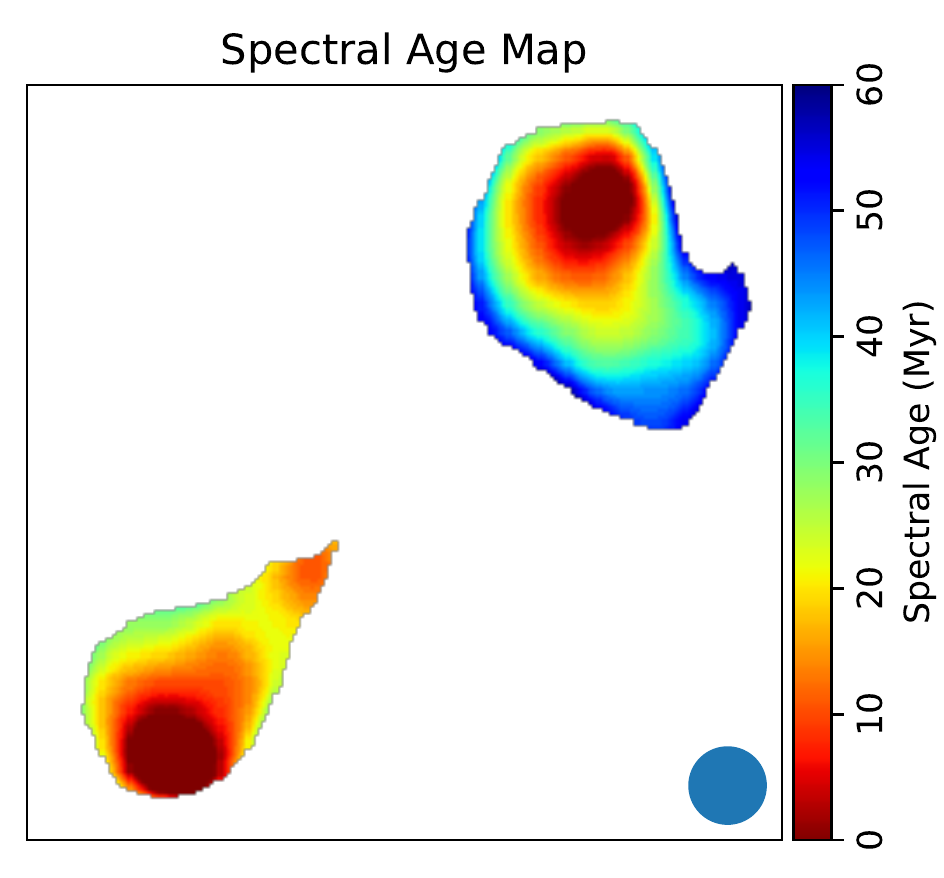} &
\includegraphics[width=0.44\textwidth, trim = {110 25 50 10}, clip=true]{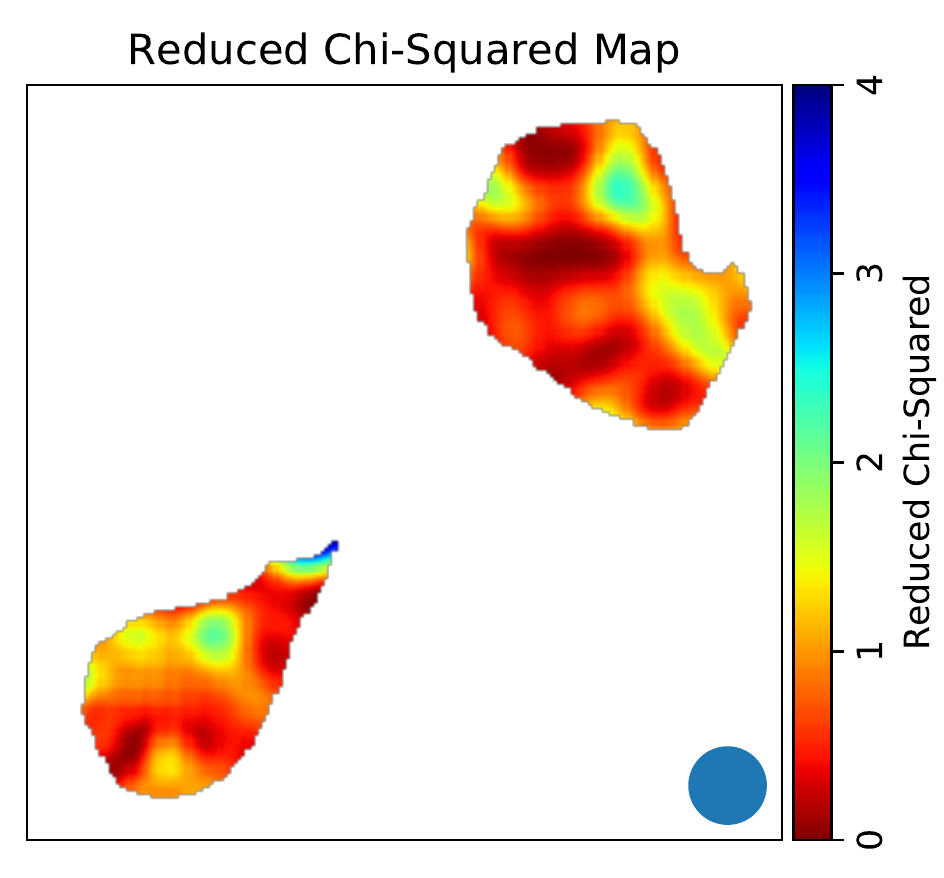} \\
\includegraphics[width=0.44\textwidth, trim = {110 25 50 10}, clip=true]{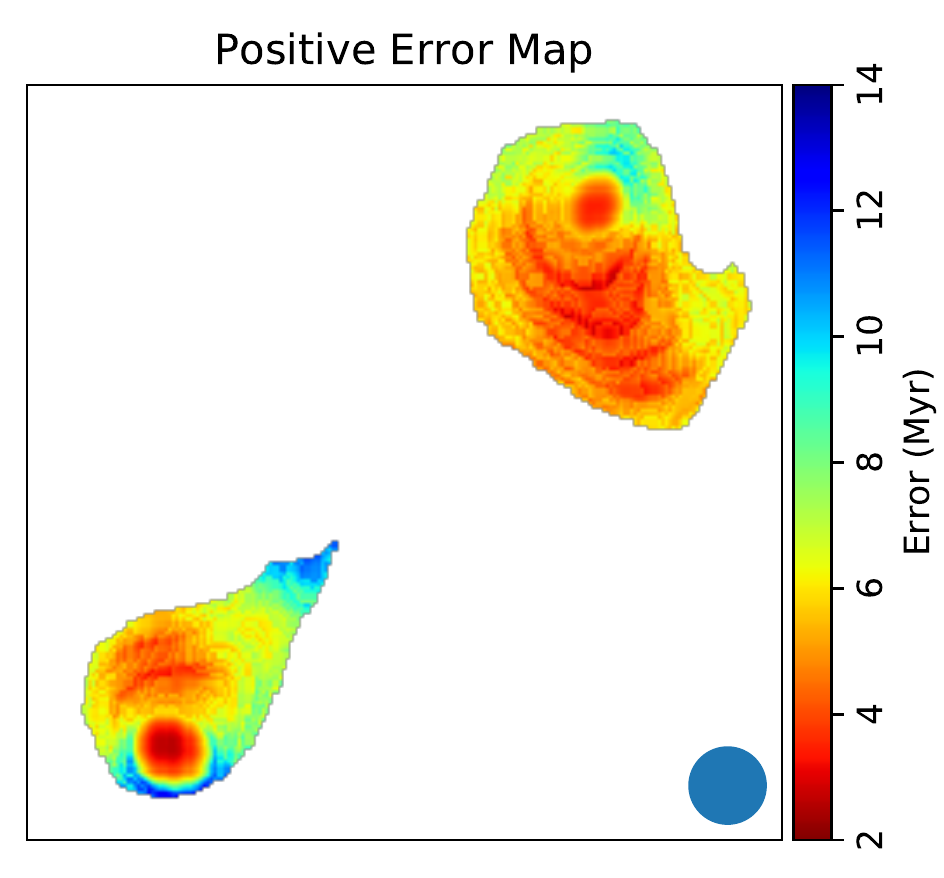} &
\includegraphics[width=0.44\textwidth, trim = {110 25 50 10}, clip=true]{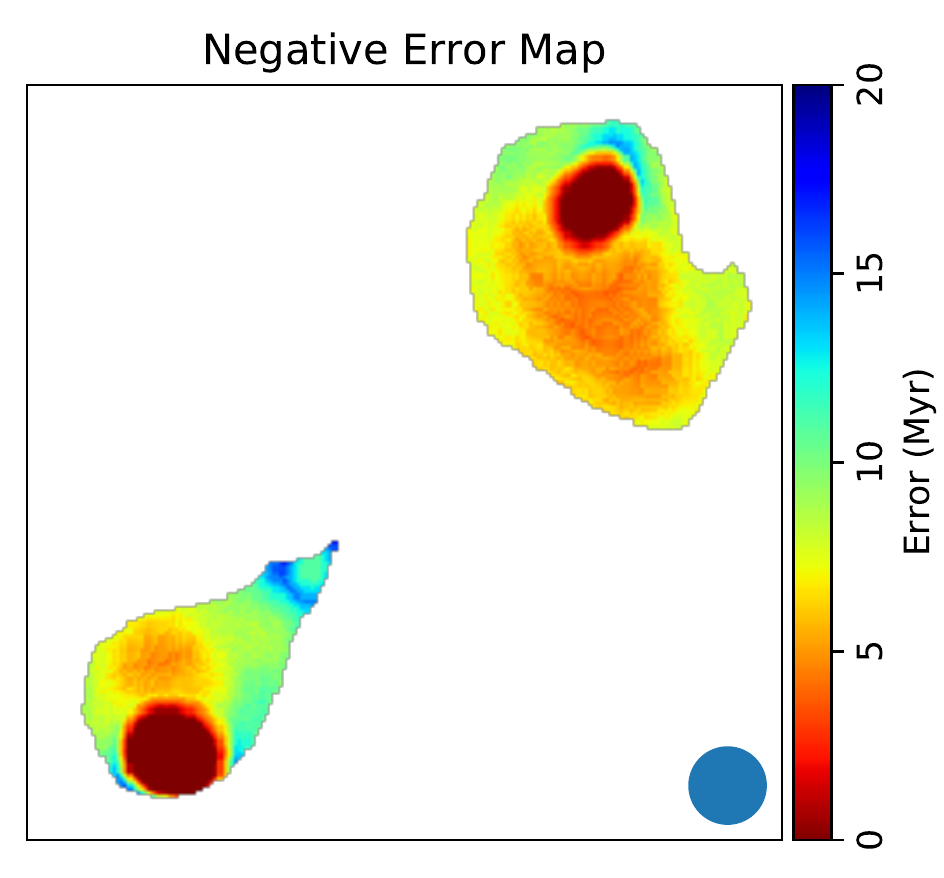}
\end{tabular}
\caption{Results of the spectral fitting for ILT J112543.06+553112.4.1. Top left shows the spectral age map and top right shows the reduced $\chi^2$ for the spectral age map. Bottom left and right images show the positive and negative errors for the spectral age map.}
\label{fig:BRATSResults-ILTJ112543}
\end{figure*}

\begin{figure*}
\centering
\begin{tabular}{cc}
\includegraphics[width=0.44\textwidth, trim = {110 25 50 10}, clip=true]{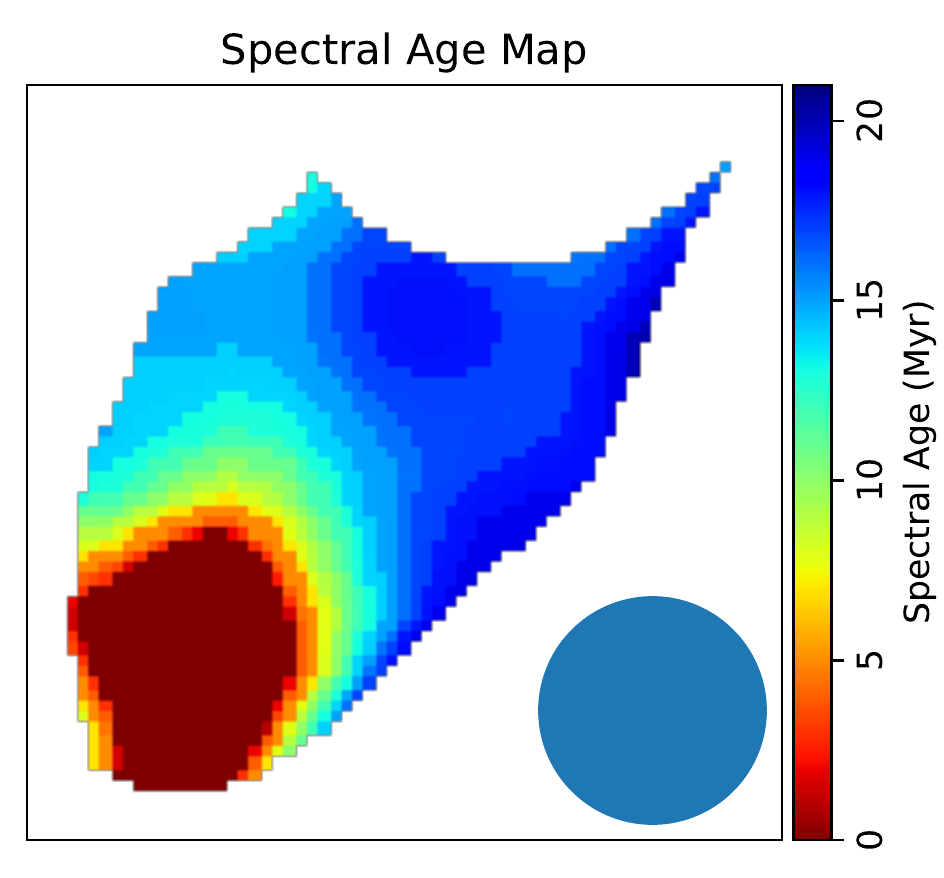} &
\includegraphics[width=0.44\textwidth, trim = {110 25 50 10}, clip=true]{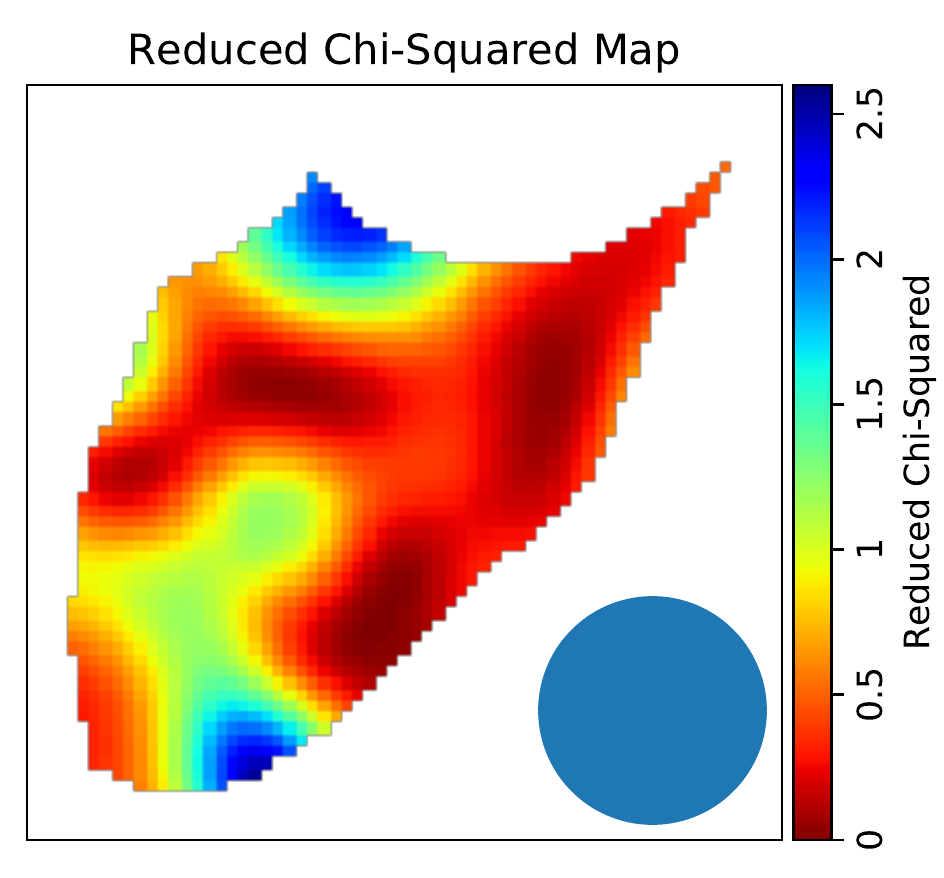} \\
\includegraphics[width=0.44\textwidth, trim = {110 25 50 10}, clip=true]{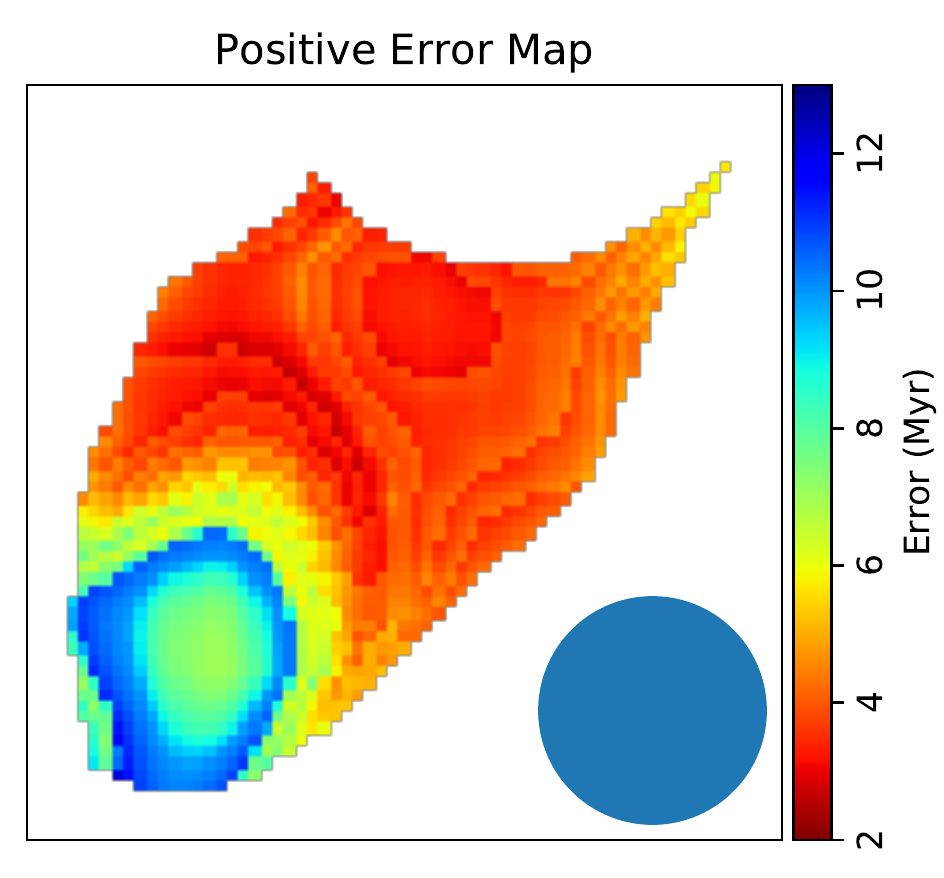} &
\includegraphics[width=0.44\textwidth, trim = {110 25 50 10}, clip=true]{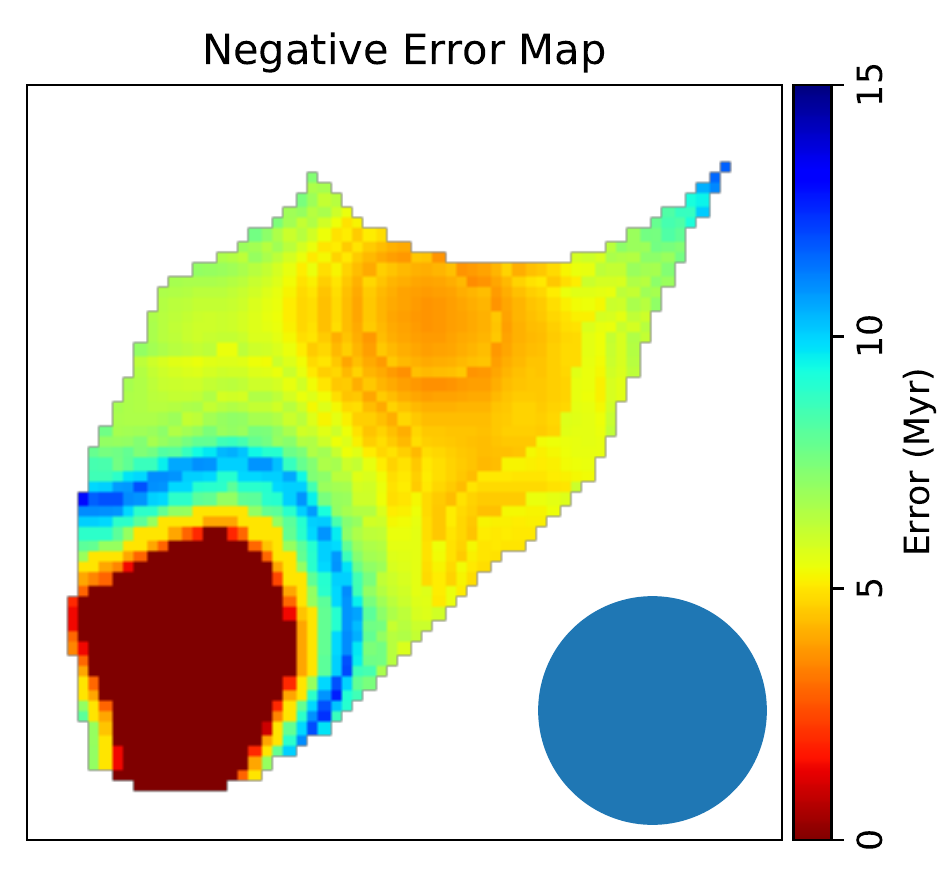}
\end{tabular}
\caption{Results of the spectral fitting for ILT J120326.64+545201.5. Top left shows the spectral age map and top right shows the reduced $\chi^2$ for the spectral age map. Bottom left and right images show the positive and negative errors for the spectral age map.}
\label{fig:BRATSResults-ILTJ120326}
\end{figure*}

\begin{figure*}
\centering
\begin{tabular}{cc}
\includegraphics[width=0.44\textwidth, trim = {110 25 50 10}, clip=true]{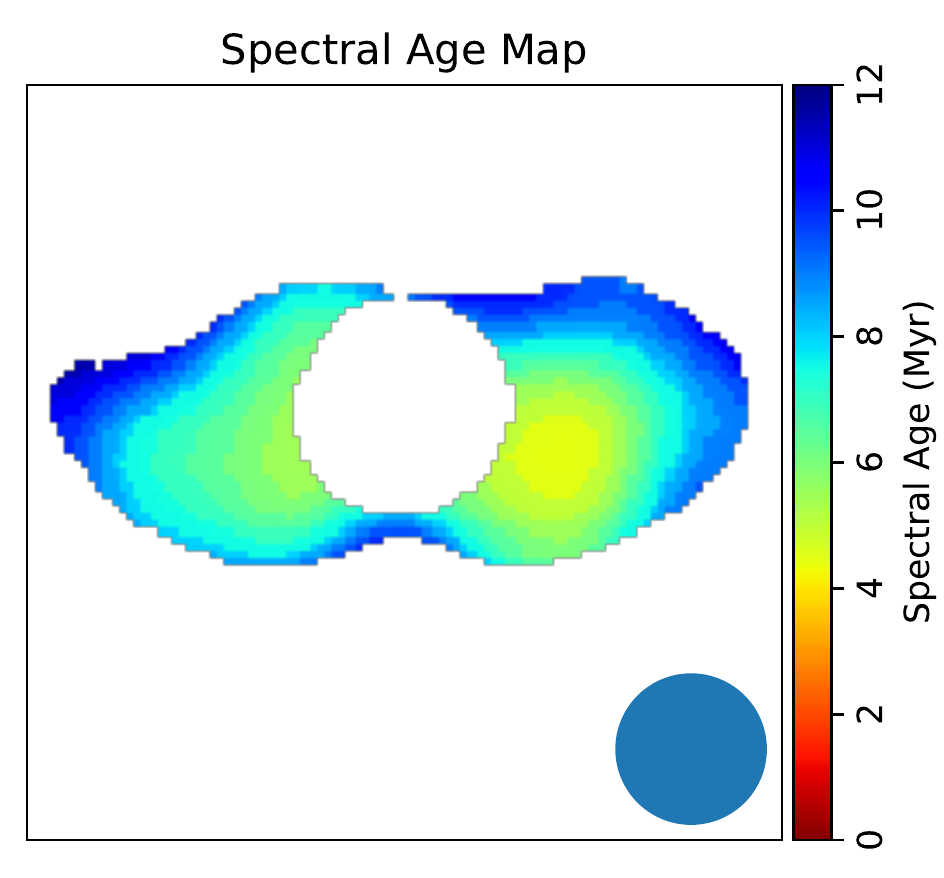} &
\includegraphics[width=0.44\textwidth, trim = {110 25 50 10}, clip=true]{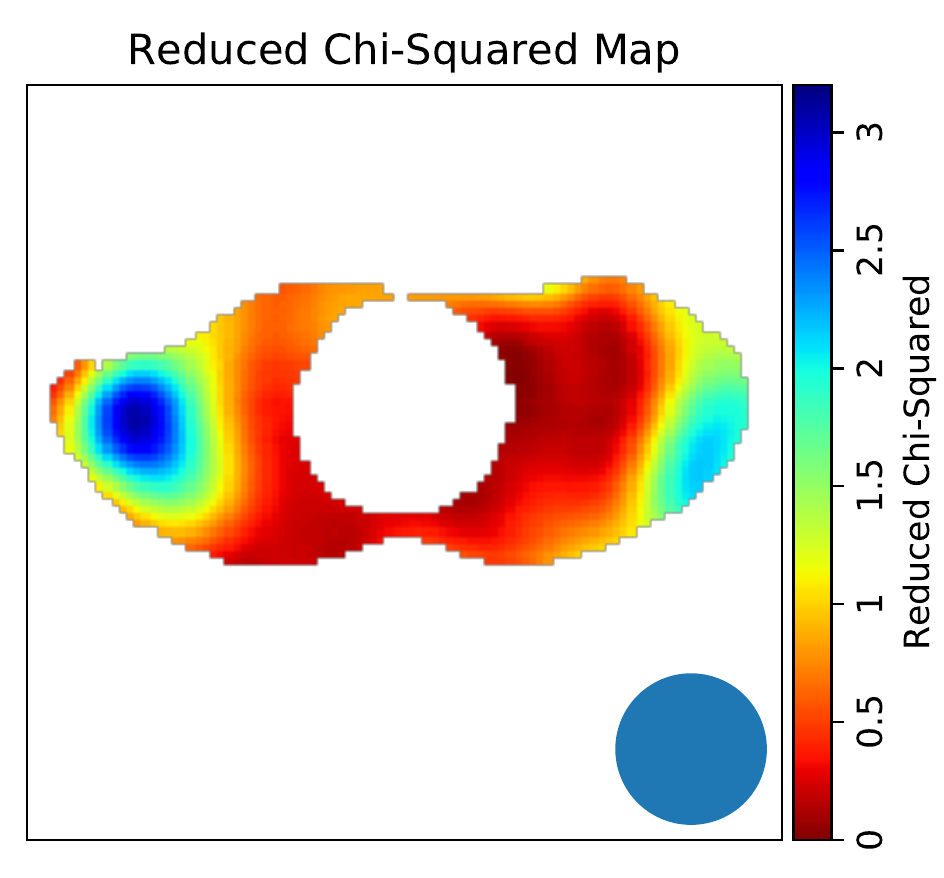} \\
\includegraphics[width=0.44\textwidth, trim = {110 25 50 10}, clip=true]{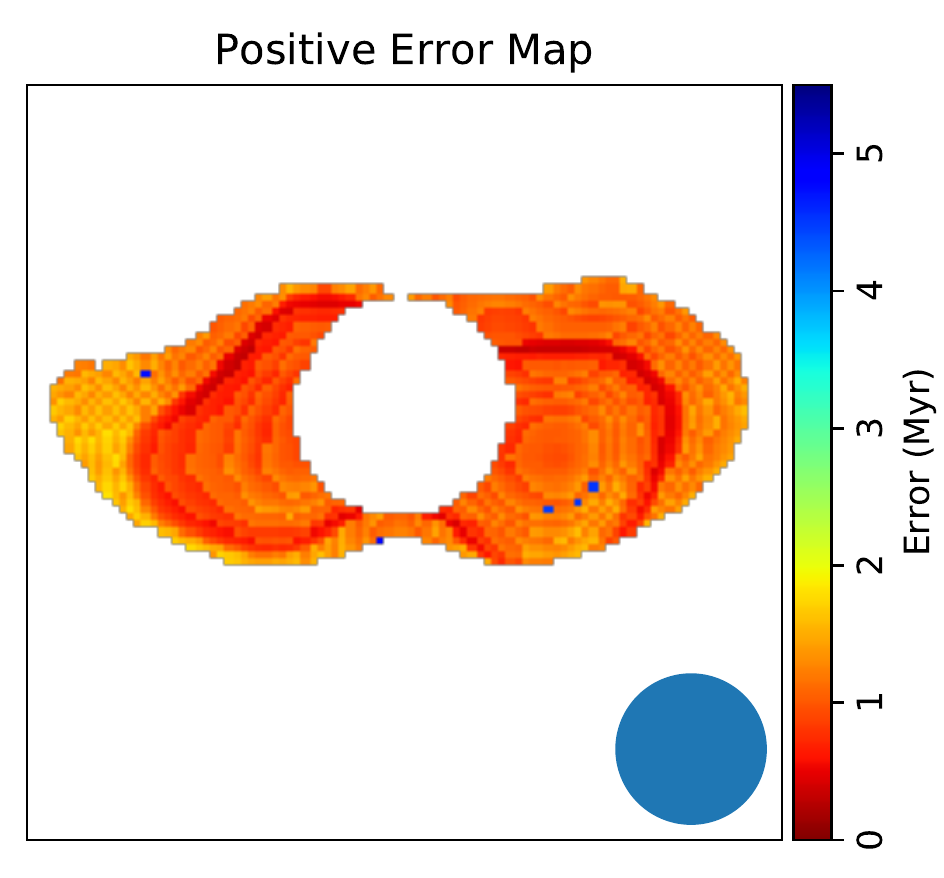} &
\includegraphics[width=0.44\textwidth, trim = {110 25 50 10}, clip=true]{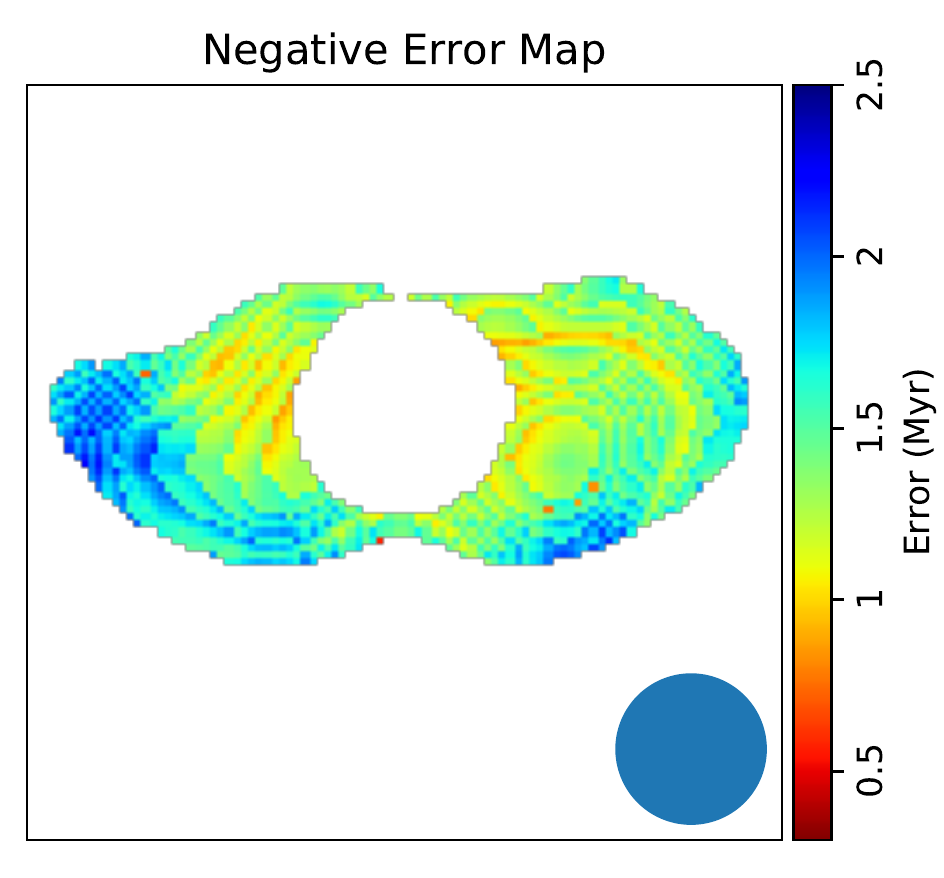}
\end{tabular}
\caption{Results of the spectral fitting for ILT J120645.20+484451.1. Top left shows the spectral age map and top right shows the reduced $\chi^2$ for the spectral age map. Bottom left and right images show the positive and negative errors for the spectral age map.}
\label{fig:BRATSResults-ILTJ120645}
\end{figure*}

\begin{figure*}
\centering
\begin{tabular}{cc}
\includegraphics[width=0.44\textwidth, trim = {110 25 50 10}, clip=true]{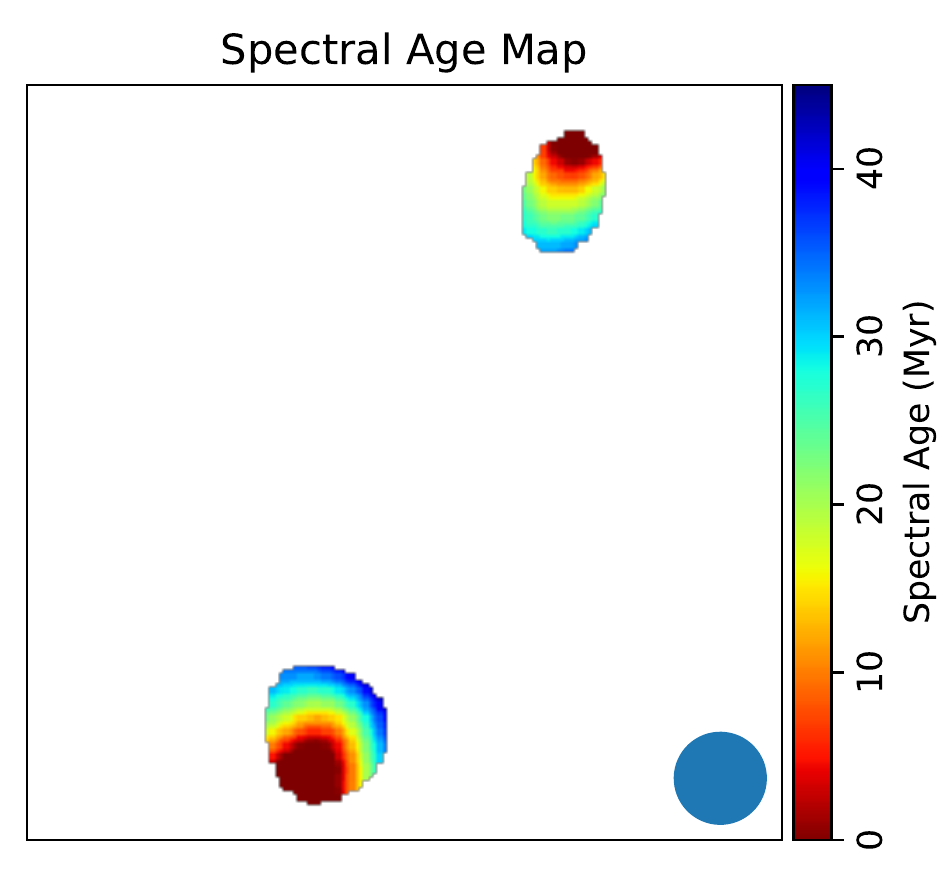} &
\includegraphics[width=0.44\textwidth, trim = {110 25 50 10}, clip=true]{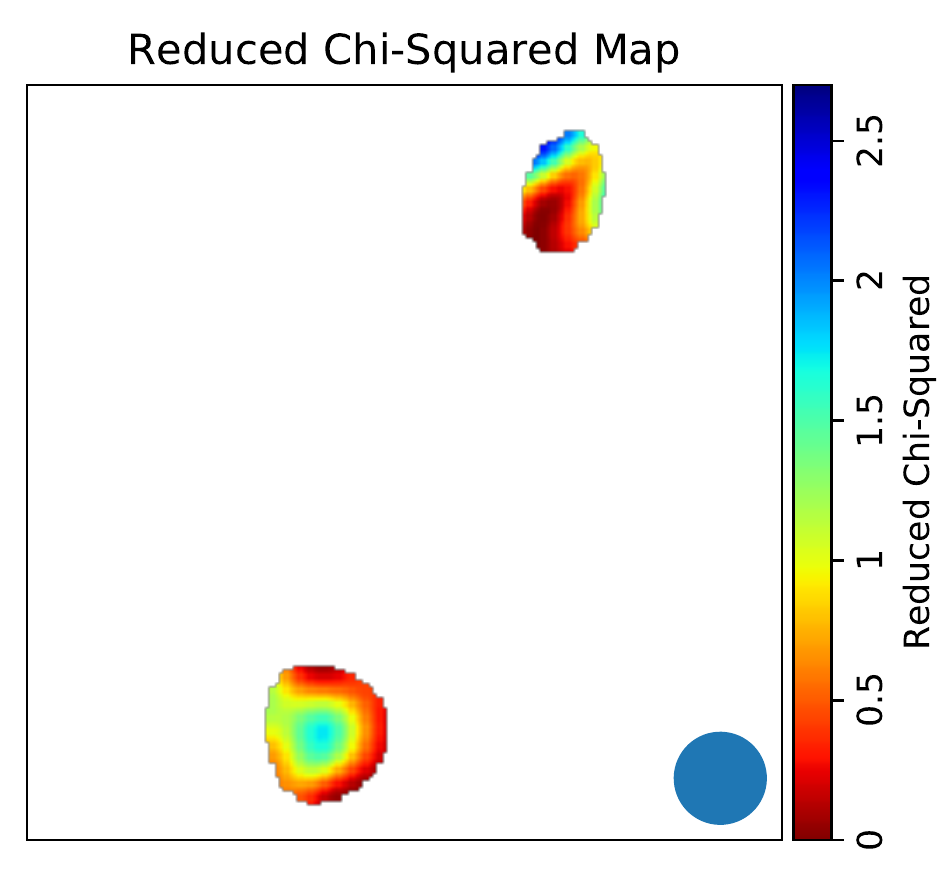} \\
\includegraphics[width=0.44\textwidth, trim = {110 25 50 10}, clip=true]{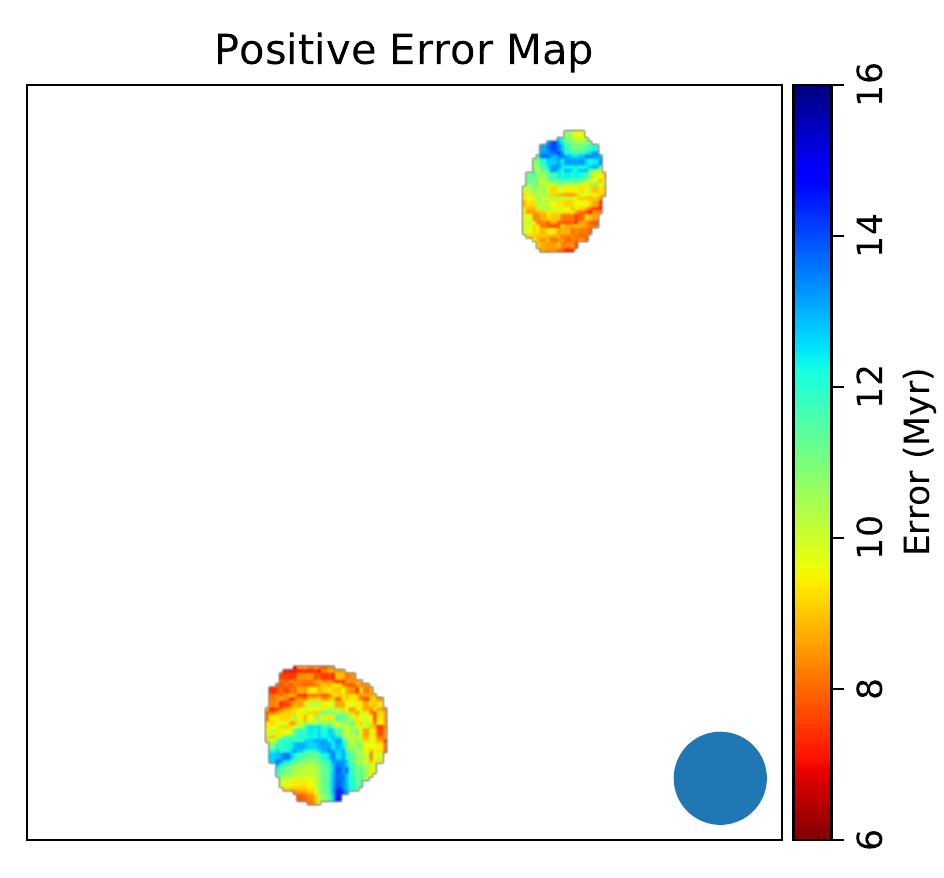} &
\includegraphics[width=0.44\textwidth, trim = {110 25 50 10}, clip=true]{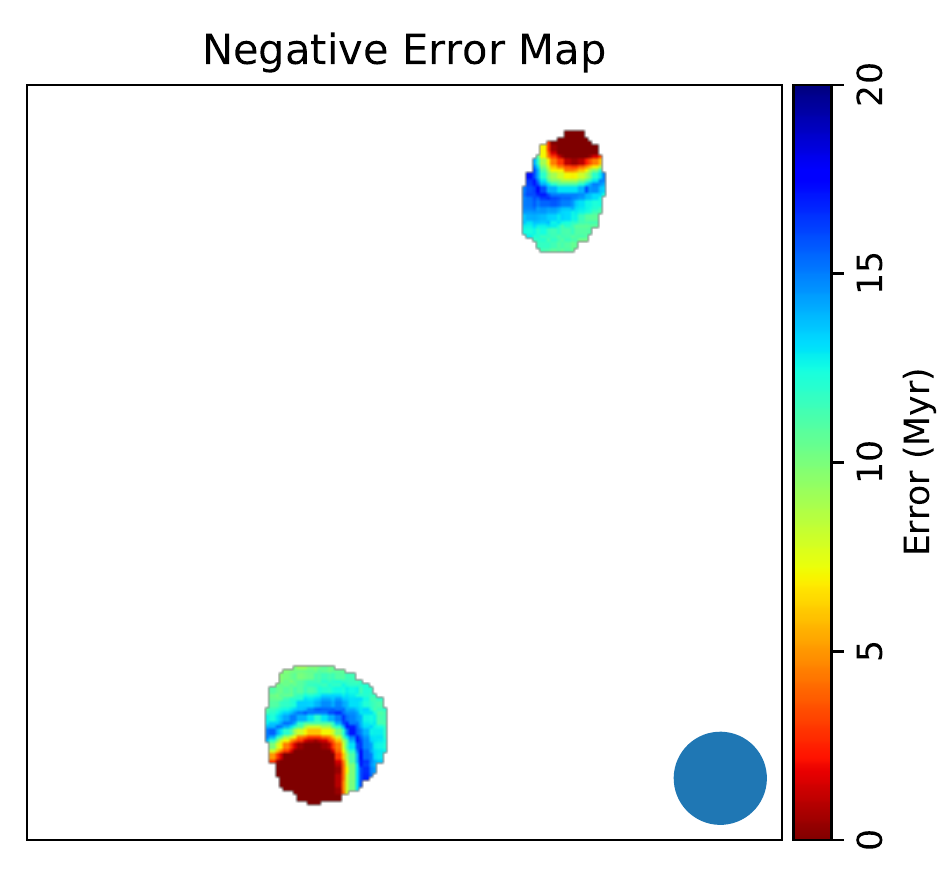}
\end{tabular}
\caption{Results of the spectral fitting for ILT J121847.41+520148.4. Top left shows the spectral age map and top right shows the reduced $\chi^2$ for the spectral age map. Bottom left and right images show the positive and negative errors for the spectral age map.}
\label{fig:BRATSResults-ILTJ121847}
\end{figure*}

\begin{figure*}
\centering
\begin{tabular}{cc}
\includegraphics[width=0.44\textwidth, trim = {110 25 50 10}, clip=true]{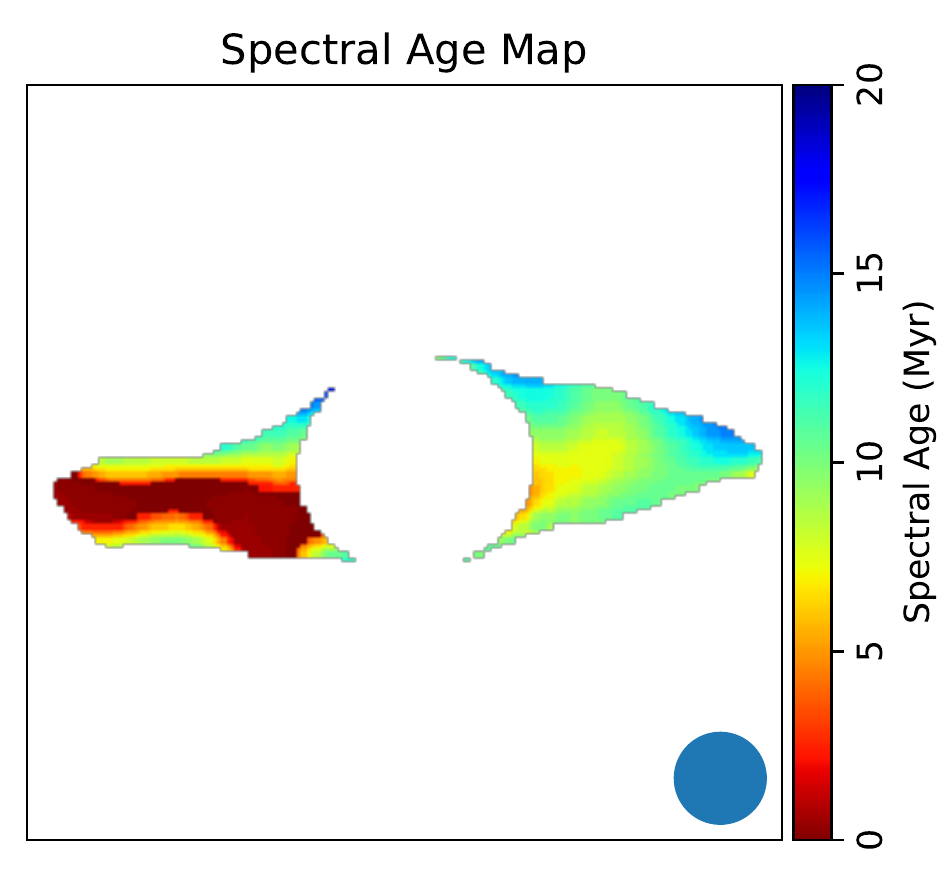} &
\includegraphics[width=0.44\textwidth, trim = {110 25 50 10}, clip=true]{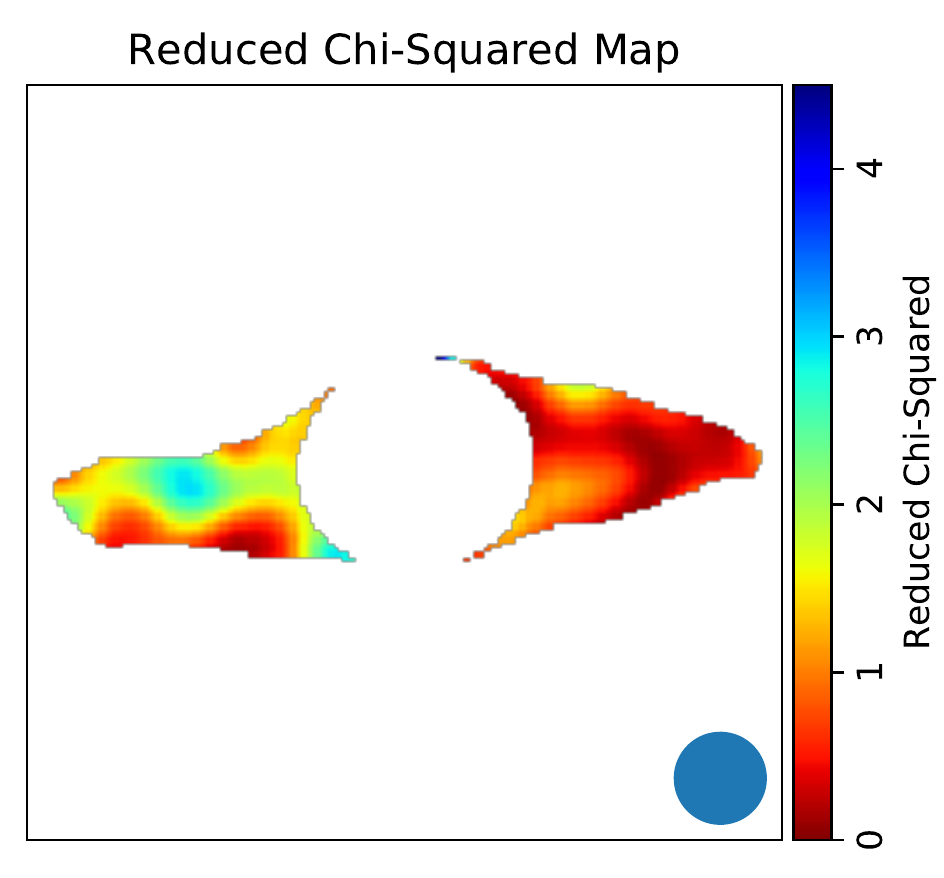} \\
\includegraphics[width=0.44\textwidth, trim = {110 25 50 10}, clip=true]{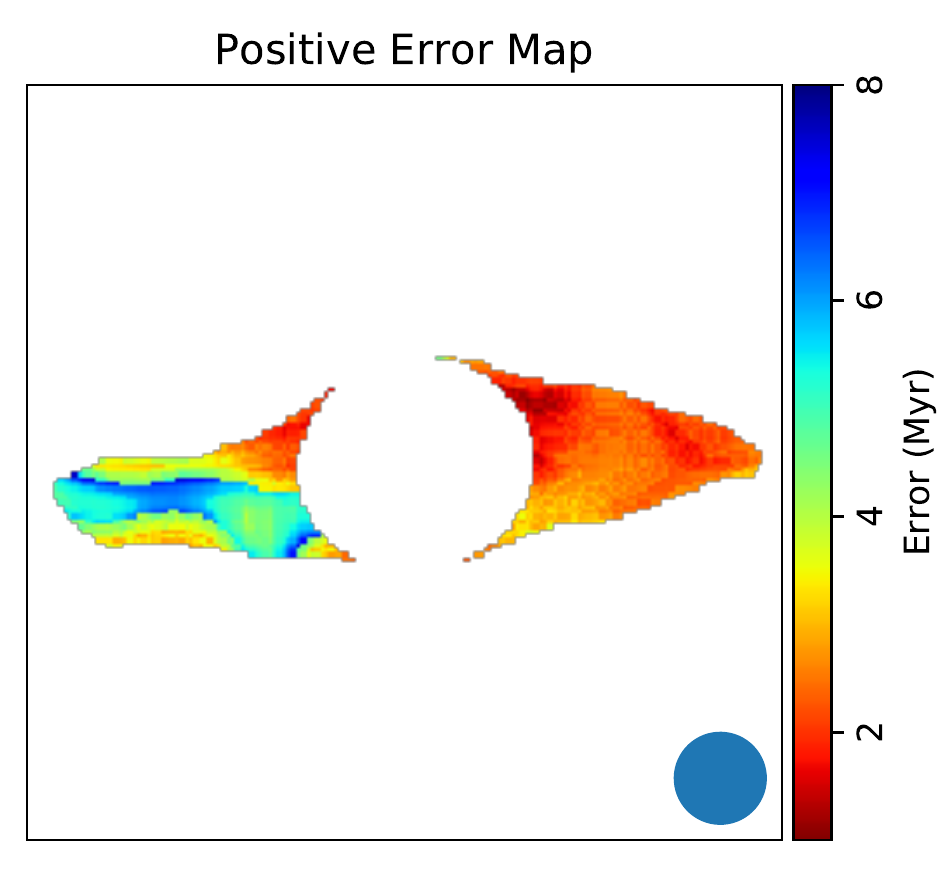} &
\includegraphics[width=0.44\textwidth, trim = {110 25 50 10}, clip=true]{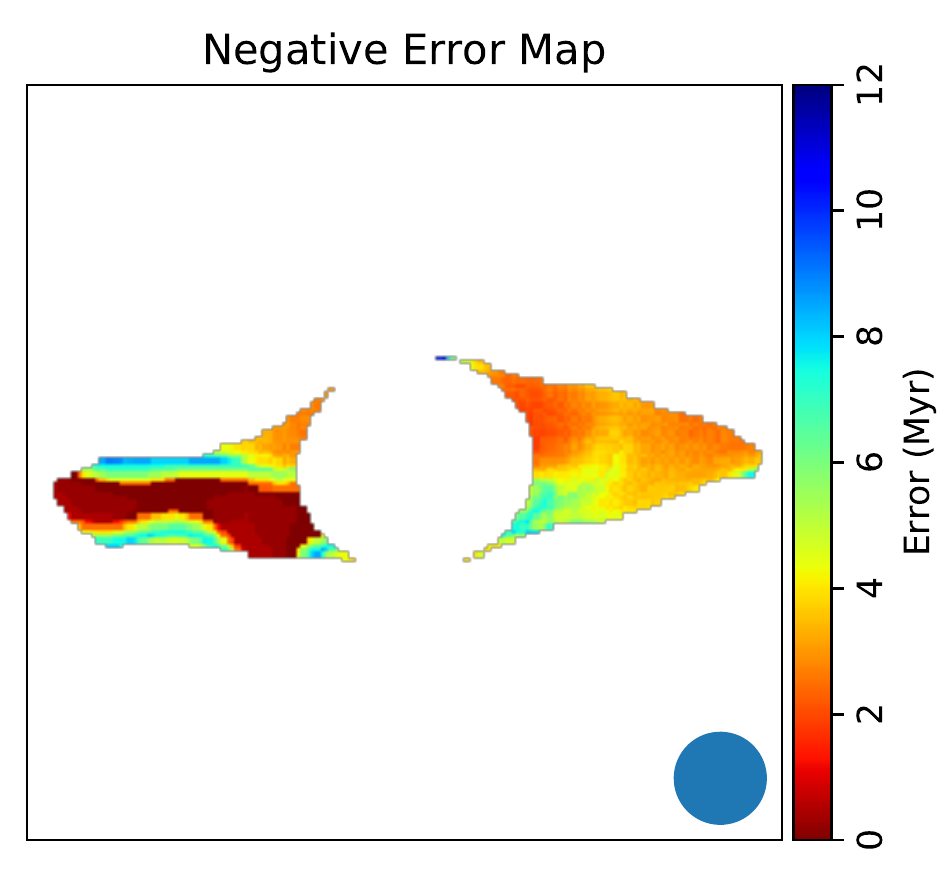}
\end{tabular}
\caption{Results of the spectral fitting for ILT J124627.85+520222.1. Top left shows the spectral age map and top right shows the reduced $\chi^2$ for the spectral age map. Bottom left and right images show the positive and negative errors for the spectral age map.}
\label{fig:BRATSResults-ILTJ124627}
\end{figure*}

\begin{figure*}
\centering
\begin{tabular}{cc}
\includegraphics[width=0.44\textwidth, trim = {110 25 50 10}, clip=true]{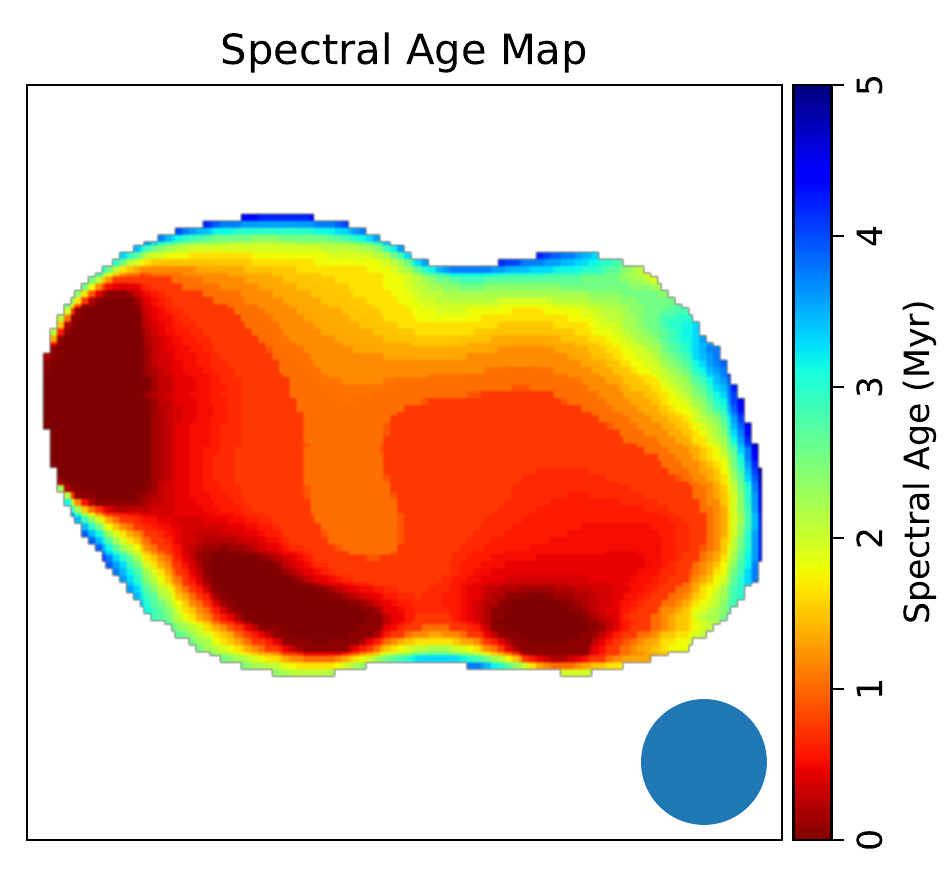} &
\includegraphics[width=0.44\textwidth, trim = {110 25 50 10}, clip=true]{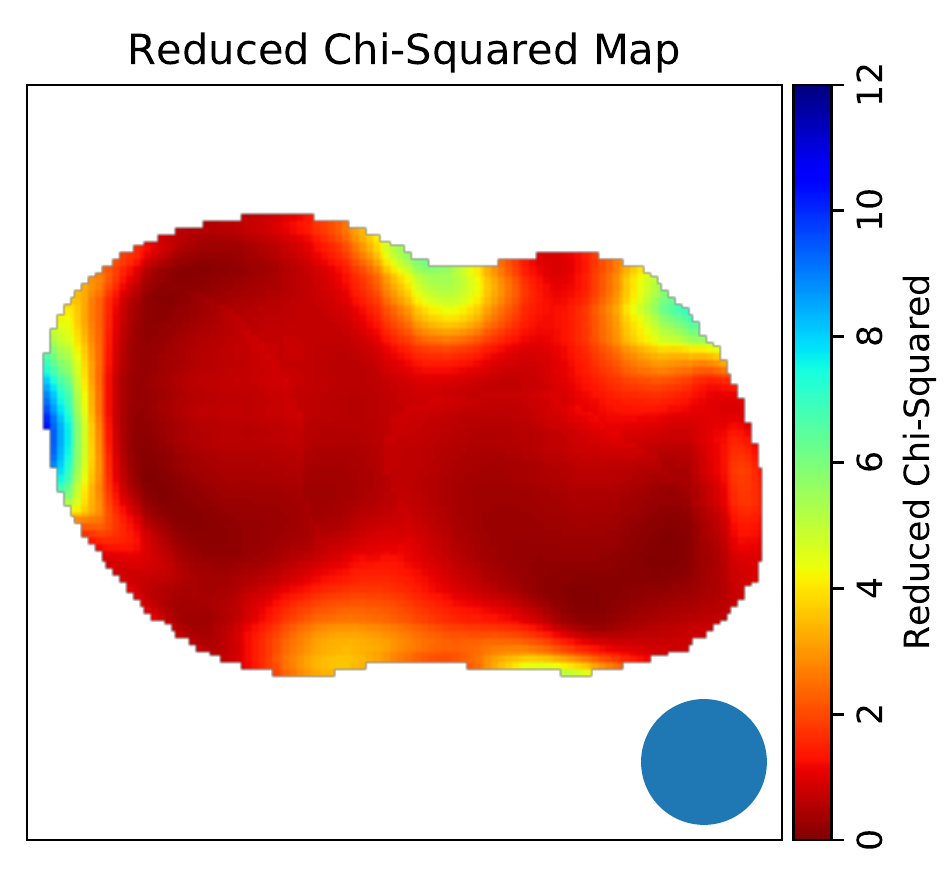} \\
\includegraphics[width=0.44\textwidth, trim = {110 25 50 10}, clip=true]{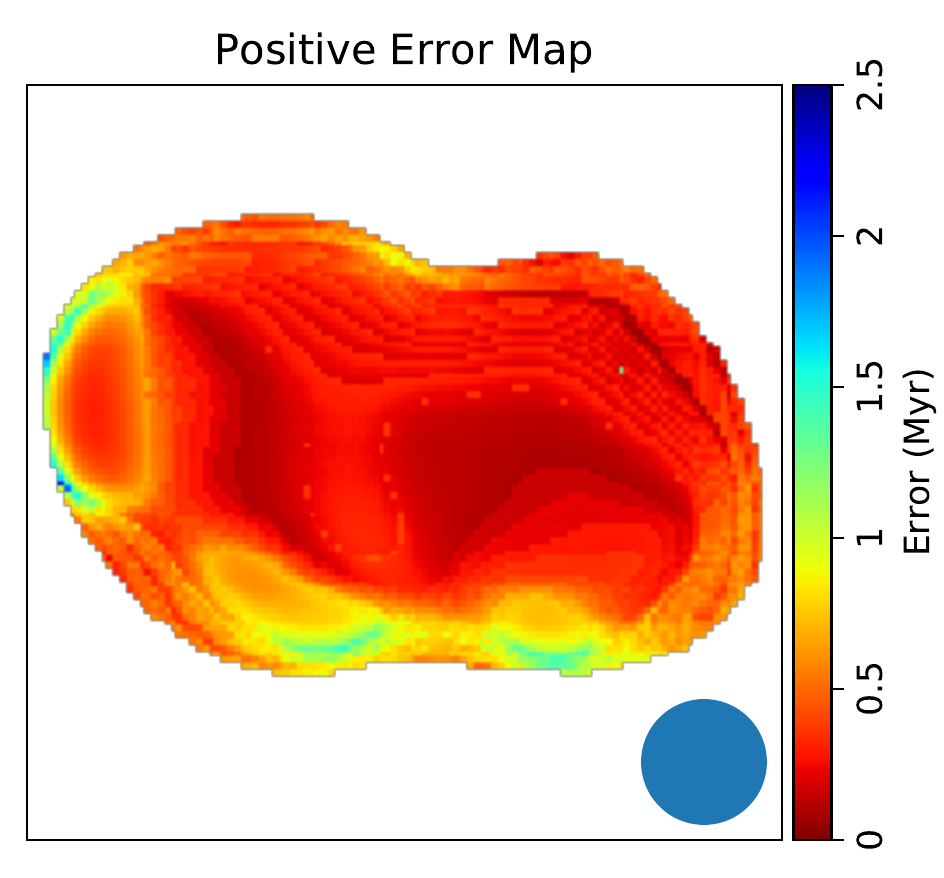} &
\includegraphics[width=0.44\textwidth, trim = {110 25 50 10}, clip=true]{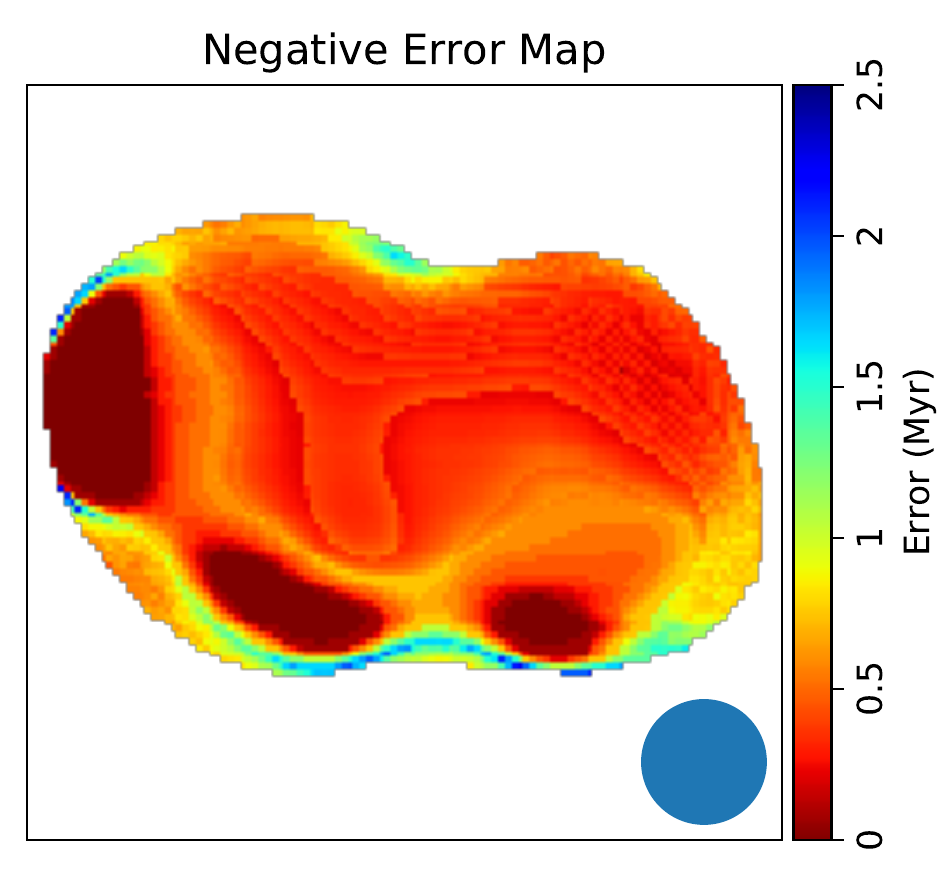}
\end{tabular}
\caption{Results of the spectral fitting for ILT J145604.90+472712.1. Top left shows the spectral age map and top right shows the reduced $\chi^2$ for the spectral age map. Bottom left and right images show the positive and negative errors for the spectral age map.}
\label{fig:BRATSResults-ILTJ145604}
\end{figure*}

The spectral ages shown in Table~\ref{tab:Spectral_Ages} and presented in Figures~\ref{fig:BRATSResults-ILTJ112543}-\ref{fig:BRATSResults-ILTJ145604} were calculated using \textsc{BRATS} with the magnetic field strengths and injection indices calculated as explained above. The FRIs in our sample are young sources with maximum ages between 5 and 20 Myr, whilst the FRIIs are older with ages between 40 and 60 Myr. We note that, for our FRIs, using equipartition field strengths (with no proton content) approximately doubles our age estimates. Similarly for the FRIIs, using equipartition field strengths approximately halves the ages listed.

None of the models fitted by \textsc{BRATS} can be rejected at even the $68$ per cent confidence level indicating that the models are a good fit to the data. The reduced $\chi^2$ maps along with average reduced $\chi^2$ values of about unity for three degrees of freedom in each case (see Table~\ref{tab:Spectral_Ages}) also confirm that the models are a good fit. However, some of the absolute errors are quite large. This is due to the large LOFAR errors and the lack of variation in the spectral index as a function of frequency observed for all of our sources (Figure~\ref{fig:TotalFluxes}) meaning that higher frequency data, where spectral curvature is most easily constrained, is needed to reduce the model errors.

\begin{table*}
\centering
\renewcommand{\arraystretch}{1.25}
\begin{tabular}{lcccccc}
    \hline
	\multicolumn{1}{c}{LOFAR}&Injection&$B_{eq}$&$B_{adj}$&Average&Minimum&Maximum\\[-2pt]
	\multicolumn{1}{c}{Source Name}&Index&/ T&/ T&$\chi^2_{\text{reduced}}$&Age / Myr&Age / Myr\\[1mm]
	\hline
	ILT J112543.06+553112.4$^a$&0.78&$1.21\times10^{-9}$&$4.84\times10^{-10}$&$0.80$& $0.00\substack{+11.58\\-0.00}$&$57.95\substack{+5.24\\-6.50}$\\
	ILT J120326.64+545201.5&0.49&$4.14\times10^{-10}$&$8.22\times10^{-10}$&$0.63$& $0.00\substack{+10.91\\-0.00}$&$20.05\substack{+4.72\\-6.19}$\\
	ILT J120645.20+484451.1&0.60&$8.42\times10^{-10}$&$1.67\times10^{-9}$&$0.87$& $4.47\substack{+1.05\\-1.27}$&$11.52\substack{+1.26\\-1.68}$\\
	ILT J121847.41+520148.4$^a$&0.77&$4.83\times10^{-10}$&$1.93\times10^{-10}$&$0.79$& $0.00\substack{+8.00\\-0.00}$&$41.98\substack{+8.42\\-11.98}$\\
	ILT J124627.85+520222.1&0.45&$6.00\times10^{-10}$&$1.19\times10^{-9}$&$1.07$& $0.00\substack{+6.44\\-0.00}$&$16.97\substack{+1.69\\-3.08}$\\
	ILT J145604.90+472712.1&0.54&$1.68\times10^{-9}$&$3.33\times10^{-9}$&$1.28$& $0.00\substack{+0.90\\-0.00}$&$4.81\substack{+0.48\\-0.54}$\\
	\hline
\end{tabular}
\caption{Summary of the results of the spectral age fitting. $B_{eq}$ is the equipartition magnetic field strength assuming no non-radiating particles, $B_{adj}$ are the adjusted magnetic field strengths used within this work to better represent the conditions of FRI/FRII galaxies (see text for details). Ages shown were calculated using the adjusted magnetic field strengths. $^a$ Host ID remains uncertain.}
\label{tab:Spectral_Ages}
\end{table*}

As expected, the regions of flatter spectral index are associated with the youngest spectral ages. Consequently, where errors are fairly well constrained, the gradients in spectral index noted in Section~\ref{sec:SpectralIndexMaps} are mirrored in the spectral ages.

\begin{figure}
    \centering
    \includegraphics[width=0.47\textwidth, , trim = {110 25 50 30}, clip=true]{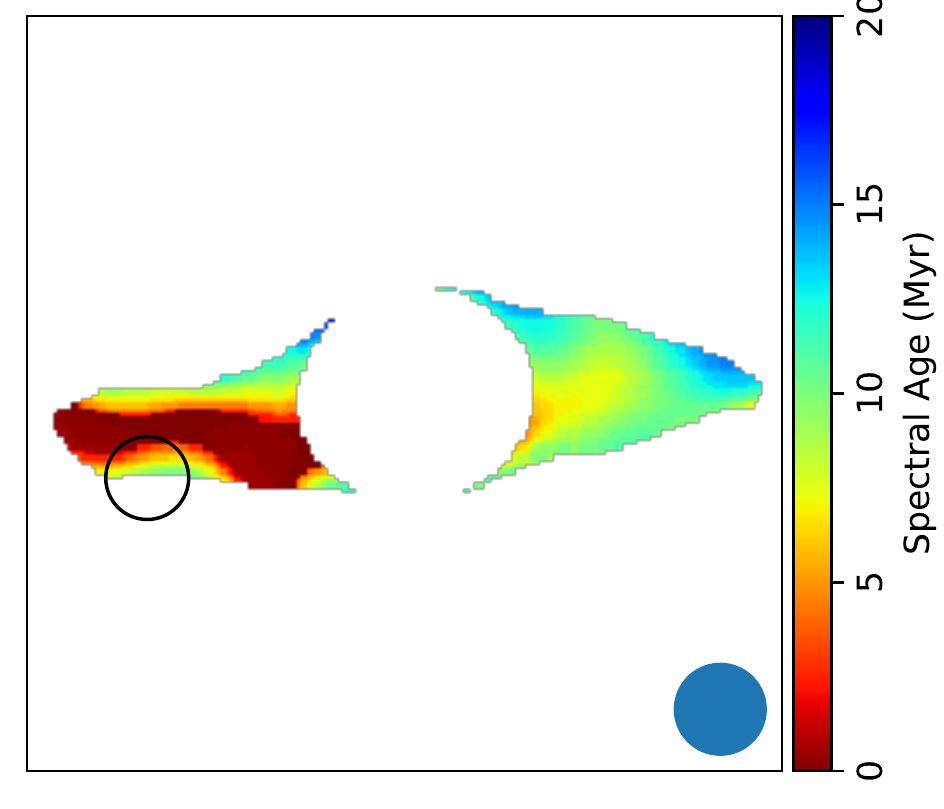}
    \caption{Spectral age map for ILT J124627.85+520222.1 with the position of the background galaxy on the eastern lobe marked by the black circle.}
    \label{fig:ILTJ124627_SpectralAgeing_BackgroundGal}
\end{figure}

Also noted in Section~\ref{sec:SpectralIndexMaps} were the anomalous regions of flat spectral indices in the eastern jet of ILT J124627.85+520222.1, which are now seen as an extended region of low spectral age (Figure~\ref{fig:BRATSResults-ILTJ124627}). This source has background galaxies visible in the optical image that overlap with the radio emission (Figure~\ref{fig:VLAImages}). Whilst there is no indication that these galaxies are emitting at radio frequencies, there is a prominent background galaxy midway along the eastern lobe that may be associated with the region of low age plasma observed. Overlaying the position of the background host on the spectral age map (Figure~\ref{fig:ILTJ124627_SpectralAgeing_BackgroundGal}), it can be seen that the galaxy lies between and slightly below the region of low spectral age. The astrometric uncertainties for LOFAR and the VLA are too small to account for the difference and so the background galaxy cannot be responsible for this low age plasma.

\citet{Harwood2015SpectralGalaxies} notes that for all FRIIs studied using \textsc{BRATS} anomalous non-physical regions showing zero age have been observed which may be due to performing spectral fitting at high resolutions on high dynamic range sources. Although not associated with FRII sources, the regions we observe in ILT J124627.85+520222.1 and at the tip of the southern lobe of ILT J120326.64+545201.5 may have a similar non-physical origin. We also note that these regions are associated with large positive errors and so may not be anomalous at all. However, we cannot discount the possibility that these regions are real and may be caused either by a region of enhanced particle acceleration within the jet and/or some interaction with the surrounding environment.

\section{Average Advance Speeds}
\label{sec:AverageAdvanceSpeeds}

We also estimated the average expansion speed for the lobes of each or our sources in order to see whether our sources could be expanding fast enough to shock the local environment. The standard methodology of \citet{Alexander1987AgeingSources} involves dividing the lobes into separate regions and estimating the speed based on the average age and separation distance of each region. However, the small size of our sources meant that this method resulted in between one and a maximum of three data points per source lobe. 

Many of our sources also show very little variation in spectral age across large portions of the lobes. This may be due to the resolution of our images with different electron populations being caught within a single beam or it may be due to physical mixing of electron populations within the sources themselves \citep[][]{Turner2018RAiSEAGN}. Regardless of the cause, the low variation in spectral age meant that applying the techniques of \citet{Alexander1987AgeingSources} resulted in unreliable estimates.

We therefore use the same methodology as \citet{Harwood2013SpectralData,Harwood2015SpectralGalaxies} where the lobe advance speed is found by dividing the source size by the age of the oldest plasma. For the FRI sources we assume that the oldest observed plasma was produced when the jet first became active and so we use the distance from the host to the edge of the observable emission as the source size. For the FRII-like sources we assume the oldest emission was produced at a time when the hot spot was located where the plasma is currently seen and so take the distance between the hot spot and the location of the oldest observed plasma as the source size. 

The results are shown in Table~\ref{tab:AverageJetSpeeds} where we also express the results in terms of the local sound speed, which we assume to be $400\text{ km s}^{-1}$ (the approximate speed of sound in the ISM at a temperature of 0.6 keV, typical of the sparse groups in which W21 found GSJ reside). For the FRIIs in our sample we find significantly sub-sonic advance speeds. For the FRIs in our sample the lobe advance speeds are supersonic. In comparison, using equipartition magnetic field strengths with no protons, this result is unchanged for the FRIIs where the lobe advance speeds become $0.2-0.5c_s$, whilst for the FRIs the advance speeds become mildly supersonic with speeds of $1.3-2.8c_s$.

\begin{table*}
    \centering
    \renewcommand{\arraystretch}{1.25}
    \begin{tabular}{lcccc}
        \hline
        \multicolumn{1}{c}{LOFAR}&Length of radio&Oldest plasma&\multicolumn{2}{c}{Lobe Advance Speeds}\\[-2pt]
        \multicolumn{1}{c}{Source Name}&emission/kpc&age/Myr&km/s&$c_s$\\
        \hline
        ILT J112543.06+553112.4(W)$^a$&$1.9$&$57.9^{+5.7}_{-6.5}$&$33^{+4}_{-3}$&$0.1^{+0.0}_{-0.0}$\\
        ILT J112543.06+553112.4(E)$^a$&$1.6$&$32.0^{+6.0}_{-7.5}$&$50^{+15}_{-8}$&$0.1^{+0.0}_{-0.0}$\\
        ILT J120326.64+545201.5(S)&$23.4$&$20.1^{+4.7}_{-5.7}$&$1143^{+456}_{-218}$&$2.9^{+1.1}_{-0.5}$\\
        ILT J120645.20+484451.1(W)&$15.9$&$11.5^{+3.2}_{-1.7}$&$1350^{+230}_{-295}$&$3.4^{+0.6}_{-0.7}$\\
        ILT J120645.20+484451.1(E)&$18.3$&$10.0^{+4.8}_{-0.6}$&$1787^{+114}_{-576}$&$4.5^{+0.3}_{-1.4}$\\
        ILT J121847.41+520128.4(N)$^a$&$3.7$&$34.0^{+7.4}_{-10.7}$&$108^{+49}_{-19}$&$0.3^{+0.1}_{-0.0}$\\
        ILT J121847.41+520128.4(S)$^a$&$2.9$&$42.0^{+8.4}_{-12.0}$&$67^{+27}_{-11}$&$0.2^{+0.1}_{-0.0}$\\
        ILT J124627.85+520222.1(W)&$25.7$&$17.0^{+1.7}_{-3.1}$&$1483^{+329}_{-135}$&$3.7^{+0.8}_{-0.3}$\\ 
        ILT J124627.85+520222.1(E)&$30.8$&$12.0^{+2.5}_{-3.9}$&$2510^{+1195}_{-433}$&$6.3^{+3.0}_{-1.1}$\\
        ILT J145604.90+472712.1(W)&$10.5$&$4.4^{+0.5}_{-0.4}$&$2361^{+257}_{-248}$&$5.9^{+0.6}_{-0.6}$\\
        ILT J145604.90+472712.1(E)&$12.9$&$4.8^{+0.5}_{-0.5}$&$2623^{+333}_{-237}$&$6.6^{+0.8}_{-0.6}$\\
        \hline
    \end{tabular}
    \caption{The oldest plasma ages and distances used to calculate the average lobe advance speeds for all sources analysed with \textsc{BRATS}. $^a$ Host ID remains uncertain.}
    \label{tab:AverageJetSpeeds}
\end{table*}

It is well-established that spectral ages are typically lower than the corresponding dynamical ages \citep[][]{Eilek1996HowNASA/ADS}, a result that has been confirmed using the high spectral resolution of \textsc{BRATS} \citep[][]{Harwood2013SpectralData}. Using the \textsc{BRATS} software Harwood et al. confirmed that dynamical ages can be up to 10 times more than the corresponding spectral age. It is therefore possible that our age estimates are too low so that our speed estimates should be considered as upper limits. 

However, \citet{Blundell2000TheLobes} note that for young radio sources with a dynamical age less than 10 Myr, spectral age fitting may be more accurate. If correct, this means that the ages, and hence lobe advance speeds, of the FRIs in our sample are, as a first-order estimate, reliable. For FRIIs, \citet{Mahatma2020InvestigatingGalaxies} found that the discrepancy between spectral and dynamic ages can be as low as a factor of $\sim2$ when using sub-equipartition magnetic fields (such as those used in this study). If correct this once again implies that the ages, and hence lobe advance speeds, of our FRIIs are reliable.

\section{Discussion}
\label{sec:Discussion}

\subsection{Remnant GSJ}
\label{sec:Discussion-Remnant}

ILT J112543.06+553112.4 and ILT J121847.41+520128.4, the two FRII spiral-hosted sources, have spectral indices towards the steeper end of the range of values found for the 3CRR sample \citep[][]{Laing1983BrightGalaxies}. Though not atypical of spectral indices found for larger AGN, recent studies have shown that these values are intermediate between sources typically classed as active and remnant AGN \citep[][]{Mahatma2018RemnantField}. Therefore, the possibility that these are remnant sources must be considered.

Like remnant sources, ILT J112543.06+553112.4 has no observable radio core. However, the lobes do not show the relaxed structure often seen in remnant sources and a bright hotspot is visible in the north west lobe with a fainter hotspot in the south east. Though the possibility of ongoing activity from a weak core cannot be excluded, it is possible that this source may have recently switched off so that the core is no longer visible but previously emitted material is still fuelling the hotspots \citep[][]{Tadhunter2016RadioEvolution}. If correct, and assuming the host has been correctly identified, the short distances from the core to the hotspots ($3.4$ and $7.2$ kpc to the North and South respectively) suggest this source may be an interesting example of a spiral AGN that has only very recently turned off.

For ILT J121847.41+520128.4, the $150$ MHz to $3$ GHz spectral index of the core is $0.72\pm0.09$ consistent with ongoing AGN activity \citep[e.g.][]{Laing1983BrightGalaxies,Sabater2019TheOn}. This, combined with the morphological evidence suggesting the core of this source is a mix of AGN and star-formation related emission (Section~\ref{sec:OpticalHosts-ILTJ121847}) means we do not believe this is a remnant source.

Unlike the two spiral-hosted sources, the extremely steep spectral index of ILT J122037.67+473857.6 means this is definitely a remnant source \citep[][]{Murgia2011DyingClusters,Mahatma2018RemnantField}  with the lack of detection in our 3 GHz VLA images lending tentative support to the idea that remnant sources fade quickly \citep[][]{Kaiser2002TheRelics,Mahatma2018RemnantField}. The small scale of the emission seen by LOFAR makes this the first known source with remnant emission of a similar size to the host. Whilst LOFAR is expected to be able to detect a large fraction of remnant sources, the outward diffusion of electrons means they are expected to be physically large \citep[e.g.][]{Hardcastle2016LOFAR/H-ATLAS:Field}, making the small size of this source unusual. Whilst we cannot directly measure the age of the source, the fact that this source never grew to large sizes does imply the radio jets were active for only a short period of time, consistent with expectations for low mass systems \citep[][]{Heckman2014TheUniverse}.

Like the two FRII spiral-hosted sources, ILT J122037.67+473857.6 does not have the morphologically relaxed appearance typical of remnant sources \citep[][]{Brienza2016LOFARRedshift,Morganti2017ArchaeologySpectrum}. Plotting this source on a BPT diagram shows ongoing AGN activity which, combined with the emission still having the appearance of being jet-related, means it is possible that this source has recently changed accretion mode, causing the radio jets to turn off.

\citet{Hardcastle2013NumericalEnvironments} showed that for large radio galaxies, half of the total energy transported by the jets would still be present in the lobes at the moment the jets turn off. Assuming a similar fraction applies to smaller sources this would mean that half of the energy transported by ILT J122037.67+473857.6 has already been transferred to its environment. Given the small physical size of the remnant this energy must have been transferred into its local environment supporting the conclusion of W21 that GSJ are capable of having a significant impact upon the evolution of their hosts.

\subsection{Source ages}

\citet{Parma1999RadiativeGalaxies} found the ages for a representative sample of 42 sources drawn from the B2 catalogue with $1.4\text{ GHz}$ luminosities between $10^{23}$ and $10^{25}\text{ W Hz}^{-1}$. These sources are predominantly FRI galaxies, though several FRII galaxies are also present. They found spectral ages ranging from $4$ to $112$ Myr with most being a few tens of Myrs old. The FRIs in our sample have maximum ages between about $5$ and $20$ Myr, consistent with the smaller, least luminous galaxies within the Parma et al. sample which are also comparable in size and luminosity to the largest most luminous GSJ. Our FRI sources can be considered as young radio galaxies. These sources are comparable in age to the 10 kpc GSJ, NGC 3801, that was found to have a spectral age between $1.8$ and $2.4$ Myr \citep[][]{Heesen2014ThePopulation}.

In contrast our FRIIs are older, both having maximum ages between $40$ and $60$ Myr. All our sources are older than the typical age of the compact CSS/GPS sources which typically have ages up to a few thousand years \citep[][]{ODea2021CompactSources}. The size and age of our sources are therefore consistent with their being located along an evolutionary path joining CSS/GPS sources and larger radio galaxies. Overall, our sources are consistent with the trend seen by \citet{Parma1999RadiativeGalaxies} suggesting that for sources of similar luminosity, larger sources will typically have older spectral ages.

\subsection{Lobe expansion speeds}

The FRII objects in our sample, ILT J112543.06+553112.4 and ILT J121847.41+520128.4, have substantially slower growth rates than the FRI objects in our sample. The growth rates of our FRII galaxies are lower than those recently found by \citet{Ineson2017AGalaxies} who, using a representative sample of 37 FRII galaxies, found only two sources with an expansion speed below Mach 1, both of which were approximately $0.7$. Whilst the sample of Ineson et al. is more luminous, and hence more powerful than ours, we note that in their study they found no relation between radio luminosity and advance speeds. It remains possible that the slow advance speeds of these two objects are related to the spiral nature of the hosts, however, these results continue to suggest that these hosts may have been misidentified and that the true source is at a higher redshift.

The majority of the sources studied by \citet{Parma1999RadiativeGalaxies} are FRIs and have lobe advance speeds between 500 and 5000 km s$^{-1}$. The advance speeds of our FRI sources are therefore similar to the slower sources in the Parma et al. sample. Previously observed GSJ have speeds between about Mach 3 and 5, which is again consistent with our results \citep[][]{Croston2007Shock3801,Croston2009High-energyA,Mingo2011MarkarianSeyfert,Mingo2012ShocksGalaxy}.

Though, as noted earlier, our speed estimates should be considered as upper limits, it appears that the FRIs in our sample are capable of driving strong shocks ($\cal{M}$ > 2). It is possible that some of the zero-age emission seen in the lobes of some of our sources is caused by shock fronts re-accelerating plasma, though further observations are needed to confirm this. 

The possibility of strong shocks means that the energy outputs estimated in W21, which assumed adiabatic expansion, should be considered as lower limits. Since W21 show that GSJ lobes potentially contain enough energy to have a significant effect on the host, the impact of these low-luminosity sources upon the evolution of their hosts could be substantial.

\section{Summary and Conclusions}
\label{sec:Conclusions}

Of the eight sources we considered to be GSJ based on the LoTSS DR2 images, our S-band VLA images have confirmed the hosts of four and rejected one. Of the remaining three, the hosts of two remain doubtful whilst one appears to be a remnant source. This clearly highlights the difficulty in identifying such small sources and emphasises the need for high-resolution data to unequivocally identify GSJ.

For our sample of GSJ we find that:
\begin{itemize}
    \item The existence of at least one galaxy-scale remnant shows that some sources never grow beyond the GSJ stage. The lobes of these sources will eventually transfer all of their energy into the host environment.
    \item GSJ have ages up to $60$ Myr with the majority between 5 and $20$ Myr. This is consistent with the evolutionary development of compact sources into larger radio galaxies.
    \item GSJ typically have lobe advance speeds a few times the local sound speed, with most predicted to be driving strong shocks into their surrounding environment.
    \item The energy stored in the lobes of GSJ is capable of having a significant effect upon the host's evolution. Our sources are therefore similar to those in the numerical simulations of \citep[][]{Mukherjee2016RelativisticDynamics,Mukherjee2018The5063} who found low-power jets are capable of affecting the host's ISM over large areas.
    \item The majority of our sources show little variation in spectral age, most likely dominated by the resolution of our images capturing different electron populations within a single beam, though physical mixing of electron population within the source cannot be ruled out.
    \item GSJ have an integrated spectral index similar to those of larger radio galaxies with no sign of a spectral turnover at frequencies above 150 MHz making our sources distinct from the GPS sources and meaning their turnover is at lower frequencies than all but the largest CSS sources.
    \item The existence of larger radio galaxies show that some GSJ must evolve beyond this stage. The youngest, most luminous source in our sample, ILT J145604.90+472712.1, has some of the fastest expansion speeds and may be evolving into a larger, FRI/FRII-type source.
\end{itemize}

This work has highlighted the potential impact of young, low-luminosity sources upon galactic evolution. Future releases of LoTSS are expected to produce larger samples of GSJ that, coupled with high-resolution, multi-wavelength data, will provide a better understanding of the full extent of the impact of these source on their hosts.

The GSJ sample of W21 did not specifically search for remnant GSJ. The presence of at least one galaxy-scale remnant within our VLA sample suggests that more will be observed by the LoTSS survey. Future releases of LoTSS should be used to find a larger selection of these interesting objects and determine their prevalence.

\section*{Acknowledgements}

We thank both the editor, Tim Pearson, and the anonymous referee for their helpful comments.

BW acknowledges a studentship from the UK Science and Technology Facilities Council (STFC). JHC and BM acknowledge support from the UK Science and Technology Facilities Council (STFC) under grants ST/R00109X/1 and ST/R000794/1 and ST/T000295/1. MJH acknowledges support from the UK Science and Technology Facilities Council under grant ST/R000905/1.

LOFAR is the Low Frequency Array designed and constructed by ASTRON. It has observing, data processing and data storage facilities in several countries, which are owned by various parties (each with their own funding sources), and which are collectively operated by the International LOFAR Telescope (ILT) foundation under a joint scientific policy. The ILT resources have benefitted from the following recent major funding sources: CNRS-INSU, Observatoire de Paris and Universit{\'e} d'Orl{\'e}ans, France; BMBF, MIWF-NRW, MPG, Germany; Science Foundation Ireland (SFI), Department of Business, Enterprise and Innovation (DBEI), Ireland; NWO, The Netherlands; the Science and Technology Facilities Council, UK; Ministry of Science and Higher Education, Poland; the Istituto Nazionale di Astrofisica (INAF), Italy.

This research has made use of the University of Hertfordshire high-performance computing facility and the LOFAR-UK computing facility located at the University of Hertfordshire and supported by STFC [ST/P000096/1].

The National Radio Astronomy Observatory is a facility of the National Science Foundation operated under cooperative agreement by Associated Universities, Inc.

\section*{Data Availability}

The data underlying this article are available from the NRAO data archive, \url{https://data.nrao.edu/portal/#/}, under project code 18B-083, the NRAO VLA Sky Survey at \url{https://www.cv.nrao.edu/nvss/} and the Westerbork Northern Sky Survey at \url{http://cdsarc.u-strasbg.fr/viz-bin/cat/VIII/62}. LoTSS DR2 data will be made available on the LOFAR Surveys website at \url{https://lofar-surveys.org}.




\bibliographystyle{mnras}
\bibliography{Mendeley} 





\bsp	
\label{lastpage}
\end{document}